%% file: 41956corr.tex
\documentclass[traditabstract]{aa} 
\usepackage{graphicx}
\usepackage{grffile}
\usepackage{siunitx}
\usepackage{txfonts}
\usepackage{graphicx}
\usepackage{caption}
\usepackage{natbib}
\usepackage{subfigure} 
\usepackage{hyperref}
\usepackage[utf8]{inputenc}
\usepackage{placeins}
\usepackage{scrextend}
\usepackage{rotating} 
\usepackage{multirow} 
\usepackage{booktabs} 

\input{Time_delay_measurement_macro}

\begin{document}

   \title{HOLISMOKES - VII. Time-delay measurement of strongly lensed Type Ia supernovae using machine learning\thanks{\url{https://github.com/shsuyu/HOLISMOKES-public/tree/main/HOLISMOKES_VII}}}

  \titlerunning{Time-delay measurement of strongly lensed SNe Ia}

   \author{S. Huber\inst{1,2}
                 \and    
          S. H. Suyu\inst{1,2,3}
                    \and
          D. Ghoshdastidar \inst{4}
                                \and
          S. Taubenberger\inst{1}
                  \and                   
             V. Bonvin\inst{5}
             \and
             J. H. H. Chan\inst{5}
             \and 
             M. Kromer\inst{6}
             \and
                        U.~M.~Noebauer \inst{1,7}  
             \and
             S. A. Sim\inst{8}
             \and
             L. Leal-Taix\'e \inst{4}
          }

   \institute{Max-Planck-Institut f\"ur Astrophysik, Karl-Schwarzschild Str. 1, 85741 Garching, Germany\\
              \email{shuber@MPA-Garching.MPG.DE}
         \and 
           Technische Universit\"at M\"unchen, Physik-Department, James-Franck-Stra\ss{}e~1, 85748 Garching, Germany
           \and
           Institute of Astronomy and Astrophysics, Academia Sinica, 11F of ASMAB, No.1, Section 4, Roosevelt Road, Taipei 10617, Taiwan
           \and
           Informatik-Department, Technische Universit\"at M\"unchen,                   
           Boltzmannstr. 3, 85748 Garching, Germany
                        \and
           Institute of Physics, Laboratory of Astrophysics, Ecole Polytechnique F\'ed\'erale de Lausanne (EPFL), Observatoire de Sauverny, 1290 Versoix, Switzerland
                        \and
                        Heidelberger Institut f\"ur Theoretische Studien,
Schloss-Wolfsbrunnenweg 35, 69118 Heidelberg, Germany
           \and
           Munich Re, IT1.6.1.1, Königinstraße 107, 80802 München, Germany
                        \and
                        Astrophysics Research Centre, School of Mathematics and Physics, Queen’s University Belfast, Belfast BT7 1NN, UK
}

   \date{Received --; accepted --}

 
  \abstract
{The Hubble constant ($H_0$) is one of the fundamental parameters in
cosmology, but there is a heated debate around the $>$4$\sigma$ tension between the local Cepheid distance ladder and the early Universe measurements. 
Strongly lensed Type Ia supernovae (LSNe Ia) are an independent and direct way to 
measure $H_0$, where a time-delay measurement between the multiple supernova (SN) 
images is required.  
In this work, we present two machine learning approaches for measuring time delays in LSNe Ia, namely, a fully connected neural network (FCNN) and a random forest (RF).
For the training of the FCNN and the RF, we simulate mock LSNe Ia from theoretical SN Ia models that include observational noise and microlensing.  We test the generalizability of the machine learning models by using a final test set based on empirical LSN Ia light curves not used in the training process, and we find that only the RF provides a low enough bias to achieve precision cosmology; as such, RF is therefore
preferred over our FCNN approach for applications to real systems. 
 For the RF with single-band photometry in the $i$ band, we obtain an accuracy better than 1\% in all investigated cases for time delays longer than 15 days, assuming follow-up observations with a 5$\sigma$ point-source depth of 24.7, a two day cadence with a few random gaps, and a detection of the LSNe Ia 8 to 10 days before peak in the observer frame. In terms of precision, we can achieve an approximately 1.5-day uncertainty for a typical source redshift of $\sim$0.8 on the $i$ band under the same assumptions. To improve the measurement, we find that using three bands, where we train a RF for each band separately and combine them afterward, helps to reduce the uncertainty to $\sim$%
1.0 day. The dominant source of uncertainty is the observational noise, and therefore the depth is an especially important factor when follow-up observations are triggered. We have publicly released the microlensed spectra and light curves used in this work.
}
   {}
   {}
   {}
   {}

   \keywords{Gravitational lensing: strong, micro - supernovae: Type Ia supernova - cosmology: distance scale}

   \maketitle
%

\section{Introduction}

The Hubble constant, $H_0$, is one of the fundamental parameters in
cosmology, but there is a tension\footnote{\url{https://github.com/shsuyu/H0LiCOW-public/tree/master/H0_tension_plots} \citep{bonvin_2020}}
 of $>$4$\sigma$ \citep{Verde:2019ivm} between the early Universe measurements inferred
from the cosmic microwave background \citep[CMB;][]{Planck:2018vks} and late
Universe measurements from the Supernova $H_0$ for the
Equation of State (SH0ES) project
\citep[e.g.,][]{Riess:2016jrr,Riess:2018byc,Riess:2019cxk,Riess:2020fzl}.\ However,
results from \cite{Freedman:2019jwv,Freedman_2020} using the tip of the red giant branch (TRGB) or from \cite{Khetan:2020hmh} using surface brightness fluctuations (SBFs) are consistent with both. An independent analysis using the TRGBs by \cite{Anand:2021sum} has derived a slightly higher $H_0$ value, bringing it closer to the results of the SH0ES project, which is based on Cepheids.  Moreover, recent results from \cite{Blakeslee:2021rqi} using SBFs that are calibrated through both Cepheids and TRGBs are in good agreement with the SH0ES project and $\sim$%
$2\sigma$ higher than the CMB values.
As an alternative to the distance ladder approach, \cite{Pesce:2020xfe} measured $H_0$ from the Megamaser Cosmology Project, which also agrees well with the SH0ES results and is about $\sim$%
2$\sigma$ higher than the Planck value. Gravitational wave sources acting as standard sirens also provide direct luminosity distances and thus $H_0$ measurements \citep[e.g.,][]{LIGOScientific:2017adf}.\ While this is a promising approach, current uncertainties on $H_0$ from standard sirens preclude them from being used to discern between the SH0ES and the CMB results.

Lensing time-delay cosmography, as an independent probe, can address
this tension by measuring $H_0$ in a single step. This method, first
envisaged by \cite{Refsdal:1964}, combines the measured time delay
from the multiple images of a variable source with lens mass modeling
and line-of-sight mass structure to infer $H_0$. The COSmological MOnitoring
of GRAvItational Lenses \citep[COSMOGRAIL;][]{2017Courbin} and H0 Lenses in COSMOGRAIL's Wellspring \citep[H0LiCOW;][]{Suyu:2016qxx}
collaborations, together with the Strong lensing at High Angular Resolution Program (SHARP)
\citep{Chen:2019ejq}, have successfully applied this method to lensed
quasar systems
\citep[e.g.,][]{Bonvin:2018dcc,Birrer:2018vtm,2019MNRAS.490..613S,Rusu:2019xrq,Chen:2019ejq}.
The latest $H_0$ measurement from H0LiCOW using physically motivated
mass models is consistent with measurements from SH0ES but is in
$>$3$\sigma$ tension with results from the CMB
\citep{Wong:2019kwg}. The STRong-lensing Insights into the Dark Energy
Survey (STRIDES) collaboration has further analyzed a
new lensed quasar system \citep{Shajib:2019toy}. The newly formed
 Time-Delay COSMOgraph (TDCOSMO) organization \citep{Millon:2019slk}, consisting of H0LiCOW,
COSMOGRAIL, SHARP and STRIDES, has recently considered a one-parameter
extension to the mass model to allow for the mass-sheet transformation
\citep[e.g.,][]{Falco:1985,Schneider:2013a, Kochanek:2020}. \citet{Birrer+2020} used
the stellar kinematics to constrain this single parameter, resulting
in an $H_0$ value with a larger uncertainty, which is statistically consistent with the previous results using physically motivated mass models. In addition to placing constraints on $H_0$, strongly lensed quasars also provide tests of the cosmological principle, especially of spatial isotropy, given the independent sight line and distance measurement that each lensed quasar yields \citep[e.g.,][]{Krishnan+2021a, Krishnan+2021b}.

In addition to strongly lensed quasars, supernovae (SNe) lensed into
multiple images are promising as a cosmological probe and are in fact
the sources envisioned by \citet{Refsdal:1964}. Even though these
systems are much rarer in comparison to quasars, they have the
advantage that SNe fade away over time, facilitating measurements of
stellar kinematics of the lens galaxy
\citep{Barnabe2011,2017:Yildirim,Shajib:2018,Yildirim:2019vlv} and surface brightness distributions of the lensed-SN host galaxy \citep{DingEtal21} to
break model degeneracies, for example, the mass-sheet transformation
\citep{Falco:1985,Schneider:2013wga}. Furthermore, strongly
lensed type Ia supernovae (LSNe Ia) are promising given that they are
standardizable candles and therefore provide an additional way to
break model degeneracies for lens systems where lensing
magnifications are well characterized
\citep{2003MNRAS.338L..25O,Foxley-Marrable:2018dzu}.

So far, only three LSNe with resolved multiple images have been observed,
namely SN ``Refsdal" \citep{Kelly:2015xvu,Kelly:2015vjq}, a
core-collapse SN at a redshift of $z = 1.49$, the LSN Ia iPTF16geu
\citep{Goobar:2016uuf} at $z=0.409$, and AT2016jka \citep{Rodney:2021keu}
at $z = 1.95$, which is most likely a LSN Ia.  Nonetheless, with the upcoming Rubin
Observatory Legacy Survey of Space and Time
\citep[LSST;][]{Ivezic:2008fe}, we expect to find $\sim$%
$10^3$ LSNe, of
which 500 to 900 are expected to be type Ia SNe
\citep{Quimby:2014,GoldsteinNugent:2017,Goldstein:2017bny,Wojtak:2019hsc}.
Considering only LSNe Ia with spatially resolved images and peak
brightnesses\footnote{of the fainter image for a double system; for a
quad system, the peak brightness of the third brightest image is considered.} brighter than 22.6 in the
$i$ band, as in the \citet[][hereafter OM10]{Oguri:2010} lens catalog, leads to 40
to 100 LSNe Ia, depending on the LSST observing strategy, of which 10
to 25 systems yield accurate time-delay measurements
\citep{Huber:2019ljb}.

To measure time delays between multiple images of LSNe Ia,
\cite{Huber:2019ljb} used the free-knot spline estimator from Python
Curve Shifting \citep[\texttt{PyCS};][]{2013:Tewesb,Bonvin:2015jia},
and therefore the characteristic light-curve shape of a SN Ia is not
taken into account. Furthermore, they do not explicitly model the
variability due to microlensing
\citep{Chang_Refsdal_1979,Irwin:1989,Wambsganss:2006nj,2016aagl.book.....M},
an effect where each SN image is separately influenced by lensing
effects from stars in the lens, leading to the additional magnification
and distortion of light curves
\citep{Yahalomi:2017ihe,Goldstein:2017bny,Foxley-Marrable:2018dzu,Huber:2019ljb,
  PierelRodney+2019,Huber:2020dxc}.
While PyCS has the advantage of being flexible without making assumptions on the light-curve forms, model-based methods are complementary in providing additional information to measure the time delays more precisely.

One such model-based time-delay measurement method was implemented by
\cite{PierelRodney+2019}, where template SN light curves are
used. Even though microlensing is taken into account in this work, it
is done in the same way for each filter. A more realistic microlensing
treatment for SNe Ia, with variations in the SN intensity distribution across wavelengths, was first introduced by \cite{Goldstein:2017bny}
using specific intensity profiles from the theoretical W7 model \citep{1984:Nomoto} calculated via the radiative transfer code
\texttt{SEDONA} \citep{Kasen:2006ce}. \cite{Huber:2019ljb,
  Huber:2020dxc} have built upon this study, but using the radiative
transfer code \texttt{ARTIS} \citep{Kromer:2009ce} to calculate
synthetic observables for up to four theoretical SN explosion
models. In this work, we follow the approach of \cite{Huber:2019ljb,
  Huber:2020dxc} to calculate realistic microlensed light curves for
LSNe Ia to train a fully connected neural network (FCNN) and a random forest (RF) model
for measuring time delays. In addition, this method also allows us to
identify dominant sources of uncertainties and quantify different follow-up strategies.

This paper is organized as follows. In Sect. \ref{sec: simulated
  light curves for LSNe Ia} we present our calculation of mock light
curves, which includes microlensing and observational uncertainties. The
creation of our training, validation, and test sets is explained with an
example mock observation in Sect. \ref{sec: Example data used for machine learning}, 
followed by an introduction to the machine learning (ML)
techniques used in this work in Sect. \ref{sec: Machine learning
  techniques}.  We apply these methods to the example mock
observation in Sect. \ref{sec: Machine learning on example
  mock-observation}, where we also test the generalizability by using empirical LSN Ia light curves not used in the training process.
  In Sect. \ref{sec: Microlensing, observational noise and choice of filters} we investigate, based on our example mock observation, potential filters for follow-up observations and the impact of microlensing and noise on the uncertainty, before we investigate more mock observations in Sect. \ref{sec: Machine learning on further
  mock-observation}. 
  We discuss our results in Sect. \ref{sec: Discussion}
  before summarizing in Sect. \ref{sec: Summary}.
  Magnitudes in this paper are in the AB system. 
  
  We have publicly released the microlensed spectra and light curves used in this work at \url{https://github.com/shsuyu/HOLISMOKES-public/tree/main/HOLISMOKES_VII}.

\section{Simulated light curves for LSNe Ia}
\label{sec: simulated light curves for LSNe Ia}

The goal is to develop a software that takes photometric light-curve observations of a
LSN Ia as input and predicts as an output the time delay between the
different images. For a ML approach, we need to
simulate a realistic data set where we account for different sources
of uncertainties. We therefore specify in Sect. \ref{sec:
  microlensing} our calculation of microlensing, and we explain in Sect.
\ref{sec: observational uncertainty} our determination of
observational uncertainties including estimates of the moon phase.

\subsection{Microlensing and SN Ia models}
\label{sec: microlensing}

To calculate light curves for a LSN Ia with microlensing, we combine
magnifications maps from {\tt GERLUMPH} \citep{Vernardos:2014lna,Vernardos:2014yva,Vernardos:2015wta} with
theoretical SN Ia models, where synthetic observables have been
calculated with \texttt{ARTIS} \citep{Kromer:2009ce}. The basic idea
is to place a SN in a magnification map and solve for the observed flux:

\begin{equation}
F_{\lambda,\mathrm{o}}(t)=\frac{1}{\lum^2(1+\sourcez)}\int \dd x \int \dd y \, I_{\lambda,\mathrm{e}}(t,p(x,y)) \, \mu(x,y),
\label{eq: microlensed flux}
\end{equation}
where $\lum$ is the luminosity distance to the source, $\sourcez$ is
the redshift of the source, $\mu(x,y)$ is the magnification factor
depending on the positions $(x,y)$ in the magnification map, and
$I_{\lambda,\mathrm{e}}(t,p)$ is the emitted specific intensity at the source
plane as a function of wavelength, $\lambda$, time since explosion, $t$,
and impact parameter, $p$ (i.e., the projected distance from the ejecta
center, where we assume spherical symmetry similar to \citealt{Huber:2019ljb, Huber:2020dxc}). Lensing magnification maps depend on three main parameters, namely the
convergence $\kappa$, the shear $\gamma$ and the smooth matter
fraction $s=1-\kappa_*/\kappa$, where $\kappa_*$ is the convergence of
the stellar component. Further, our maps have a resolution of 20000
$\times$ 20000 pixels with a total size of 20 $\Rein$ $\times$ 20 $
\Rein$, where the Einstein radius, $\Rein$, is a characteristic size of the map
that depends on the source redshift $\sourcez$, lens redshift
$\lensz$, and masses of the microlenses. As in \cite{Huber:2020dxc}, we follow \citet{ChanEtal21} for generating the microlensing magnification maps and assume a Salpeter initial mass
function (IMF) with a mean mass of the microlenses of $0.35 M_\odot$; the specifics of the assumed IMF 
have negligible impact on our studies. From the flux we obtain the AB magnitudes via
\begin{equation}
\scalebox{1.13}{$
m_\mathrm{AB,X}(t_i) = -2.5 \log_{10} \left(\frac{\int \dd\lambda \, \lambda S_\mathrm{X}(\lambda) \, F_{\lambda,\mathrm{o}}(t)  }{\int \dd \lambda \, S_\mathrm{X}(\lambda) \, c \, / \lambda} \times \si{\square\cm\over\erg} \right)  - 48.6$}
\label{eq: microlensed light curves for ab magnitudes}
\end{equation}
\citep{Bessel:2012}, where $c$ is the speed of light and
$S_\mathrm{X}(\lambda)$ is the transmission function for the filter X
(that can be \textit{u}, \textit{g}, \textit{r}, \textit{i},
\textit{z}, \textit{y}, \textit{J}, or \textit{H} in this work). This
calculation is discussed in much greater detail by
\cite{Huber:2019ljb}, which was initially motivated by the work of
\cite{Goldstein:2017bny}.

The calculation of microlensing of LSNe Ia requires a theoretical
model for the SN Ia that predicts the specific intensity. To increase
the variety of different light-curve shapes we use four SNe Ia
models computed with \texttt{ARTIS} \citep{Kromer:2009ce}. These models have also been used in
\cite{Suyu:2020opl} and \cite{Huber:2020dxc}, and are briefly
summarized in the following: i) the parameterized 1D deflagration model
W7 \citep{1984:Nomoto} with a Chandrasekhar mass ($M_\mathrm{Ch}$) carbon-oxygen (CO) white
dwarf (WD), ii) the delayed detonation model N100
\citep{Seitenzahl:2013} of a $M_\mathrm{Ch}$ CO WD, iii) a
sub-Chandrasekhar (sub-Ch) detonation model of a CO
WD with $1.06 M_\odot$ \citep{Sim:2010}, and iv) a merger model of
two CO WDs of $0.9 M_\odot$ and $1.1 M_\odot$
\citep{Pakmor:2012}.

\begin{figure}
\centering
\subfigure{\includegraphics[width=0.45\textwidth]{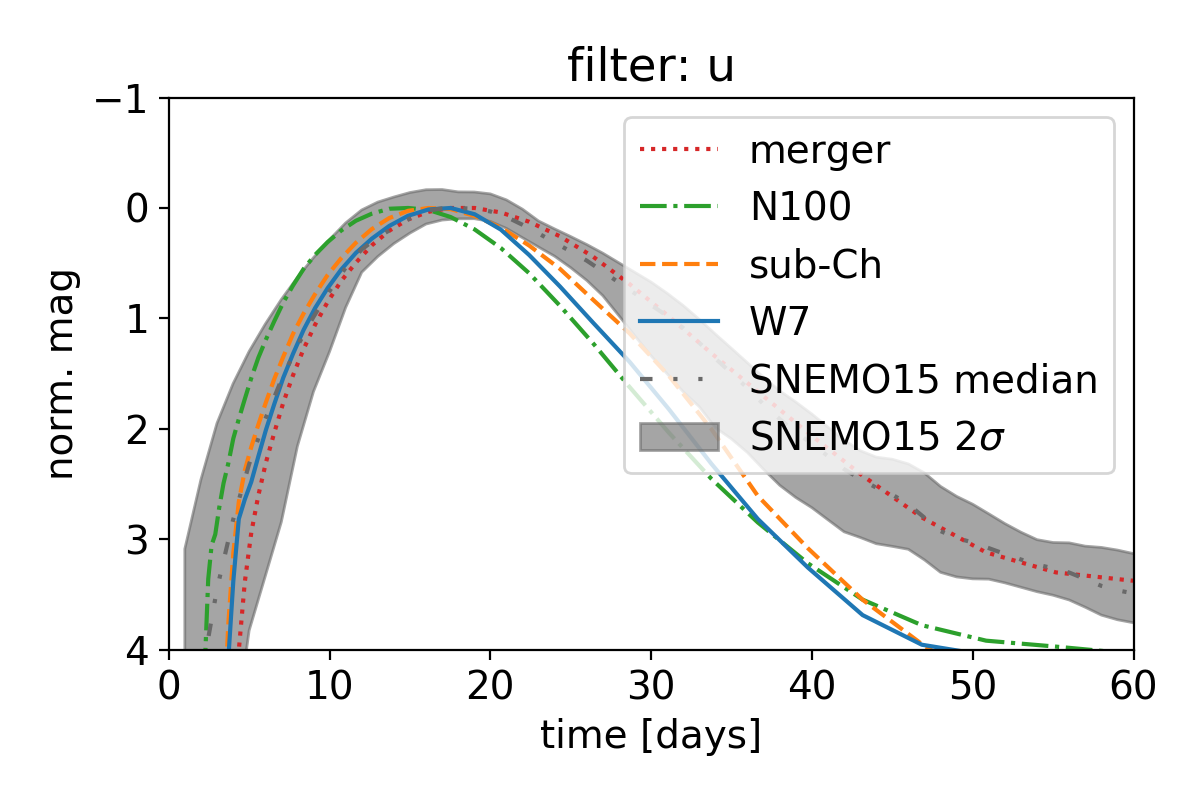}}
\subfigure{\includegraphics[width=0.45\textwidth]{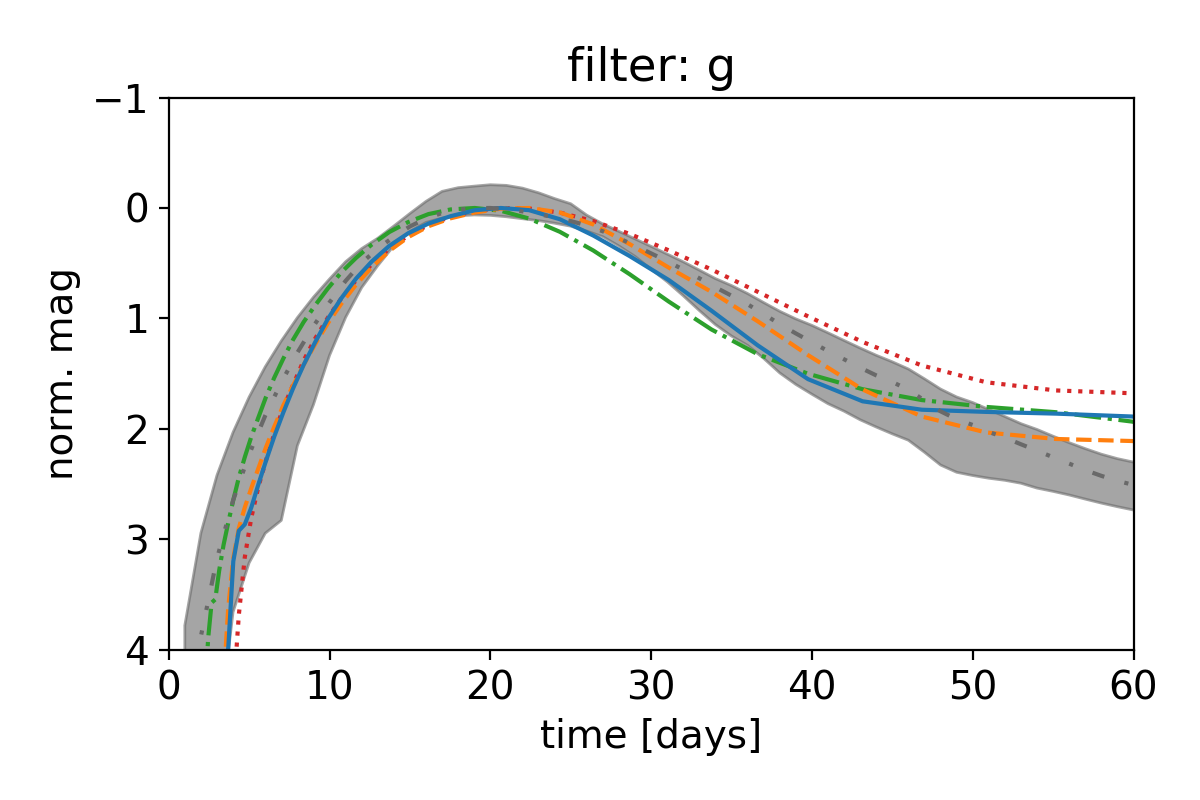}}
\caption{Normalized LSST $u$- and $g$-band rest-frame light curves for four theoretical SN Ia models (merger, N100, sub-Ch, and W7) in comparison to the empirical model \texttt{SNEMO15}.}
\label{fig: SN Ia models vs SNEMO}
\end{figure}

Figure \ref{fig: SN Ia models vs SNEMO} shows the light curves for the
four SN Ia models in comparison to the empirical \texttt{SNEMO15} model
\citep{Saunders:2018rjn}. The light curves are normalized by the peak.  
Magnitude
differences between SN Ia models are within 1 magnitude. To produce
the median and $2\sigma$ (97.5th percentile $-$ 2.5th percentile)
light curves of \texttt{SNEMO15}, we consider all 171 SNe Ia from
\cite{Saunders:2018rjn}. Data of the empirical models cover only $
\SI{3305}{\angstrom}$ to $\SI{8586}{\angstrom}$ and therefore the
\textit{u} band, starting at $\SI{3200}{\angstrom}$, is only an
approximation, but an accurate one since the filter transmission in
the missing region is low. The rest-frame $u$ and $g$ cover
approximately the observed $r$, $i$, and $z$ bands for a system with redshift of
0.76, which we investigate in Sects. \ref{sec: Example data used for
  machine learning} and \ref{sec: Machine learning on example
  mock-observation}. Light curves from theoretical and empirical models
show the same evolution, although there are quite some differences in
the shapes. The variety of different theoretical models is helpful to
encapsulate the intrinsic variation of real SNe Ia.
In building our training, validation and test
sets for our ML methods, we also normalize the light curves after the calculation of the observational noise, which we describe next.

\subsection{Observational uncertainty and the moon phase}
\label{sec: observational uncertainty}

Magnitudes for filter X including observational uncertainties can be calculated via
\begin{equation}
m_\mathrm{data,X} = m_{\mathrm{AB,X}} + r_\mathrm{norm} \sigma_{1,\mathrm{X}},
\label{eq:noise realization random mag including error LSST science book}
\end{equation}
where $m_{\mathrm{AB,X}}$ is the intrinsic magnitude without observational noise, $r_\mathrm{norm}$ is a random Gaussian number with a standard deviation of unity, and $\sigma_{1,\mathrm{X}}$ is a
quantity that depends mainly on $m_{\mathrm{AB,X}}$ relative to the
$5\sigma$ depth $m_5$. This calculation is based on
\cite{2009:LSSTscience} (for more details, see also Appendix
\ref{sec:Appendix LSST uncertainty}).

In order to calculate $m_\mathrm{data,X}$, the $5\sigma$ depth of the
corresponding filter X is needed. In this work we consider eight filters, namely the
six LSST filters, \textit{u, g, r, i, z,} and \textit{y}, as well as
two infrared bands, \textit{J} and \textit{H}. To estimate the moon
phase dependence of filter X, we used the exposure time calculator (ETC)
of the European Southern Observatory (ESO) with a flat template spectrum. For \textit{ugriz} we used the
ETC of
OmegaCAM\footnote{\url{https://www.eso.org/observing/etc/bin/gen/form?INS.NAME=OMEGACAM+INS.MODE=imaging}},
and for \textit{yJH} we used the ETC of
VIRCAM\footnote{\url{https://www.eso.org/observing/etc/bin/gen/form?INS.NAME=VIRCAM+INS.MODE=imaging}},
where we assume an airmass of 1.2. Further, we used the typical fixed
sky model parameters with seeing $\le$%
$1''$ as provided by the ETC, which we found to be a
conservative estimate of the $5\sigma$ depth by testing other sky
positions. We investigated one cycle phase (25 August 2020 to 24
September 2020) to obtain relative changes of the $5\sigma$ depth with
time and matched these relative changes to the typical mean of the single-epoch LSST-like 
$5\sigma$ depth plus one magnitude, given by (23.3+1, 24.7+1, 24.3+1,
23.7+1, 22.8+1, and 22.0+1) for (\textit{u, g, r, i, z,} and \textit{y}), respectively, assuming a fixed exposure time.
These mean values 
take into account that in typical LSST observing strategies, redder
bands are preferred around full moon, while bluer bands are used more
around new moon. Going one magnitude deeper than the LSST $5\sigma$
depth provides a better quality of photometric measurements for time-delay measurements, and is feasible even
for a 2\,m telescope \citep{Huber:2019ljb}. The absolute values for
\textit{J} and \textit{H} bands are set by the ETC of VIRCAM in
comparison to the \textit{y} band.

\begin{figure}
\includegraphics[width=0.48\textwidth]{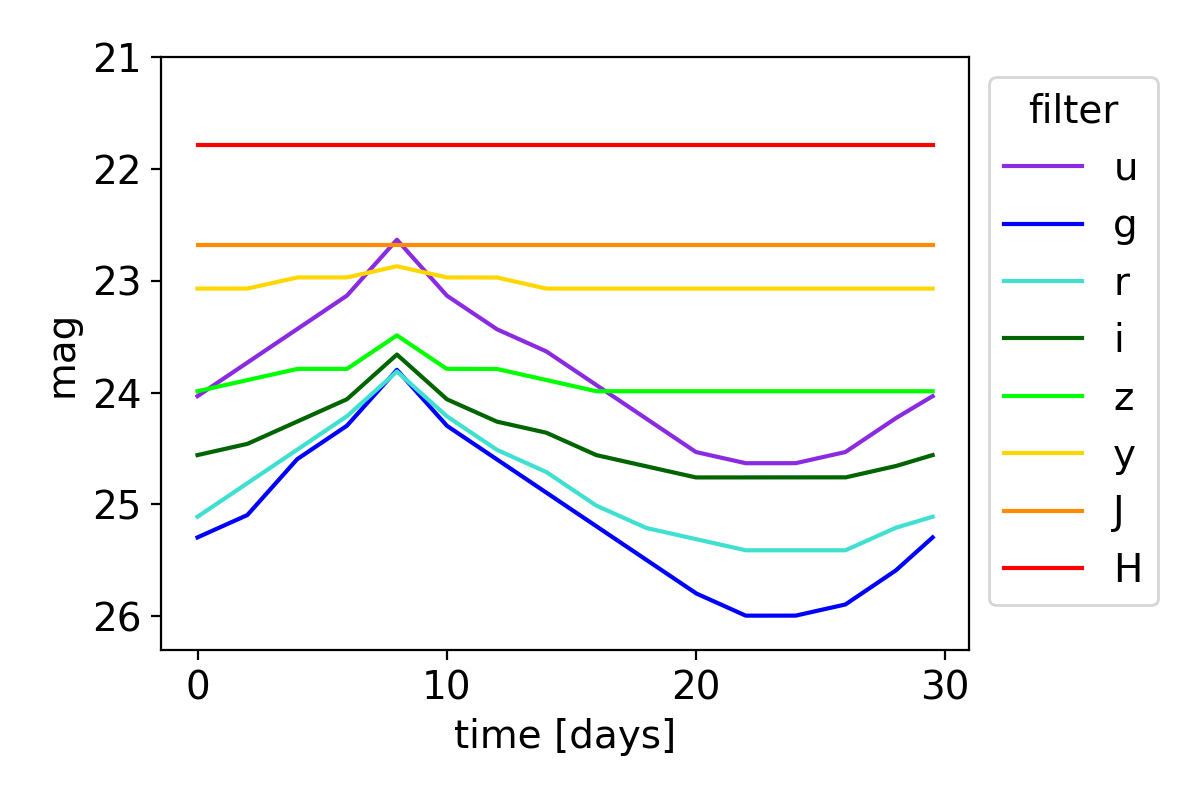}
\caption{Estimated $5\sigma$ depth for eight different filters, \textit{u, g, r, i, z, y, J,} and \textit{H,} accounting for the moon phase. Day 0 corresponds to the first quarter in the moon phase.  Full moon is around day 8, and new moon is on day 23.}
\label{fig: moon phase}
\end{figure}

The results for one cycle phase are
shown in Fig. \ref{fig: moon phase}, where we find full moon around
day 8 and new moon around day 23. As expected, bluer bands are much
more influenced by the moon phase in comparison to redder bands. As we
are typically interested in getting LSNe Ia with time delays greater
than 20 days \citep{Huber:2019ljb}, it is important to take the moon
phase into account.
Furthermore, we note that our approach on the $5\sigma$ depth assumes
an isolated point source, where in reality we also have contributions
from the host and lens light, which are the lowest for faint hosts and
large image separations. Even though these are the systems we are
interested in targeting, our uncertainties are on the optimistic side. The
construction of light curves in the presence of the lens and host is
deferred to future work, although LSNe have the advantage that the SNe
fade away and afterward an observation of the lensing system without
the SN can be taken and used as a template for
subtraction.

\section{Example mock observation and data set for machine learning}
\label{sec: Example data used for machine learning}

In this section we present a specific mock observation as an example,
to explain the data structure required for our ML
approaches.

\subsection{Mock observation}
\label{sec: mock observation 187}

\begin{figure}
\includegraphics[width=0.48\textwidth]{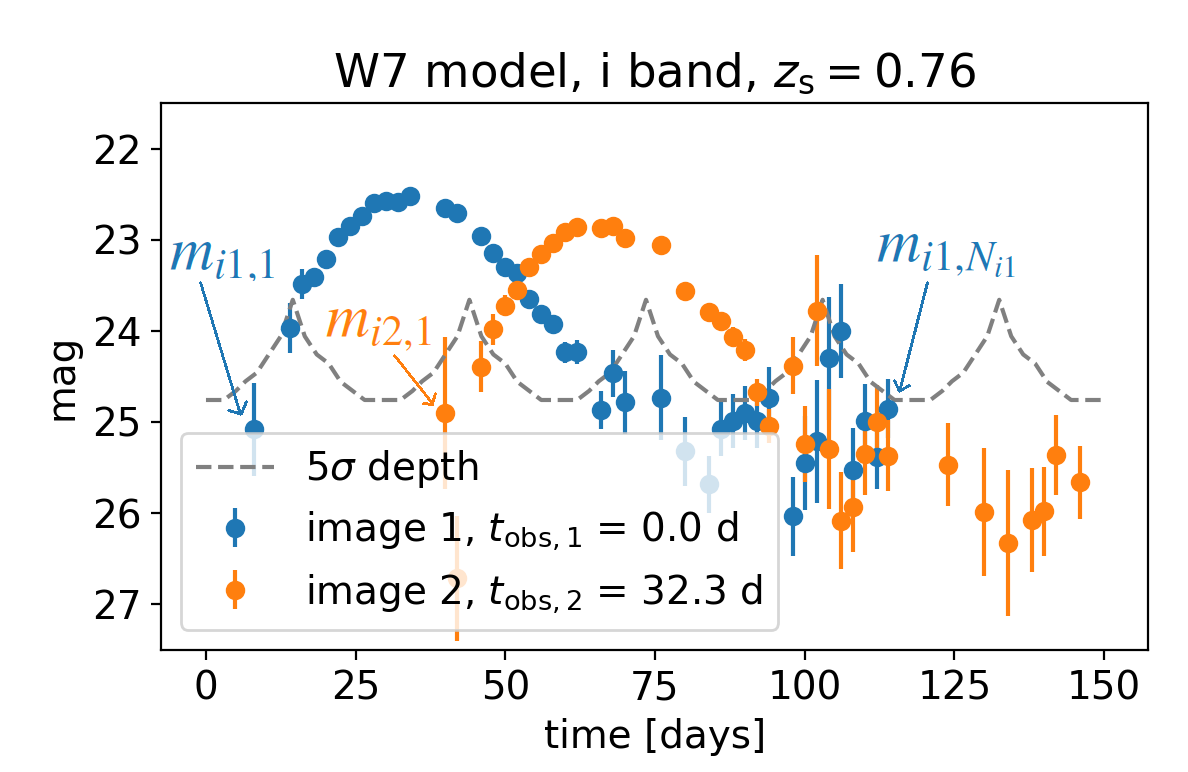}\caption{Simulated
  observation for which ML models will be trained to measure
  the time delay. The gray dashed curve marks the 5$\sigma$
  point-source depth, that accounts for the moon phase. The marked data points are also listed in Eq. (\ref{eq: data structure}).}
\label{fig: example light curve 187 system}
\end{figure}

As an example, we take a LSN Ia double system of the OM10
catalog \citep{Oguri:2010}, which is a mock lens catalog for strongly
lensed quasars and SNe. The parameters of the mock LSN Ia are
given in Table \ref{tab: Example double LSNe Ia}, where we have picked
a system with a source redshift close to the median source redshift
$\sourcez = 0.77$ of LSNe Ia in OM10 \citep{Huber:2020dxc}.  The
corresponding mock light curves are produced assuming the W7 model, where the $i$ band is shown in Fig. \ref{fig: example light
  curve 187 system} and all bands ($ugrizyJH$) together are shown in Appendix
\ref{sec:Appendix further bands of mock observation}. To calculate magnitudes with observational noise we use Eq. (\ref{eq:noise realization random mag
  including error LSST science book}). For the moon phase we assume a configuration where the $i$-band light curve peaks around new moon. 
Further configurations in the moon
phases will be discussed in Sect. \ref{sec: moon phases}. 
To avoid
unrealistic noisy data points $m_\mathrm{data,X}$ for our mock system in Fig. \ref{fig: example light curve 187 system}, we only take points $m_\mathrm{AB,X}$ brighter than $m_5
+ 2 \mathrm{mag}$ into account, before we add noise on
top. Furthermore, we assume a two day cadence with a few random gaps.


\begin{table}
\centering
\caption{Mock system of the OM10 catalog for generating mock light curves to train our 
ML techniques.}
\begin{tabular}{cccccc}
$\sourcez$ & $\lensz$ & image 1 $(\kappa,\gamma)$ & image 2 $(\kappa,\gamma)$ & time delay [days] \\
\midrule
0.76 & 0.252 & (0.251, 0.275) & (0.825, 0.815)  & 32.3
\end{tabular}

  \vspace{1ex}
     {\raggedright \textbf{Notes}. We assume $s=0.6$, similar to \cite{Huber:2020dxc}. The image separation for this double system is 1.7 arcsec and therefore typically resolvable with certain ground-based telescopes under most seeing conditions. \par}
\label{tab: Example double LSNe Ia}
\end{table}

\subsection{Data set for machine learning}
\label{sec: Data set for machine learning}



Our data of the mock LSN Ia contain measurements of light curves in one or more filters of the two SN images.  
The input data for our ML approaches are ordered, such that for a
given filter, all magnitude values from image 1 are listed (first
observed to last observed), followed by all magnitude values from
image 2. This structure is illustrated in the following definition and
will be referred to as a single sample,

\begin{equation}
m_{i1, 1} \, \, ... \, m_{i1, N_{i1}} \, m_{i2, 1} \, \, ... \, m_{i2, N_{i2}} \, \, \equiv \, \, d_{1} \, d_{2} ... \, d_{N_\mathrm{d}}
\label{eq: data structure}
,\end{equation}



%
\noindent for an example of a double LSN Ia with observations in the \textit{i}
band.  There are $N_{i1}$ photometric measurements in the light curve for SN image 1, and $N_{i2}$ photometric measurements for SN image 2. The magnitude value of the first data point in the $i$ band from
the first image in Fig. \ref{fig: example light curve 187 system} is
denoted as $m_{i1, 1}$, and the last data point is $m_{i1, N_{i1}}$. The first data point of the second image in the $i$ band is denoted as $m_{i2, 1}$. For simplification, we define $N_{\rm d} = N_{i1} + N_{i2}$, and $d_j$ as the $j$-th magnitude value in Eq. (\ref{eq: data structure}).
If multiple filters are available, then a ML model
can be trained per band, or multiple bands can be used for a single
ML model, which will be explored in Sect. \ref{sec: filters used for
  training}.

We introduce our FCNN and RF methods in detail in Sect. \ref{sec: Machine learning techniques}; we describe here the data set required for these two approaches in the remainder of this section. Important to note is that both methods always require 
the same input structure as defined in Eq. (\ref{eq: data structure}), with exactly the same number of data points\footnote{To avoid unrealistically noisy data points, we limit the maximum amount of noise allowed, as described in Appendix \ref{sec:Appendix LSST uncertainty}.}. 
From this input, we can then build a FCNN or a RF that
predicts the time delay. As additional
information, the $5\sigma$ depth is required for each data
point, to create noise in a similar way as in our mock observation.
Furthermore, microlensing uncertainties are taken into account 
by using the $\kappa, \gamma,$ and $s$ values of each LSN Ia image. 
The weakness of this approach is that we need to train a ML model
for each observed LSNe Ia individually, but the advantage is that we can
train our model very specifically for the observation pattern, noise
and microlensing uncertainties such that we expect an accurate result
with a realistic account of the uncertainties. Given that the data
production and training of such a system take less than a week and
multiple systems can be trained in parallel, this approach is easily
able to measure the delays of the expected 40 to 100 potentially promising LSNe
 in the 10 year LSST survey
\citep{Huber:2019ljb}.

Our ML approaches
require the same number of data points in each sample. We therefore produce our data
set, for training, validation and testing of the ML models,
such that the number of data points is always the same as in our mock
observation in Fig. \ref{fig: example light curve 187 system}. We calculate the light curves for the SN images via Eqs. (\ref{eq: microlensed flux}) and
(\ref{eq: microlensed light curves for ab magnitudes})
where we use random microlensing map positions. We then shift
the light curves for each SN image randomly in time around a first
estimate of the delay. In our example, we use the true observed time
values of the mock observation $t_{\rm obs,1}= 0.0 \, \mathrm{d}$ and $t_{\rm obs,2}= 32.3 \, \mathrm{d}$ as the first estimate for the SN images 1 and 2, respectively. For a real
system, we do not know these time values exactly and therefore probe a range of values around these first estimates in our training, validation and test sets.  In particular, for each sample in the training set, we pick random values between
$t_\mathrm{obs} - 10 \, \mathrm{d}$ and $t_\mathrm{obs} + 12 \,
\mathrm{d}$ as the ``ground truth'' (input true time value) for that specific sample. Different samples in the training set have different ground truth values.
We also tested more asymmetric ranges with
$t_\mathrm{obs} - 10 \, \mathrm{d}$ and $t_\mathrm{obs} +
t_\mathrm{est}$, where $t_\mathrm{est} = 16, 18, 22, 30 \,
\mathrm{d}$, and find results in very good agreement, with no
dependence on asymmetries in the initial estimate.

Data points are then created at the same epochs as the initial
observation. Using the $5 \sigma$ depth of each data point of our
observation, we calculate for each random microlensing position
10 random noise realizations following Eq. (\ref{eq:noise
  realization random mag including error LSST science book}). Since we
are not interested in the overall magnitude values we normalize the resulting light curve
by its maximum. Our total data set used for training has a size of
400000 samples 
coming from 4 theoretical SN Ia models, 10000 microlensing map positions and 10 noise
realizations.  Each sample has the data structure of Eq. (\ref{eq: data structure}).  For the validation and test sets, we calculate two
additional microlensing maps with the same $\kappa$, $\gamma$ and $s$
values as the training set, but with different microlensing patterns from random realizations of the stars. This provides ``clean'' validation and test sets that the ML methods have not encountered during training
in order to fairly assess the performance of the methods.
Our validation and test
set have each a size of 40000 samples, from 4 models, 1000 microlensing map
positions and 10 noise realizations.

\begin{figure}
\subfigure{\includegraphics[trim=14 17 22 16,width=0.45\textwidth]{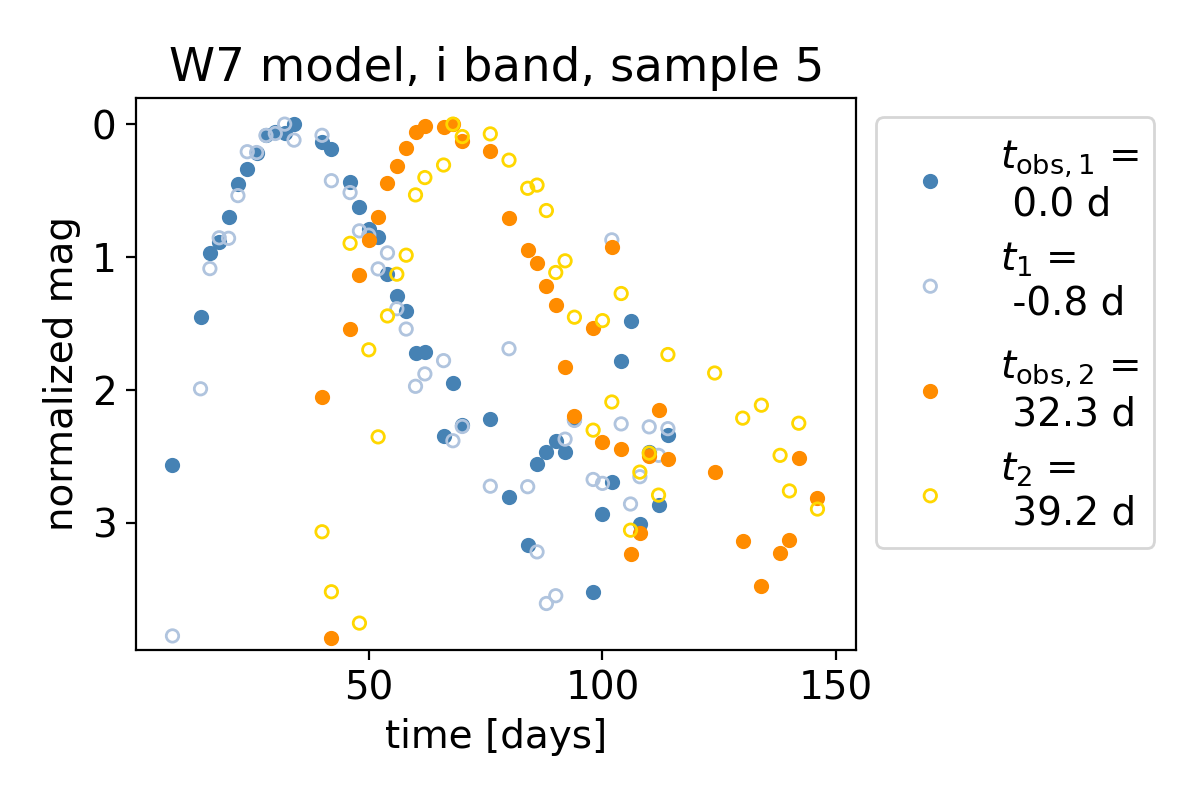}}
\subfigure{\includegraphics[trim=14 17 22 16,width=0.45\textwidth]{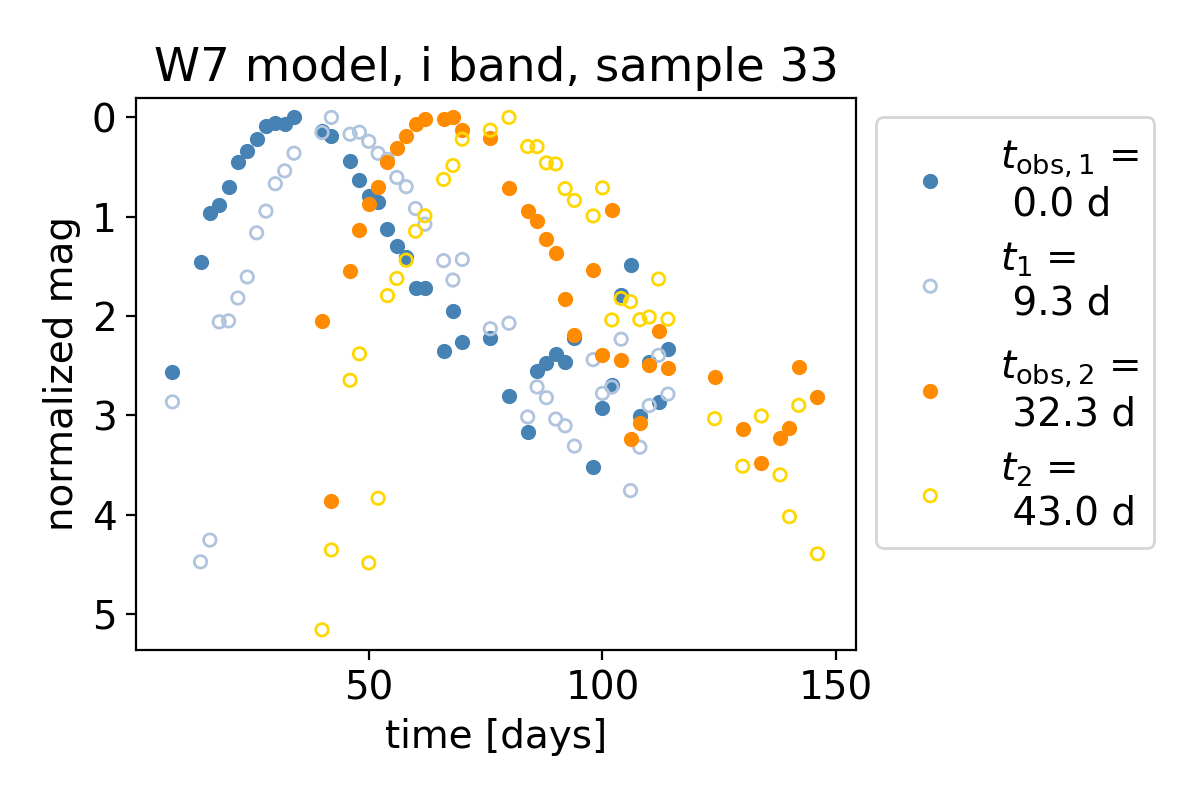}}\caption{Simulated data to train a ML model. The filled dots correspond to the mock observation shown in Fig. \ref{fig: example light curve 187 system}. The open dots represent the simulated training samples, where two out of the 400000 are shown for the $i$ band in the top and bottom panels.}
\label{fig: data to train NN}
\end{figure}

Two examples of our training data are shown in Fig. \ref{fig: data
  to train NN} in open circles. The first panel (sample 5) shows for the first SN image a
good match to the initial mock observation (in solid circle). The simulated training data are
therefore almost the same as the mock observation. Differences in fainter
regions (higher normalized magnitudes) come from observational
noise. For the second image, the time value $t_2$ of the simulated
training data is larger than the true value $t_{\rm obs,2}$ and therefore we find the peak  a few data points later. The general idea of providing data
in such a way is that the ML model learns to translate the
location of the peak region into the time value $t$. The difference
between the two time values from the first and second image is then
the time delay we are interested in. The second panel (sample 33) in
Fig. \ref{fig: data to train NN} is a nice example illustrating why going
directly for the time delay is not working that well in this
approach. We see that both simulated images for training are offset to
the right by almost the same amount. This would in the end lead to a very similar time delay as the initial mock observations,
 even though the input values are very different from those of the initial observations. 

Our described approach can be seen as a fitting process that
has the weakness that if the models for training are very
different in comparison to a real observation, our approach will
fail. From Fig. \ref{fig: SN Ia models vs SNEMO} we see that the
four SN Ia models predict different shapes of the light curves and
locations of peaks. Therefore, to compensate for different peak locations,
we randomly shift the four SN Ia models in time by $-5$ to 5
days. Furthermore, to make the noise level more random and compensate
for different peak brightness, we vary also the overall magnitude values by
$-0.4$ to 0.4. The random shifts in time and magnitude are the same for
a single sample, and therefore this approach creates basically a new
model with the same light-curve shape, but slightly different peak
location and brightness. Since the ML models do not know the 
actual values of the random shifts in time or magnitude the location
of the peak for a certain SN Ia model is smeared out. 
Therefore, this approach introduces a much
larger variety in the SN Ia models and Appendix \ref{sec:Appendix Bias
  training just on 3 models} shows that this helps to generalize to
light curves from sources that were not used in training the ML model. We also tested 
random multiplication factors to stretch or squeeze the light curves
in time (instead of the random constant shift in time as just described), but our approach with the random shifts works slightly better
as discussed in Appendix \ref{sec:Appendix Bias training just on 3 models}.  We therefore use the random shifts for the rest of this paper.

\section{Machine learning techniques}
\label{sec: Machine learning techniques}

In this section we explain the two different ML models
used in this work, namely a deep learning network using fully connected layers and a RF. We use these simple ML approaches to get started,
because if they work well, then more complicated models might not be necessary.
Results from these simple approaches would also serve as a guide for the development of more complex ML models.
 The techniques all use the input data structure as
described in Sect. \ref{sec: Example data used for machine learning},
and provide for each image of the LSN Ia a time value $t$ as shown in
Fig. \ref{fig: example light curve 187 system}. For the first
appearing image, the (ground truth) time $t = 0$ is the time of explosion and for the
next appearing image it is the time of explosion plus the time delay
$\Delta t$. Given our creation of the data set, which is done like a
fitting process for each light curve, we do not train the system to predict only the time
delay, but instead we have as output one time value per image as
described in Sect. \ref{sec: Data set for machine learning}.

\subsection{Deep learning - Fully connected neural network}
\label{sec: Deep Learning Network}

\begin{figure}
\centering
\includegraphics[width=0.49\textwidth]{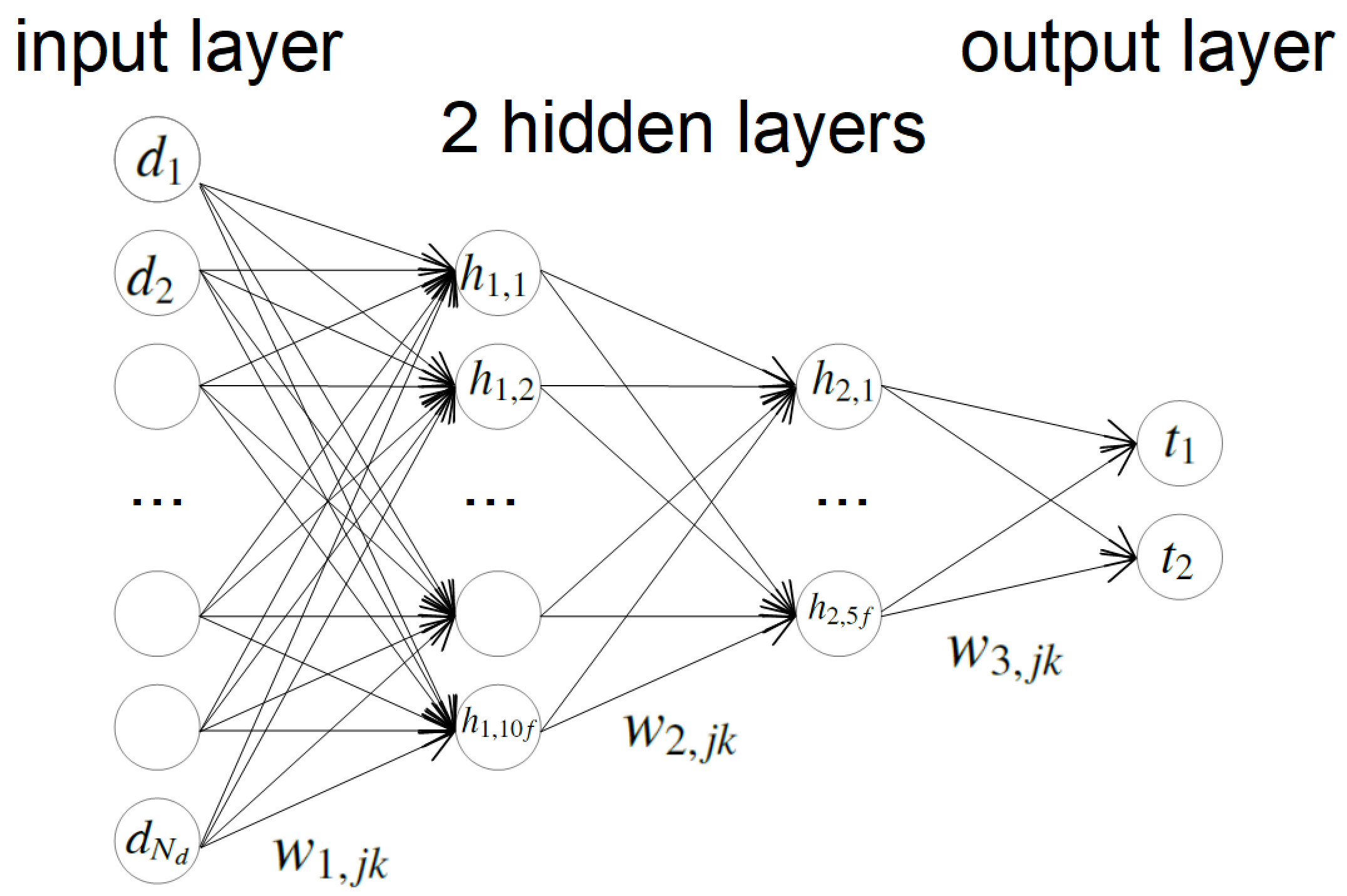}
\caption{FCNN, where the input layer has $N_{\rm d}$ data points and $d_{j}$ stands for the magnitude value of the $j$-th data point in Eq. (\ref{eq: data structure}). The size of the
  two hidden layers scales by a factor $f$, and the outputs are two
  (four) time values for a double (quad) LSN Ia.}
\label{fig: fully connected neural network}
\end{figure}

Neural networks are a powerful tool with a broad range of
applications. 
To solve our regression problem, we used a FCNN,
consisting of an input layer, two hidden layers, and one output layer,
as shown in Fig. \ref{fig: fully connected neural network}. Although universal 
approximation results \citep{Cybenko1989ApproximationBS,HORNIK1989359} suggest that a FCNN with only one hidden layer of arbitrarily large 
width can approximate any continuous function, FCNNs with finite widths but more layers have 
shown to be more useful in practice. We therefore used two hidden layers instead of one and 
tested different widths of the networks by introducing the scaling factor $f$ for a variable 
number of nodes in the hidden layers in order to optimize the number of hidden nodes. 

In our FCNN, each
node of the input layer corresponds to a magnitude value of a single
observation for a given filter and image, sorted as in the example of
Eq. (\ref{eq: data structure}). Each node of the input layer ($d_j$) is
connected by a weight ($w_{1,jk}$) to each node of the first hidden layer ($h_{1,k}$).
In addition, a bias ($b_{1,k}$) is assumed and we introduce non linearities, by using a rectified linear units
(ReLU) activation function \citep[e.g.,][]{Glorot:2011,Maas:2013},
which is 0 for all negative values and the identity function for all
positive values. Therefore, the nodes of the first hidden layer can be calculated via

\begin{equation}
h_{1,k} = \mathrm{ReLU}\big(\sum_{j=1}^{N_{\rm d}} w_{1,jk} \, d_j + b_{1,k}\big), \qquad k=1,2,...10f.
\end{equation}

Further, all nodes in the first hidden layer are connected to all nodes in the second hidden layer in a similar manner:

\begin{equation}
h_{2,k} = \mathrm{ReLU}\big(\sum_{j=1}^{10 f} w_{2,jk} \, h_{1,j} + b_{2,k}\big), \qquad k=1,2,...5f.
\end{equation}
The nodes from the second hidden layer are then finally connected to the output
layer to produce the time values
\begin{equation}
  t_{k} = \sum_{j=1}^{5 f} w_{3,jk} \, h_{2,j} + b_{3,k}, \qquad
  k=\left\{ \begin{array}{ll}
                  1,2 \ (\mathrm{double\ system})\\
                  1,2,3,4 \ (\mathrm{quad\ system}).\\
  \end{array} \right.
\end{equation}

The output layer consists of two nodes for a double LSN Ia and four nodes for a quad LSN Ia. 
We tested also other
FC network structures such as using a different network for each image, using three hidden layers, or
using a linear or leaky ReLU activation function, but our default approach described above works best.

We train our system for a certain number of epochs $N_\mathrm{epoch}$,
where we use the ML library \texttt{PyTorch}
\citep{NEURIPS2019_9015}. At each epoch, we subdivide our training data
randomly into mini batches with size $N_\mathrm{batch}$. Each mini batch is propagated through
our network to predict the output
that we compare to the ground-truth values by using the mean squared error
(MSE) loss. To optimize the loss function,  we use the Adaptive Moment
Estimation (Adam) algorithm \citep{Kingma:2014} with a learning rate
$\alpha$ on the MSE loss to update the weights in order to improve the
performance of the network\footnote{For the other hyperparameters of
the Adam optimizer, we used the PyTorch default values.}. Per epoch, we calculate the MSE
loss of the validation set from our FCNN, and store in the end the network at the epoch with the lowest validation loss. By selecting the epoch with the lowest validation loss, we minimize the chance of
overfitting to the training data. Typically we reach the lowest validation loss around epoch 200 and an example for the training and validation curve for our FCNN is shown in Appendix \ref{sec:Appendix train and validation loss}.

The test data set is used in the end to compare different FCNNs,
which have been trained with different learning rates $\alpha$, sizes
$f$ and mini-batch sizes $N_\mathrm{batch}$.

\subsection{Random forest}
\label{sec: Random Forest network}

The RF \citep{breiman2001random} is a method used for
classification and regression problems, by constructing many random
decision trees. In this section we give a brief introduction on the
idea of a RF and explain the setup we are using.

\begin{figure}
\centering
\includegraphics[width=0.49\textwidth]{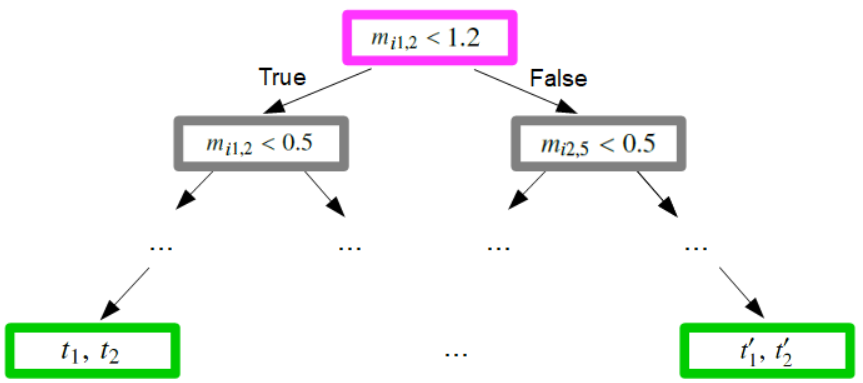}
\caption{Schematic example of a regression tree for a double system
  that predicts two time values for certain input data as in Eq.
  (\ref{eq: data structure}). The root node is represented by the
  magenta box, the internal nodes by gray boxes, and the leaf nodes by
  green boxes.}
\label{fig: regression tree}
\end{figure}

To build a RF, we construct many random regression trees, which are a
type of decision trees, where each leaf represents numeric values (for the outputs). 
For our case, we create a total of $N_\mathrm{trees}$ random regression trees where a schematic example for a single regression tree is shown in Fig.
\ref{fig: regression tree}.  The root node is shown in magenta, the
internal nodes in gray and the leaf nodes in green. 
The root node splits our whole data set containing samples as defined by Eq. (\ref{eq:
  data structure}),
into two groups based on a certain criterion (e.g., $m_{i1,2} < 1.2$): first where the criterion is true, and second where it is not. The internal nodes split the data in the same manner, until no
further splitting is possible and we end up at a leaf node to predict the two time values $t_1$ and $t_2$ as output. 

To create random regression trees we use a bootstrapped data set, 
which draws randomly samples
from the whole training data (400000 samples) until it reaches a given size
$N_\mathrm{max \, samples}$. Importantly, an individual sample of
the original training data can be drawn multiple times and each random regression tree
is built from an individual bootstrapped data set, which is used to create the root, internal and leaf nodes. However, only a random subset of the features (e.g., just
$m_{i1,2}$, $m_{i2,5}$, and $m_{i2,9}$) is considered to construct the root node or a single internal node, where the splitting criterion (e.g., $m_{i1,2} < 0.5$) of a single feature is defined based on the mean value (e.g., $\bar{m}_{i1,2} = 0.5$) from all samples under investigation ($N_\mathrm{max \, samples}$ for the root node and fewer samples for the internal nodes depending on how the data set was split before). The number of available features we pick randomly from all features for the creation of a node is $N_\mathrm{max \, features}$\footnote{A single feature
such as $m_{i1,2}$ can be picked multiple times.}. 

In the following, we demonstrate the construction of the root node for a regression tree, as shown in 
Fig. \ref{fig: regression tree}, for the example of $N_\mathrm{max \, features} = 3$. Therefore, we randomly pick three features from Eq. (\ref{eq: data structure}), which we assume to be $m_{i1,2}$, $m_{i2,5}$, and $m_{i2,9}$. From a bootstrapped data set with $N_\mathrm{max \, samples}$ samples of our training data set, we assume to find the mean values $\bar{m}_{i1,2} = 1.2$, $\bar{m}_{i2,5} = 1.0$, and $\bar{m}_{i2,9} = 0.6$. Therefore, we investigate the three criteria $m_{i1,2} < 1.2$, $m_{i2,5} < 1.0$, and ${m}_{i2,9} < 0.6$ as potential candidates for the root node, where each of the criteria splits the $N_\mathrm{max \, samples}$ training samples into two groups. We select the best splitting criterion as the one that results in the lowest variance in the predictions within each of the groups created by the split. In other words, we can compute through this comparison a residual for $t_1$ and $t_2$ for each sample. From this, we can 
calculate the sum of squared residuals for each candidate criterion, and the criterion that predicts the 
lowest sum of squared residuals will be picked as our root node, which would be $m_{i1,2} < 1.2$ in our schematic example.  
For each of the resulting two groups,
we follow exactly the same procedure to construct
internal nodes that split the data further and further until no
further splitting is possible or useful\footnote{Further splitting is not useful if none of the investigated splitting criteria would lead to further improvements of the sum of squared residuals in comparison to not splitting the remaining samples.} and we end up at a leaf node to predict
the output. To avoid a leaf node containing just a single training sample, we used two parameters, namely, $N_\mathrm{msl}$,
the \textbf{m}inimum number of \textbf{s}amples required to be in a
\textbf{l}eaf node, and $N_\mathrm{mss}$, the \textbf{m}inimum number of
\textbf{s}amples required to \textbf{s}plit an internal node. From the multiple training samples in a leaf node, the $t_1$ and $t_2$ values of a leaf node 
are the average of all samples in the leaf node.

Following the above
procedure, many random regression trees are built; to create an output for a
single (test) sample, all regression trees are considered and the final output
is created from averaging over all trees.


For this approach we used the object
\texttt{sklearn.ensemble.RandomForestRegressor} of the software \texttt{scikit-learn}
\citep{scikit-learn,sklearn_api}, where we assume the
default parameters except for the previously mentioned
$N_\mathrm{msl}$, $N_\mathrm{mss}$, $N_\mathrm{trees}$, $N_\mathrm{max
  \, samples}$ and $N_\mathrm{max \, features}$.

\section{Machine learning on example mock observation}
\label{sec: Machine learning on example mock-observation}

In this section we apply the ML techniques
from Sect. \ref{sec: Machine learning techniques} to our example
mock observation of a double LSNe Ia described in Sect. \ref{sec:
  Example data used for machine learning}. In Sect. \ref{sec: Best fit, DL vs RF} we 
  find the best FCNN and RF and compare results from
  the corresponding test sets based on the four theoretical models also used in the training process.
  In Sect. \ref{sec: Evaluation on SNEMO test set} we use the best FCNN and RF and apply it
  to an empirical data set not used in the training process to test the generalizability of both models. This final test is very important since in reality we can never assure 
  that our assumed light-curve shapes in the training process will fully match a real observation.

\subsection{Best fit: Fully connected neural network versus random forest}
\label{sec: Best fit, DL vs RF}
To find a FCNN and a RF that provide the best fit to our mock
observation from Fig. \ref{fig: example light curve 187 system}, we
explore a set of hyperparameters as listed in Table \ref{tab: DL
  varied parameter for training} for the FCNN and Table
\ref{tab: RF varied parameter for training} for the RF.

To find the best ML model for our mock observation, we used the test set to evaluate each set of
hyperparameters. This is just to find an appropriate set of hyperparameters, which we will use for the sake of simplicity from 
here on throughout the paper\footnote{Such hyperparameter optimizations are usually performed using the validation set, whereas we are using the test set because we have the SNEMO15 data set for ultimate performance test.}. Our final judgment of the performances of the ML models will be based on
the ``SNEMO15 data set'' where
light curves will be calculated using an empirical model (see Sect. \ref{sec: Evaluation on SNEMO test set}). The distinctions between the various data sets for our ML approaches are summarized in Table \ref{tab: explanation traing, val, test, snemo data set}.
For each sample $i$ of the test set, we get two time
values, $t_{1,i}$ and $t_{2,i}$, from which we can calculate the time delay
$\Delta t_i = t_{1,i} - t_{2,i}$, which we compare to the true time delay
$\Delta t_{\mathrm{true},i}$ to calculate the ``time-delay deviation'' of the sample
as

\begin{equation}
  \tau_i = \Delta t_i - \Delta t_{\mathrm{true},i}.
  \label{eq:td_deviation}
\end{equation}

We investigate here the absolute time-delay deviation instead of the relative one 
($\tau_i / \Delta t_{{\rm true},i}$), because this allows us to draw conclusions 
about the minimum time delay required to achieve certain goals in precision and accuracy.
From our results, we do not find a dependence on the absolute time delay
(e.g., 32.3 d for Fig. \ref{fig: data to train NN}) used in the training process, 
which is what we expect from the setup of the FCNN and the RF and is demonstrated in Sect. \ref{sec: quad lsne ia and higher microlensing uncertainties}

For the FCNN, we find that ($\alpha, f, N_\mathrm{batch},
N_\mathrm{epoch}) = (0.0001, 40, 256, 400)$ provides the best result,
meaning that the median of $\tau_i$ of the whole test set is lower
than 0.05 days (to reduce the bias) and the 84th$-$16th percentile (1$\sigma$ credible interval) of the test set is
the lowest of all networks considered. For the RF the
hyperparameters $(N_\mathrm{trees}, N_\mathrm{mss}, N_\mathrm{msl},
N_\mathrm{max \, samples}, N_\mathrm{max \, features}) = (800, 4, 1,
200000, \sqrt{N_\mathrm{all \, features}})$ provide the best
result. In the following we always use these two sets of
hyperparameters for the FCNN or the RF, unless specified
otherwise. We note that $N_\mathrm{trees} = 800$ is on the upper 
side of what we investigated, but increasing the number of trees 
further makes the computation even more costly. Nevertheless, we 
tested also $N_\mathrm{trees} = 1000, 1200, 1600, 2000$ and 
$N_\mathrm{trees} = 3000$ with $(N_\mathrm{mss}, N_\mathrm{msl},
N_\mathrm{max \, samples}, N_\mathrm{max \, features})$ from the 
best fit as listed above. We find results that are basically the 
same as for $N_\mathrm{trees} = 800$ or slightly worse (0.02 d at most)
and therefore we stick with $N_\mathrm{trees} = 800$, which is sufficient.
The comparison between the FCNN and the RF is shown in
Fig. \ref{fig: best fit DL vs RF for example light curve 187
  system}, where we
quote 
the median (50th percentile), with the 84th-50th percentile (superscript) and 16th-50th percentile (subscript) of the whole sample of light curves from the corresponding test set. The results include microlensing and observational
uncertainties as described in Sect. \ref{sec: simulated light curves
  for LSNe Ia}. For the training and testing, we considered the four SN
Ia models, merger, N100, sub-Ch and W7 (therefore we use the description ``corresponding test set'' in the title of Fig. \ref{fig: best fit DL vs RF for example light curve 187
  system}). Further, the results are based
on using just the \textit{i} band, assuming the data structure as
defined in Eq. (\ref{eq: data structure}). 

\begin{figure}
\includegraphics[width=0.49\textwidth]{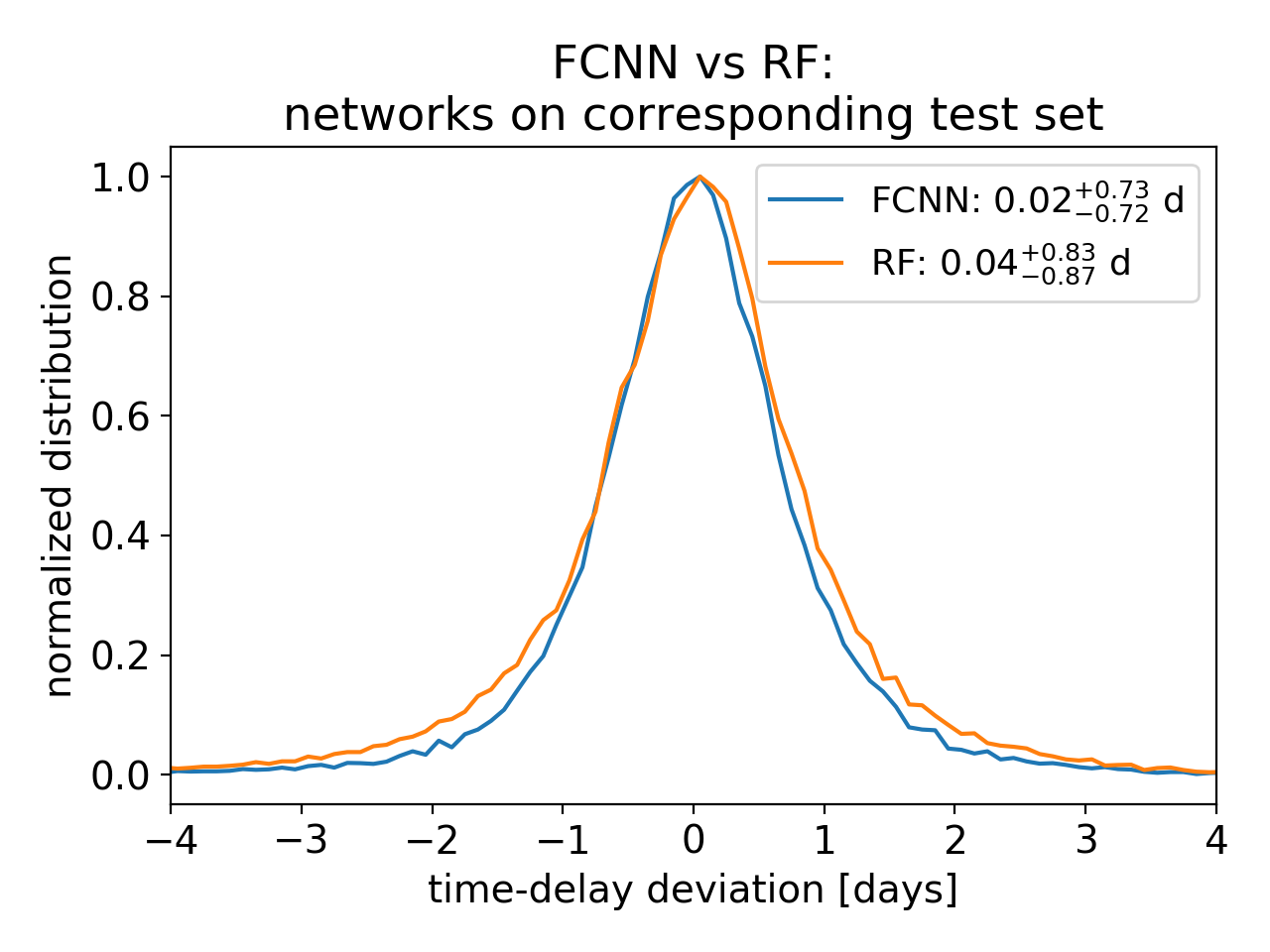}
\caption{FCNN and RF on the whole sample of light curves from the specified test set for the mock observation in Fig. \ref{fig: example light curve 187 system}.  The ML models' hyperparameters are set to the values at which the test set yields a bias below 0.05 days and the smallest 68\% credible interval of the time-delay deviation (in Eq. (\ref{eq:td_deviation})).}
\label{fig: best fit DL vs RF for example light curve 187 system}
\end{figure}

Instead of looking at the whole sample of light curves from the test set at once, we show in Appendix \ref{sec:Time-delay deviation as function of time delay} how the time-delay deviation $\tau_i$ depends on the time delay of the test samples. We find for both networks a slight trend that time delays far away from the true time delay of the mock observation yield larger deviations, where the effect is stronger for the RF in comparison to the FCNN.
However, this is not surprising as very long time delays come from rare scenarios where the $t_1$ value of the first image is highly underestimated and the $t_2$ value of the second image is highly overestimated.  Similarly, very short time delays tend to have $t_1$ that is highly overestimated and $t_2$ that is underestimated.
Given that these scenarios are rare in the training set, it is more difficult to learn these cases. Still, the FCNN compensates for these edge effects better, which explains the better performance of the FCNN in comparison to the RF on the corresponding test set as shown in Fig. \ref{fig: best fit DL vs RF for example light curve 187 system}.

However, we see that both ML models provide accurate measurements of the time
delay with the $1 \sigma$ uncertainty for the FCNN around 0.7
days and the RF around 0.8 days, where both have low bias ($\leq$%
$0.04$ days). Nevertheless, the training and test set is produced by using the
same SN Ia models.  If light curves in the
test sets are different from the ones used for training, this can lead to broadened
uncertainties, and more critically, also to biases (see Appendix \ref{sec:Appendix
  Bias training just on 3 models}). Further, we
learn from Appendix \ref{sec:Appendix Bias training just on 3 models}
that, as soon as the different light curves used for training cover a
broad range, the trained ML model can be used for light-curve shapes it
has never seen. Therefore, in Sect. \ref{sec: Evaluation on SNEMO
  test set}, we evaluate the RF and the FCNN trained on four
theoretical models on a data set based on the empirical
\texttt{SNEMO15} model.

\begin{table}
\centering
\caption{Investigated parameters for the training process of the FCNN 
(see Fig. \ref{fig: fully connected neural network}) for
  the system listed in Table \ref{tab: Example double LSNe Ia}.}
\begin{tabular}{cc}
$\alpha$ & 0.01, 0.001, 0.0001, 0.00001 \\
\midrule
$f$ & 5, 10, 20, 40, 80, 160\\
\midrule
$N_\mathrm{batch}$  & 64, 128, 256, 512 \\
\midrule
$N_\mathrm{epoch}$ & 400 \\

\end{tabular}
  
  \vspace{1ex}
     {\raggedright \textbf{Notes}. We
  vary the learning rate, $\alpha$, the size of the hidden layers by a
  factor $f$, and the size of the mini batches,
  $N_\mathrm{batch}$. Furthermore, 400 training epochs
  ($N_\mathrm{epoch}$) are sufficient given that the minimum loss of
  the validation set is typically reached around 200 training
  iterations. \par}
\label{tab: DL varied parameter for training}
\end{table}

\begin{table}
\centering
\caption{Investigated parameters for the training process of the RF
(see Fig. \ref{fig: regression tree} for a single
  regression tree) for the system listed in Table \ref{tab: Example
    double LSNe Ia}.}
\begin{tabular}{cc}
$N_\mathrm{trees}$ & 200, 400, 800 \\
\midrule
$N_\mathrm{mss}$  & 2, 4, 8\\
\midrule
$N_\mathrm{msl}$ & 1, 2, 4 \\
\midrule
$N_\mathrm{max \, samples}$ & 50000, 100000, 200000, 300000, 400000 \\
\midrule
$N_\mathrm{max \, features}$ & 1, $\sqrt{N_\mathrm{all \, features}}, N_\mathrm{all \, features}$ 
\end{tabular}

  \vspace{1ex}
     {\raggedright \textbf{Notes}. We vary the number of trees, $N_\mathrm{trees}$,
  the minimum number of samples required to split an internal node,
  $N_\mathrm{mss}$, the minimum number of samples required to be in a
  leaf node, $N_\mathrm{msl}$, the size of the bootstrapped data set,
  $N_\mathrm{max \, samples}$, and the maximum number of features,
  $N_\mathrm{max \, features}$, considered to create a root or internal
  node. \par}
\label{tab: RF varied parameter for training}
\end{table}

\begin{table}
\centering
\caption{Explanation of the different types of data sets used for our ML approaches.}
\begin{tabular}{l|l}
Data set type & Description and purpose \\ 
\hline
\multirow{2}{*}{training set} & To train the ML models. Light curves of \\ & four  theoretical SNe Ia models are used. \\
\hline
\multirow{3}{*}{validation set} & To find the training epoch for the FCNN \\ & that has lowest validation loss (four \\ & SNe Ia models as in training process). \\
\hline
\multirow{10}{*}{(corresponding) test set} & To evaluate the performance of a ML \\ & model using four theoretical models as \\ & in the training process. The term \\ & ``corresponding'' is used if all \\ & parameters (e.g., $\kappa$, $\gamma$, ...) for the \\ & production of the test set are the same \\ & as for the training set. \\ & This data set does not test the \\ & generalizability to different SN Ia \\ & light-curve shapes.\\
\hline
\multirow{5}{*}{SNEMO15 data set} & Final test set using light curves from the \\ & empirical \texttt{SNEMO15} model not used \\ & in the training process, which most \\ & importantly tests the generalizability of \\ & the trained ML models.\\
\end{tabular}
\label{tab: explanation traing, val, test, snemo data set}
\end{table}

\subsection{Generalizability of ML models: Evaluation on \textit{SNEMO15} data set}
\label{sec: Evaluation on SNEMO test set}

To test if the ML models trained on four SN Ia models with the random shifts in time and
magnitude as introduced in Sect. \ref{sec: Data set for machine
  learning} can generalize well enough to real SN Ia data, we created a data
set based on the empirical \texttt{SNEMO15} model, which is shown in
Fig. \ref{fig: SN Ia models vs SNEMO}. The empirical model covers
only a wavelength range from 3305 $\AA$ to 8586 $\AA$, and 
with $\sourcez = 0.76$ (Table \ref{tab: Example double LSNe Ia}), the $i$ band is the bluest band we can
calculate.

To account for macrolensing and brightness deviations for the
\texttt{SNEMO15} model in comparison to the theoretical SN models, we
set the median \texttt{SNEMO15} light curve equal to the mean value of
the four macrolensed SN Ia models. Since the light curves are
normalized before the training process, this is only
important to avoid over- or underestimations of the observational
noise. Furthermore, to include microlensing, we use microlensed light
curves from the four theoretical models, initially created for the
corresponding test set, and subtract the macrolensed light curve, assuming
$\mu_\mathrm{macro} = 1/((1-\kappa)^2-\gamma^2)$. Therefore, we get
from our 4 models 4000 microlensing contributions for the light
curves, which the FCNN or the RF have not seen in its training
process.  For each of the microlensing contributions, we then draw
randomly one of the 171 \texttt{SNEMO15} light curves to create a
microlensed \texttt{SNEMO15} light curve.  From the 4000 microlensing contributions, we have a sample of 4000 microlensed light curves.  For each light curve, we then draw 10
random noise and time-delay realizations to create a data set, as described
in Sect. \ref{sec: Data set for machine learning}.  We call this the \texttt{SNEMO15} data set. 

\begin{figure}
\subfigure{\includegraphics[width=0.49\textwidth]{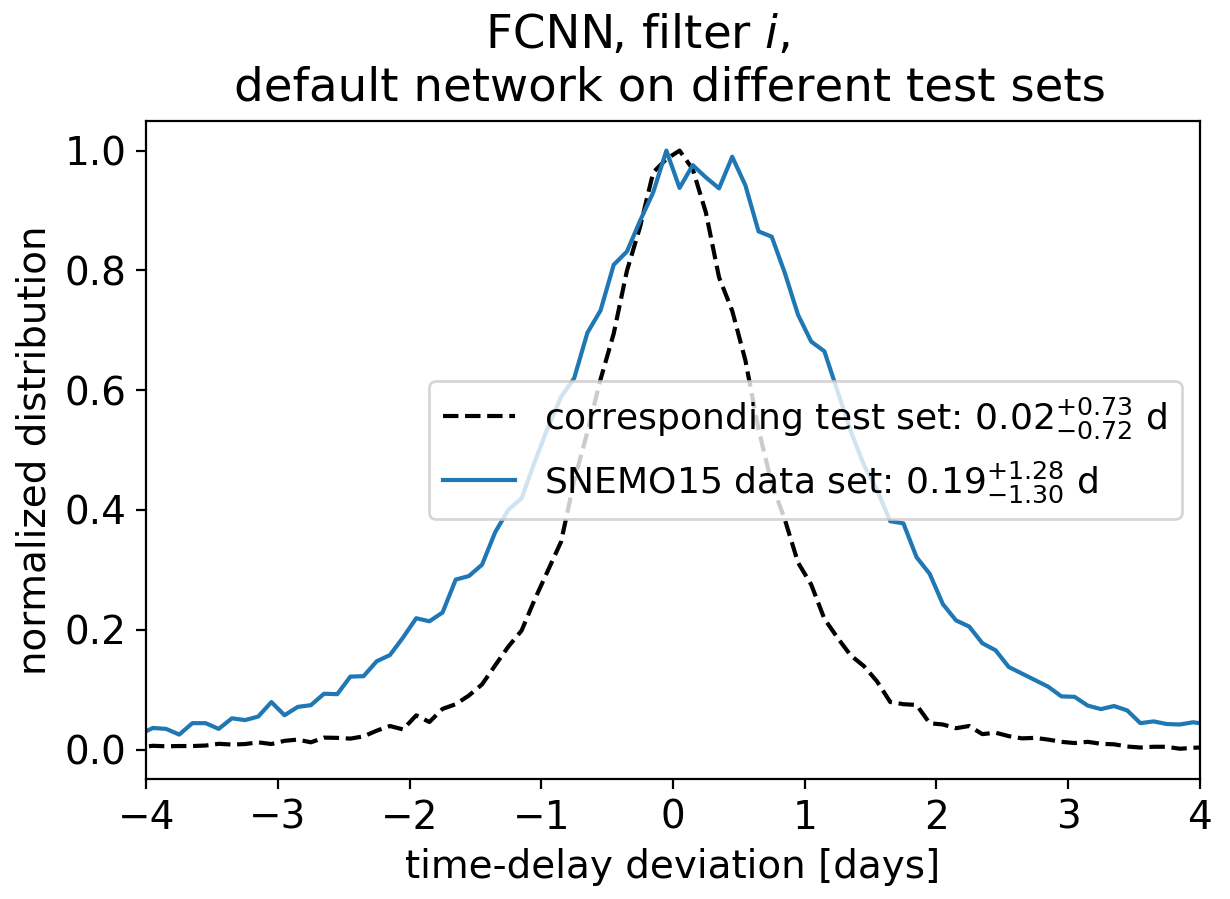}
}
\subfigure{\includegraphics[width=0.49\textwidth]{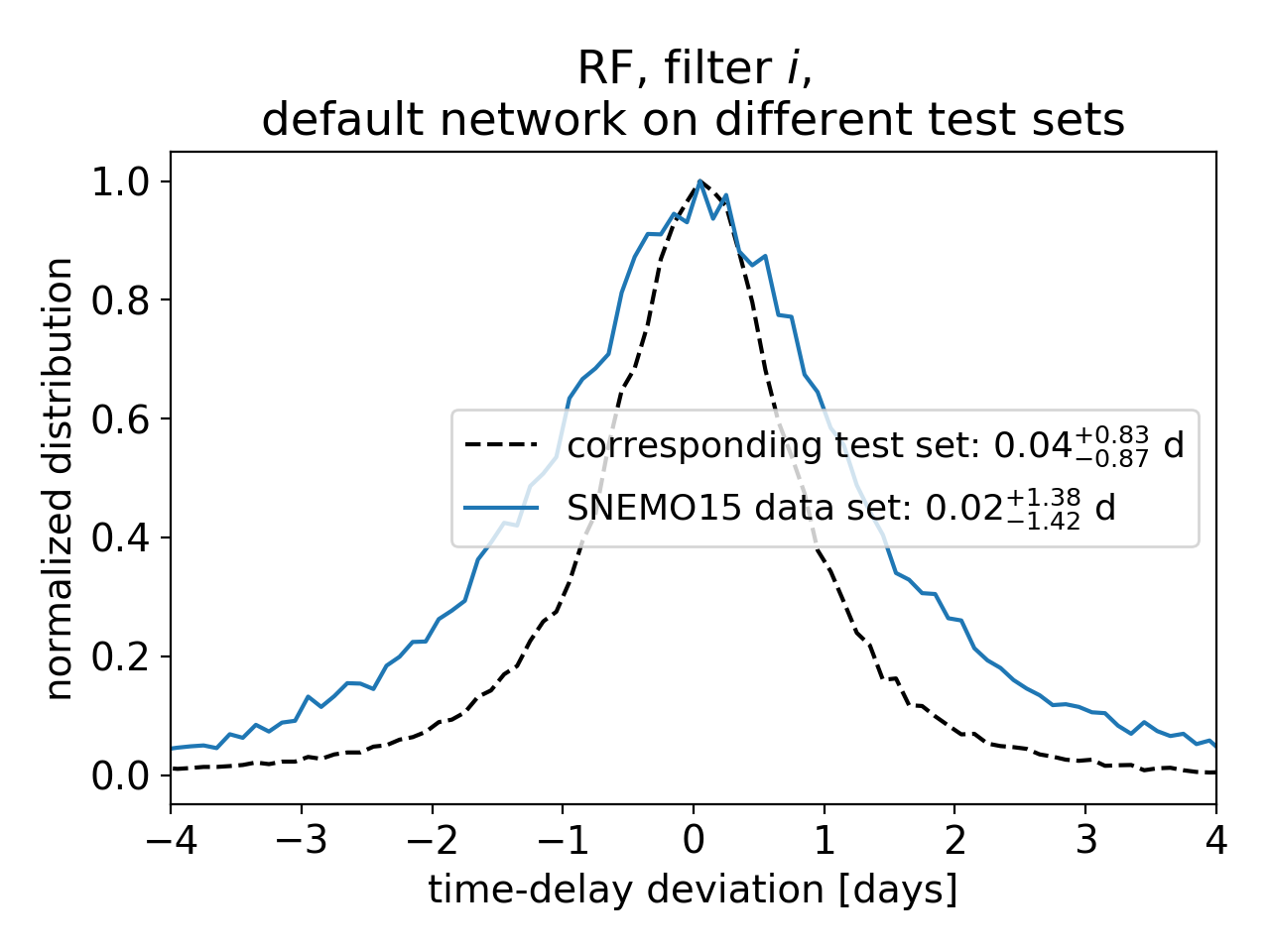}
}
\caption{FCNN and RF trained on four theoretical models for the
  \textit{i} band evaluated on the whole sample of light curves from the two specified data sets. The dashed black line
  represents the corresponding test set based on the four theoretical
  models, and the data set of the blue line is based on the empirical
  \texttt{SNEMO15} model.}
\label{fig: SENOM15 test for DL and RF using filters iz}
\end{figure}

Figure \ref{fig: SENOM15 test for DL and RF using filters iz} shows
the results where we evaluate the FCNN and the RF from Fig. \ref{fig: best fit DL vs RF for example light curve 187 system}, trained on four
theoretical SN Ia models, on the corresponding test set (built from the same four theoretical SN Ia models) and on the \texttt{SNEMO15} data set. The first important thing we note is that the RF shows almost
no bias, whereas the FCNN has a higher bias when evaluated on the \texttt{SNEMO15} data set. To investigate this
further, we look at results from the RF and the FCNN for the set of
hyperparameters as listed in Tables \ref{tab: DL varied parameter for
  training} and \ref{tab: RF varied parameter for training} for three
different cases using  the $i$ band, $z$ band, or $y$ band.

We find
that the absolute bias of the FCNN for the different hyperparameters and bands ($i,z,$ and $y$) 
is mostly below 0.4 days but
higher values are also possible. The problems are that these
variations in the bias in the \texttt{SNEMO15} data set are not related to biases we see in the
corresponding test sets or due to a specific set of hyperparameters. As a result, we cannot identify the underlying source of the bias, apart from that it is due to suboptimal generalization of the theoretical SN Ia models to \texttt{SNEMO15} in the FCNN framework. 

The RF works much better in this context, as the absolute bias is always lower than 0.12
days for the $i$, $z$, and $y$ bands. Only the hyperparameter $N_\mathrm{max
  \, features} = 1$ can lead to a higher bias up to 0.22 days, but
this hyperparameter is excluded because of its
much worse performance in precision on the corresponding test set 
in comparison to $N_\mathrm{max \,
  features} = \sqrt{N_\mathrm{all \, features}}$ or $N_\mathrm{max \,
  features} = N_\mathrm{all \, features}$. Therefore, as long as we
restrict ourselves to LSNe Ia with delays longer than 12 days we can
achieve a bias below 1 percent, which allows accurate measurements of
$H_0$. Furthermore, the
bias is not the same in all filters. While the absolute bias in the
$y$ band goes up to 0.12 days, we have a maximum of 0.08 days in the $z$
band and 0.03 days in the $i$ band. The comparison of multiple bands
therefore helps to identify some outliers.

\begin{figure}
\subfigure{\includegraphics[width=0.45\textwidth]{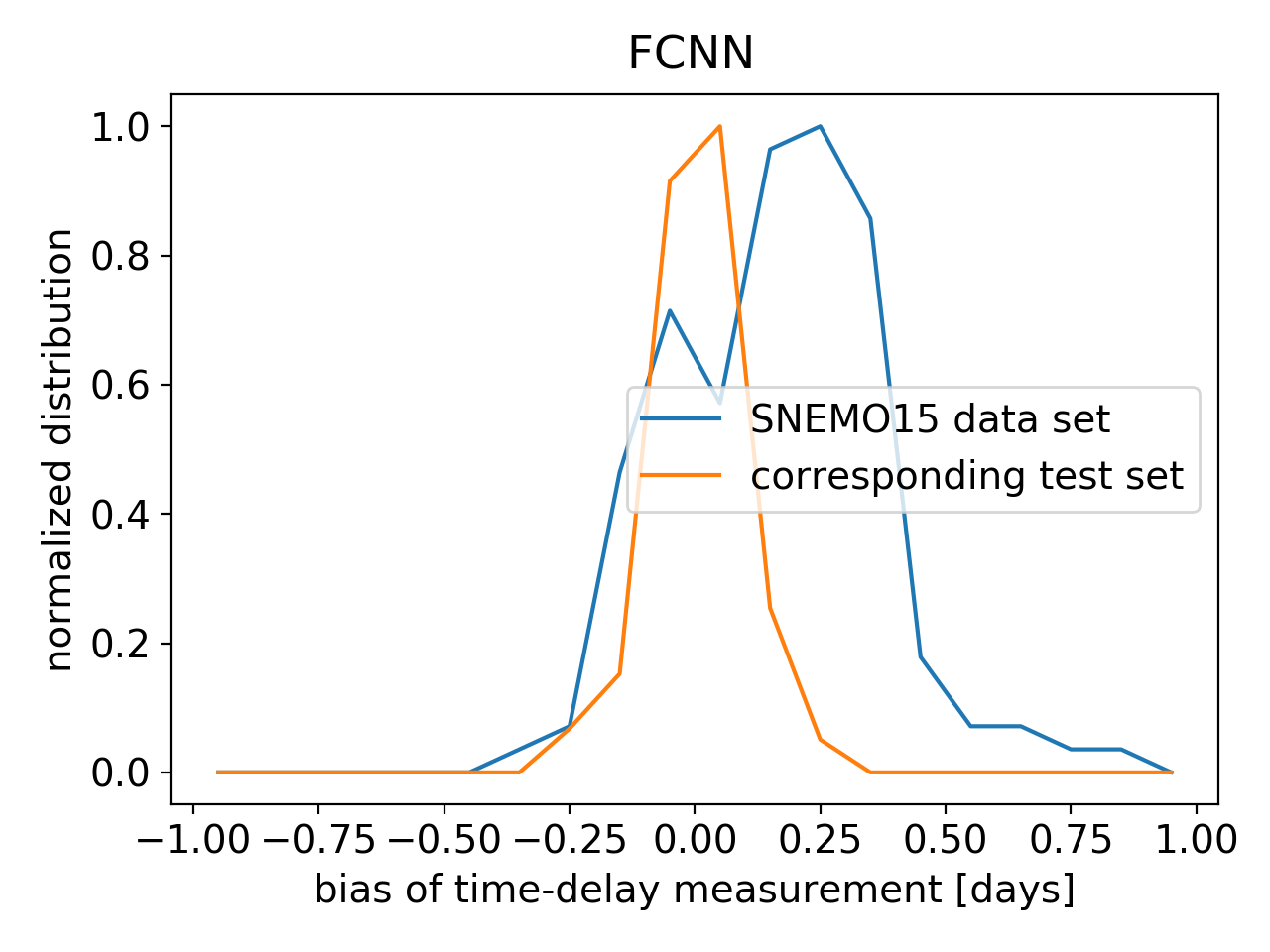}}
\subfigure{\includegraphics[width=0.45\textwidth]{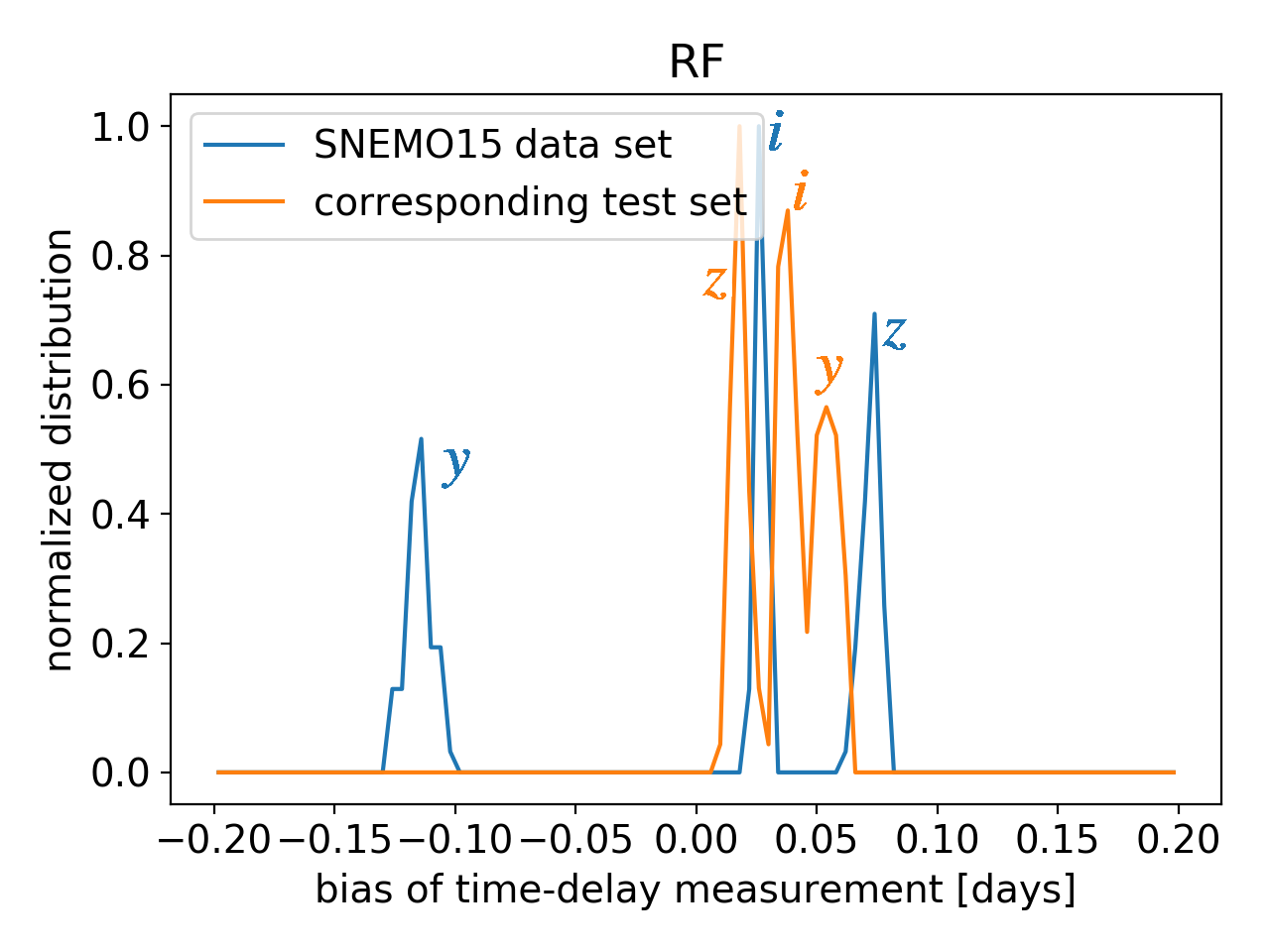}}
\caption{Bias of FCNN and RF on the corresponding test set, composed of four theoretical SN Ia models used for training, and a data set based on the empirical \texttt{SNEMO15} model, not used during training, for a variety of different hyperparameters and filters ($i, z,$ and $y$), i.e., from model averaging. The large biases on the \texttt{SNEMO15} data set up to 1 day in our FCNN approach come from the different hyperparameters even though the corresponding test set provides biases within 0.25 days. The RF provides much lower biases in all cases; it depends only weakly on the hyperparameters and is instead mostly set by the filters under consideration.} 
\label{fig: DL and RF bias on corresponding and SNEMO15 test set}
\end{figure}

The bias investigation of the FCNN and the RF is summarized in Fig. \ref{fig: DL and RF bias 
on corresponding and SNEMO15 test set} using all hyperparameters (except $N_\mathrm{max
  \, features} = 1,$ which is excluded because of its bad performance on the corresponding test set) 
  and the $i, z,$ and $y$ bands. From the upper panel we see that the large biases of our FCNN 
  on the \texttt{SNEMO15} data set are not related to biases we see in the corresponding test 
  set and therefore identifying a set of hyperparameters just from the corresponding test set which works 
  also well on the \texttt{SNEMO15} data set is not possible. From the lower panel of Fig. 
  \ref{fig: DL and RF bias on corresponding and SNEMO15 test set} we see that also the biases in the 
  RF from the corresponding test set and the \texttt{SNEMO15} data set are not directly related with 
  each other but this is not a problem as the biases on the \texttt{SNEMO15} data set are low enough for 
  precision cosmology. From this example we see that the RF is able to generalize to a new kind of data not used in the training process, which 
  does not work well for our FCNN. 
  In principle this was already suggested by the 
  investigation done in Appendix \ref{sec:Appendix Bias training just on 3 models}, but with the 
  random shifts in time we introduced, it seemed to significantly improve the generalizability, but it 
  was still not enough for the final test on the \texttt{SNEMO15} data set. Investigating the 
  importance of all the input features as listed in Eq. (\ref{eq: data structure}),
  we find that the FCNN focuses mostly on the peak directly whereas for the RF the features before 
  and after the peak are the most important ones.
  More about this is discussed in Appendix \ref{sec:Appendix Feature importance}.

In the remainder of the paper, we proceed to present results based on the RF,
because the significant bias in our FCNN makes accurate
cosmology difficult to achieve especially for LSN Ia systems with short delays. 
Using deeper networks would not be enough to improve our FCNN,
as this would just allow a better fit to the
training data but does not ensure any improvement on
the generalizability of the network.
Therefore, it would be necessary
to provide more realistic input light curves for the training process,
as it has problems to generalize to light-curve shapes it has not
seen. Such an improvement could be achieved by using the
\texttt{SNEMO15} light curves as well in the training process, but then
a test set with light-curve shapes it has never seen would be
missing. Another approach would be to incorporate regularization 
or dropout into our FCNN or by constructing a network that 
outputs in addition to the time values
the associated uncertainties, but given that this was not necessary
for the corresponding test set to perform well, it would be some kind of fine tuning
to our \texttt{SNEMO15} data set, because all tests before were encouraging to proceed to
the final test.
Therefore, we postpone further investigations of FCNNs to future studies, especially
since other network architecture, such as recurrent neural networks, long
short-term memory networks \citep{Sherstinsky_2020}, or Gaussian processes, could potentially reduce model complexity while having lower inductive bias\footnote{Inductive bias refers to the bias coming from assumptions that a ML model has to make to generalize based on training samples.} \citep{2020arXiv200208791W}.

Another thing we learn is that the distribution of the recovered time delays from the \texttt{SNEMO15} data
set is $\sim$0.5 days broader than that of the corresponding test
sets. This is not surprising as the RF and the FCNN have never seen
such light curves in the training process. A $\sim$1.4 day precision
on a single LSN Ia is still a very good measurement and allows us to
conduct precision cosmology from a larger sample of LSNe
Ia. Nevertheless, we see in this section that even though the
uncertainties for the RF are larger than that of the FCNN, the RF
provides low bias when used on empirical data and is therefore
preferred.

\section{Microlensing, noise, and choice of filters}
\label{sec: Microlensing, observational noise and choice of filters}

In this section we use the RF from Sect. \ref{sec: Best fit, DL vs RF} and apply it 
to the mock observation from Sect. \ref{sec: Example data used for machine learning}, for hypothetical assumptions about microlensing and noise to find sources of uncertainties (Sect. \ref{sec: Microlensing map parameters kappa, gamma, s} and \ref{sec: Uncertainties due to microlensing and noise}).
We further investigate potential bands to target for follow-up observations (Sect. \ref{sec: filters used for training}). In this section all results presented are based on the RF on test sets from the four theoretical models. The conclusions drawn in this section would be the same if the results from the FCNN would be presented.

\subsection{Microlensing map parameters $\kappa, \gamma, s$}
\label{sec: Microlensing map parameters kappa, gamma, s}
To investigate uncertainties in the microlensing characterization,
we use the RF from Sect. \ref{sec: Best
  fit, DL vs RF}, but evaluate it on different test sets with varying
$\kappa, \gamma$, and $s$ values, which deviate from the original
training data.

Figure \ref{fig: evaluated on different kappa, gammas} shows the RF
evaluated on different test sets. The black dashed line
represents the evaluation of the RF on the corresponding test
set, which is calculated according to Sect. \ref{sec: Data set for
  machine learning}. The blue and orange lines represent very similar
test sets, but calculated on a different microlensing map. Instead of
the $\kappa$ and $\gamma$ values listed in Table \ref{tab: Example
  double LSNe Ia}, we assume for the first image $(\kappa, \gamma) =
(0.201, 0.225)$ and for the second image $(\kappa, \gamma) = (0.775,
0.765)$ to calculate the test set corresponding to the blue line. The
orange line represents a LSNe Ia where we have for the first image
$(\kappa, \gamma) = (0.301, 0.325)$ and for the second image $(\kappa,
\gamma) = (0.875, 0.865)$. Even though the RF has never seen $(\kappa, \gamma)$ configurations as represented by the orange and blue line
in the training process, the results are
very similar to the corresponding test set of the RF and given
that typical model uncertainties are around $0.05$
\citep[e.g.,][]{More:2016sys}, uncertainties in $\kappa$ and $\gamma$ are not
critical for our procedure.

\begin{figure}
\includegraphics[width=0.49\textwidth]{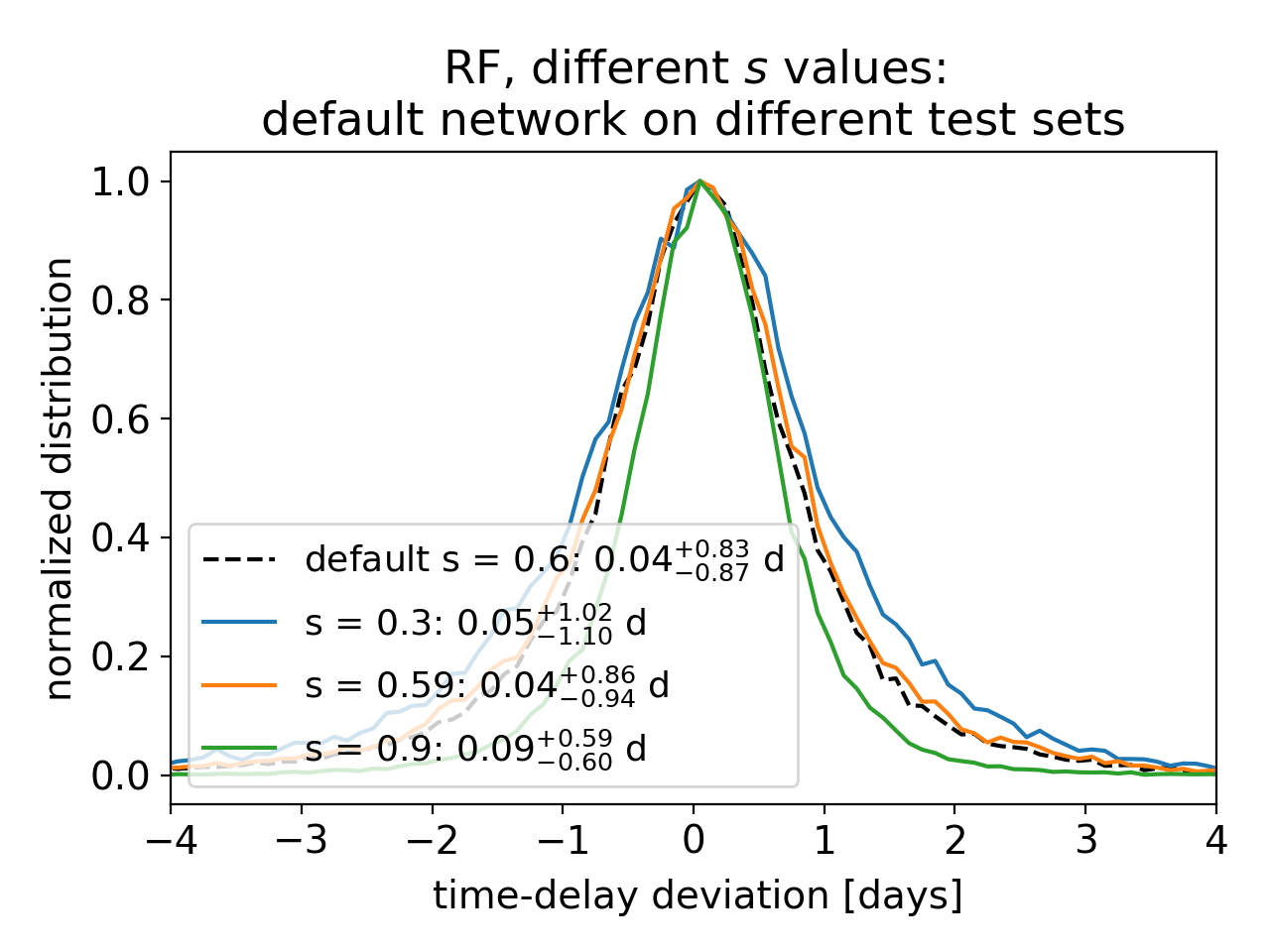}
\caption{RF evaluated on all samples from its corresponding test set (black dashed
  line, where training and test sets have the same $s$ value) and on all samples from three
  other test sets (blue, orange, and green), with different
  $s$ values of the microlensing map in comparison to that of the training
  data.}
\label{fig: evaluated on differen s}
\end{figure}

In Fig. \ref{fig: evaluated on differen s} we do a similar investigation,
but this time we vary the $s$ value of
the microlensing maps. From the comparison of the black dashed line to
the orange line, which represents almost the same $s$ value, we see
that the uncertainties are almost comparable. Therefore, the much wider
uncertainty for $s=0.3$ (blue line) is not due to variations from
different microlensing maps for the same parameter set, but from the
fact that lower $s$ values provide more micro caustics in the map,
which leads to more events where these caustics are crossed and
therefore to more microlensing events and higher uncertainties. This
also explains the much tighter uncertainties of $s=0.9$, which
corresponds to a much smoother microlensing map. These results are in
good agreement with those of \cite{Huber:2020dxc}, who also showed that higher $s$
values lead to lower microlensing uncertainties.

For a real observation, the $s$ value is often not known very
precisely, which is no problem as the RF still works very
well. The only thing one has to be careful about is that an underestimation
of the $s$ value leads to an overestimation of the overall
uncertainties. Therefore, going for a slightly lower $s$ value as one
might expect is a good way to obtain a conservative estimate of the
uncertainties.

\begin{figure}
\includegraphics[width=0.49\textwidth]{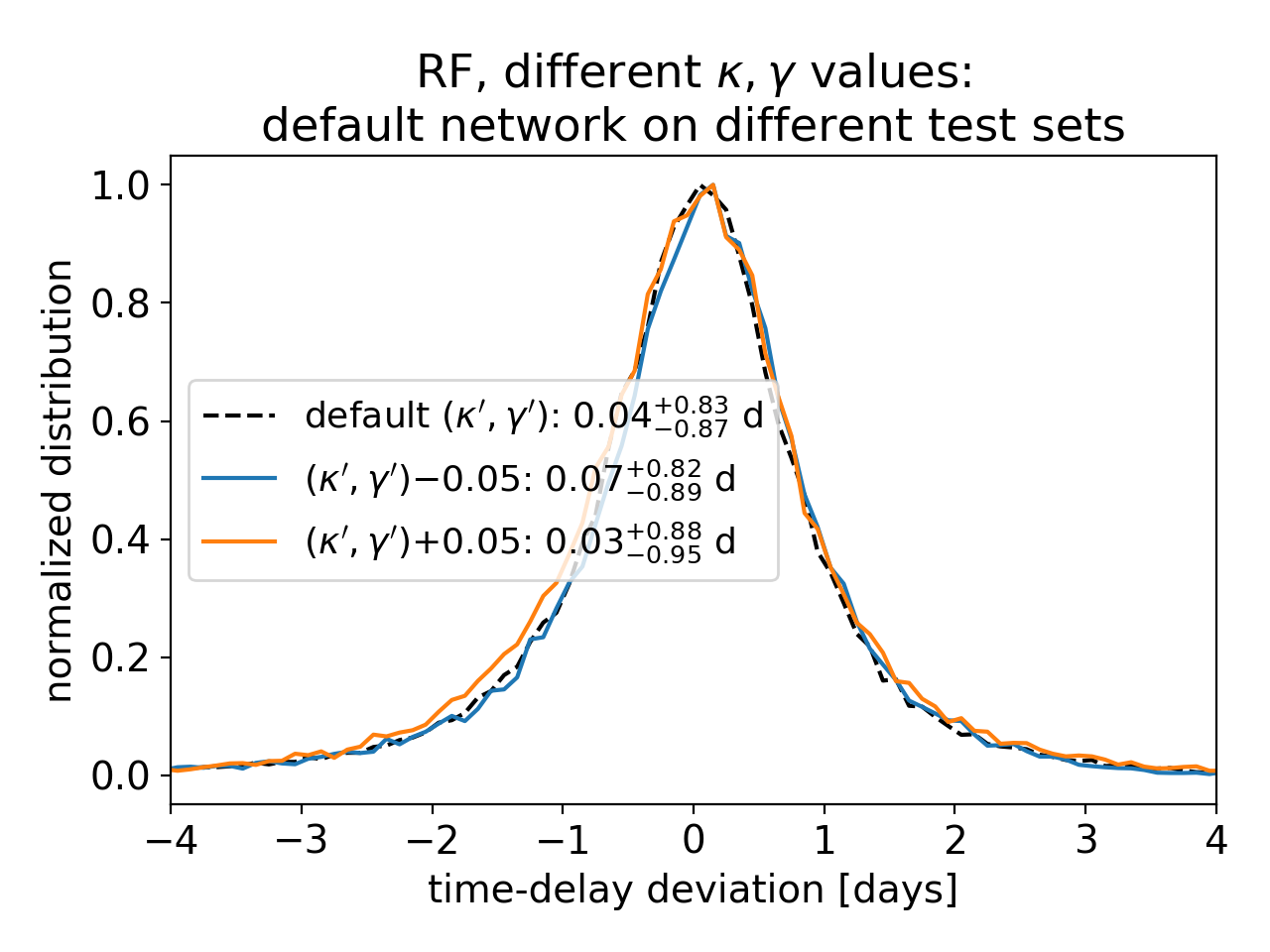}
\caption{RF evaluated on all samples from its corresponding test set (black dashed
  line, where training and test sets have the same $\kappa$ and $\gamma$ values)
  and on all samples from two other test sets (blue and orange), with slightly
  different $\kappa$ and $\gamma$ values of the microlensing map in
  comparison to that of the training data.}
\label{fig: evaluated on different kappa, gammas}
\end{figure}

\subsection{Uncertainties due to microlensing and noise}
\label{sec: Uncertainties due to microlensing and noise}
In this section we compare the RF from Sect. \ref{sec:
  Best fit, DL vs RF} to other RF models with various 
assumptions about microlensing and noise as shown in Fig. \ref{fig:
  no micro, no noise, no micro noise}.

From the two cases containing microlensing in comparison to the two
cases without microlensing, we find that microlensing increases the
uncertainties almost by a factor of two. Although this is quite
substantial, we see that the contribution of the observational noise
is much higher and is the dominant source of uncertainty in the time-delay measurement. Therefore, to
achieve lower uncertainties, deeper observations with smaller photometric uncertainties are required. This is
in agreement with \cite{Huber:2019ljb}, who found that a substantial
increase in the number of LSNe Ia with well measured time delays can be
achieved with greater imaging depth.

\begin{figure}
\includegraphics[width=0.49\textwidth]{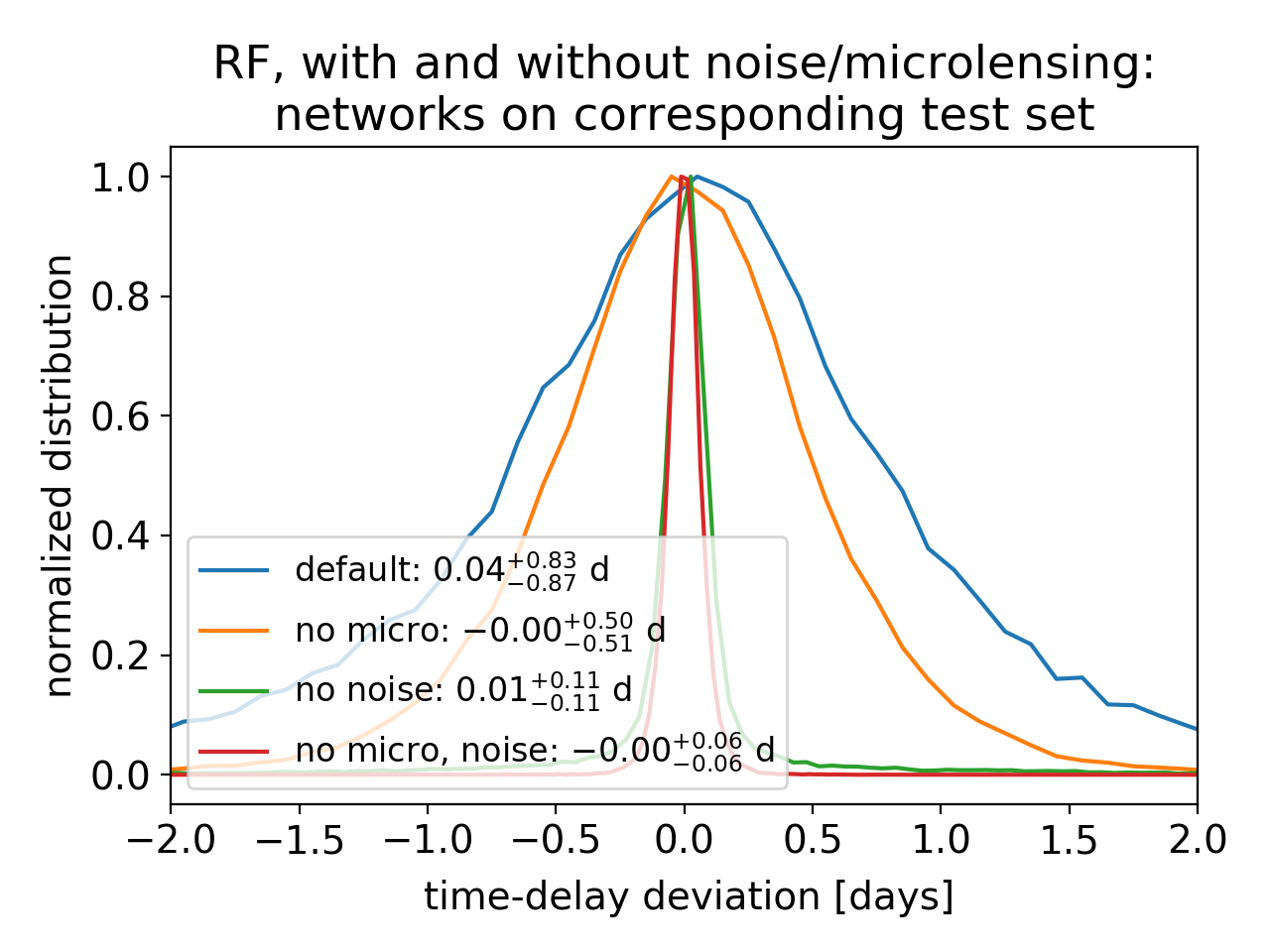}
\caption{Comparison of the RF model from Sect. \ref{sec:
    Best fit, DL vs RF} to three other RF models with hypothetical
  assumptions about noise and microlensing. Each histogram is based on the whole sample of light curves from their corresponding test set. For our realistic mock observation, the noise in the light curves dominates over microlensing as the main source of uncertainty for measuring the time delays.}
\label{fig: no micro, no noise, no micro noise}
\end{figure}

\subsection{Filters used for training}
\label{sec: filters used for training}

In this section we investigate eight different filters
(\textit{ugrizyJH}) and possible combinations of them to get more
precise measurements. Figure \ref{fig: 8 different bands, each one
  seperatly} shows eight RF models where each is trained and
evaluated on a single band. The $i$ band, presented first in Sect.
\ref{sec: Best fit, DL vs RF} provides the most precise 
measurement. The next promising filters are $r$, $z$, $g$, and $y$ in
that order. For the bands $u$, $J,$ and $H$, the precision of the
measurement is poor and therefore almost not usable. The reason
for the strong variation between different bands is the quality of the
light curve, which becomes clear from Fig. \ref{fig: appendix further bands of mock
  observations}, where only the $g$ to $y$ bands provide observations where
the peak of the light curves can be identified. Light curves with the
best quality are the $r$ and $i$ bands, which therefore work best for
our RF.

There are different ways to combine multiple filters to measure the
time delay. The first possibility would be to construct color curves
to reduce the effect of microlensing in the so-called achromatic phase
\citep{Goldstein:2017bny,Huber:2020dxc}. However, as pointed out by
\cite{Huber:2020dxc} our best quality color curve $r-i$ would be not
ideal as there are no features for a delay measurement within the
achromatic phase. Further, we saw in Sect. \ref{sec: Uncertainties
  due to microlensing and noise} that our dominant source of
uncertainty is the observational noise instead of
microlensing. Therefore, using color curves for this mock example is
not practical. We further see that even though color curves are in
theory a good way to reduce microlensing uncertainties, in a real
detection it might fail because not enough bands with high quality
data are available.

Another way of combining multiple filters is to train a single RF model
for multiple filters.
Generalizing Eq. (\ref{eq: data structure}) for the $r$ and $i$ bands,
we used as input structure
\begin{equation}
m_{r1, 1} \, m_{r1, 2} \, .. \, m_{r1, N_{r1}} \, m_{r2, 1} \, .. \, m_{r2, N_{r2}} \, m_{i1, 1} \, .. \, m_{i1, N_{i1}} \, m_{i2, 1} \, .. \, m_{i2, N_{i2}},
\label{eq: data structure ri}
\end{equation}
and more bands will be attached in the same way. The results are
summarized in Fig. \ref{fig: different bands using in a single
  network}, where we see that combining the two most promising bands
improves the uncertainty by about $0.1$ days, but adding more bands
does not help. Comparing these results to Fig. \ref{fig: different
  bands multiplying distributions from singe networks}, where
different distributions from Fig. \ref{fig: 8 different bands, each
  one seperatly} are multiplied with each other\footnote{We assume that different filters have independent detector noise.}, we see that a single
RF model for multiple filters does not profit much from multiple
bands. Therefore, it is preferable to use a single RF model per band and
combine them afterward. Using three or more filters can also help to
identify potential biases in a single band as pointed out in Sect.
\ref{sec: Evaluation on SNEMO test set}. Combining the $r$, $i$, and $z$
bands via multiplication helps to reduce the uncertainty by more than a
factor of two in comparison to using just the $i$ band for our system
with $\sourcez = 0.76$. Further bands that might be considered for
follow-up observations are the $g$ and $y$ bands. 

The choice of the ideal filters depends on the source redshift and
therefore we show in Fig. \ref{fig: appendix filters for different
  redshifts} a similar plot as in Fig. \ref{fig: 8 different bands,
  each one seperatly} but for $\sourcez = 0.55$ and $\sourcez = 0.99$,
which corresponds to the 16th and 84th percentile of the source
redshift from LSNe Ia in the OM10 catalog. From this we learn that the
three most promising filters are the $g, r,$ and $i$ bands for $\sourcez \lesssim 0.6$,
whereas for $\sourcez \gtrsim 0.6$ the $r, i,$ and $z$ bands are preferred. 
The main reason for this behavior is the low rest-frame UV flux of SNe Ia due to line blanketing, 
which gets shifted more and more into the $g$ band for higher $\sourcez$.
If four filters could be used, then we have $g, r, i,$ and $z$ for $\sourcez \lesssim 0.8$ and
$r, i, z,$ and $y$ for $\sourcez \gtrsim 0.8$. If resources for five filters are available, we recommend $g, r, i, z,$ and $y$; the $J$ band might be preferred over the $g$ band for high source redshifts ($\sourcez > 1.0$).\ However, given the poor precision in the $g$ and $J$ bands at such high redshifts, it is questionable how useful the fifth band is in these cases.

\begin{figure}
\includegraphics[width=0.49\textwidth]{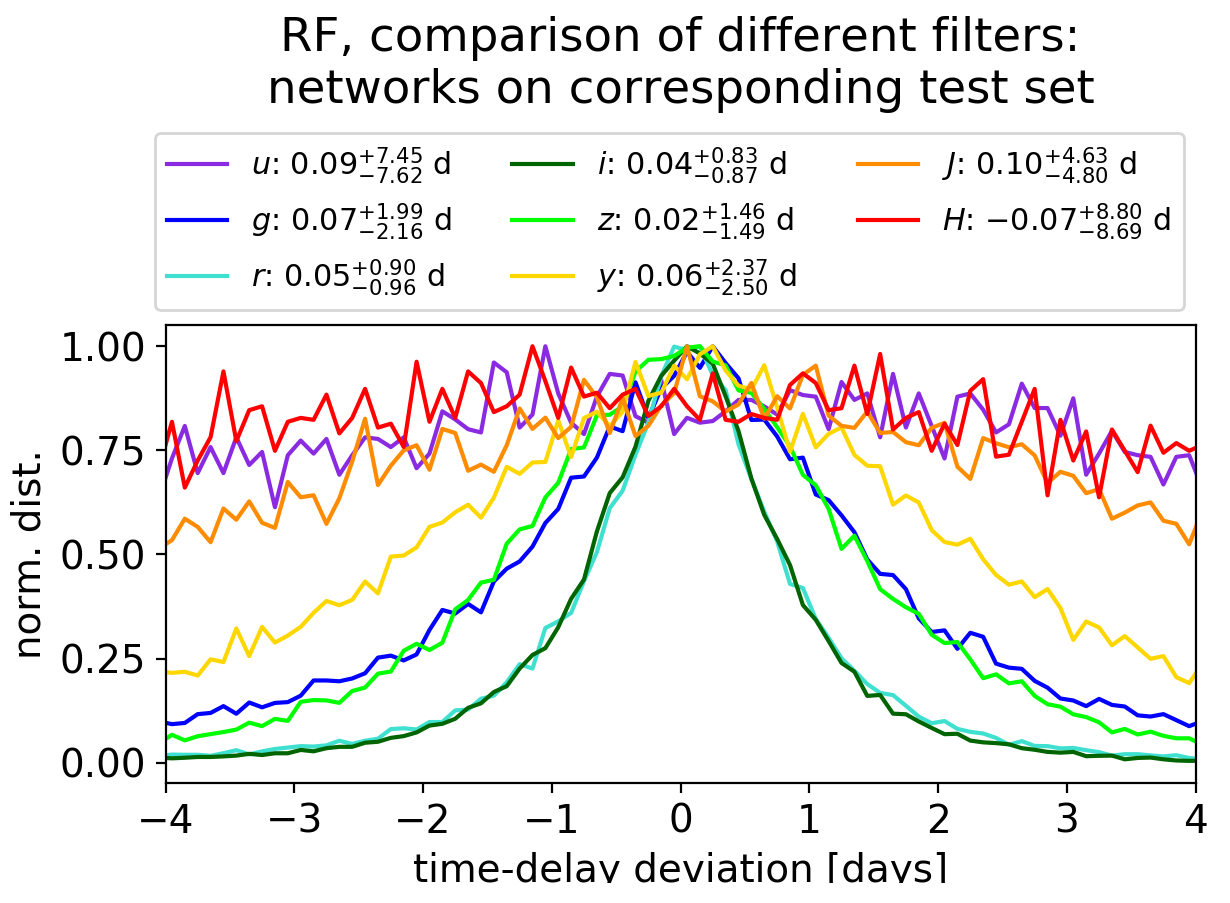}
\caption{Eight different RF models, each trained on a data set from a single band (as indicated in the legend) and evaluated on the whole sample from the corresponding test set, similar in procedure to Sect. \ref{sec: Best fit, DL vs RF}.}
\label{fig: 8 different bands, each one seperatly}
\end{figure}

\begin{figure}
\includegraphics[width=0.49\textwidth]{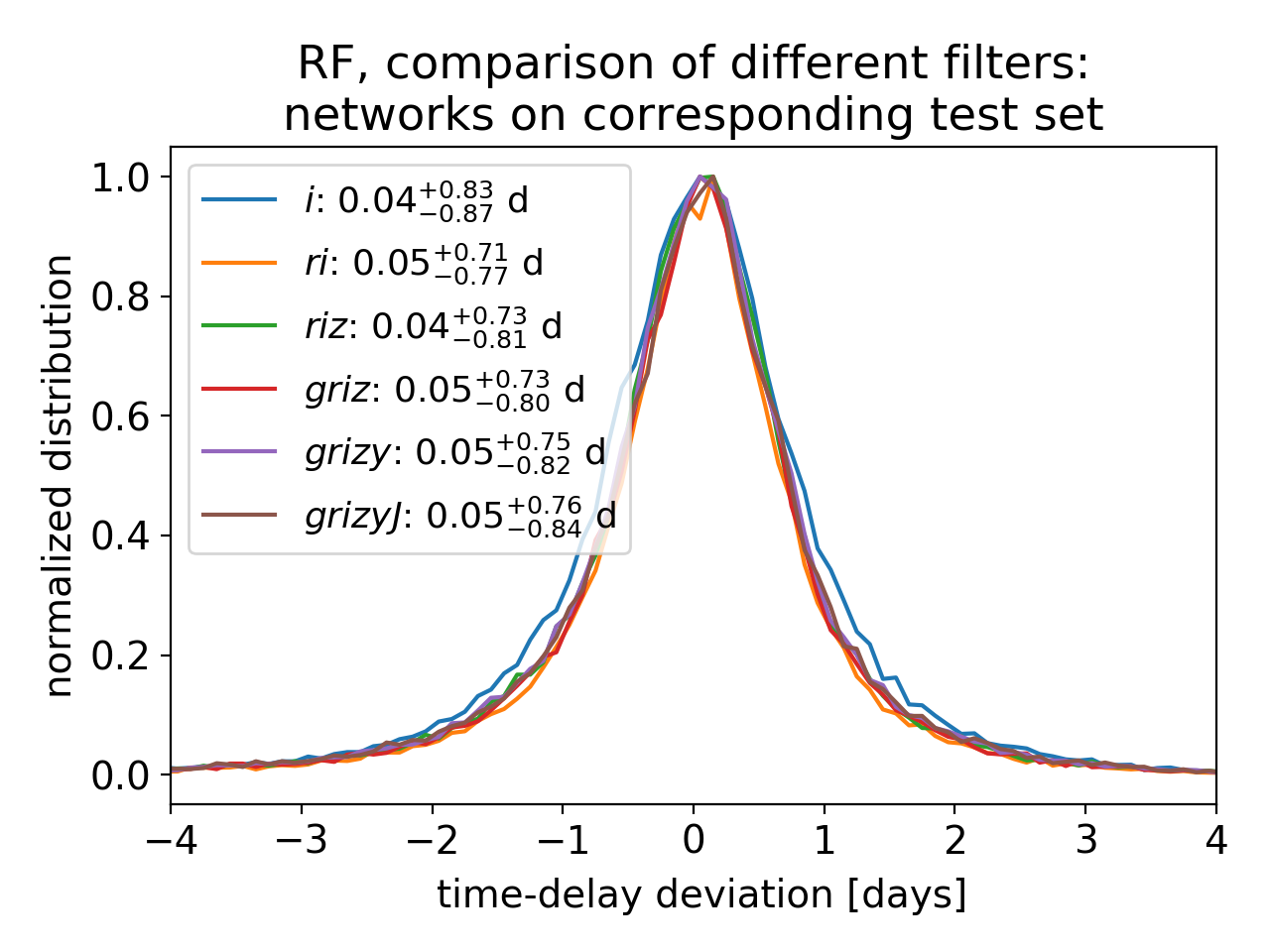}
\caption{Multiple filters used to train a single RF following the data structure as defined in Eq. (\ref{eq: data structure ri}) (example for $ri$). Each histogram is based on the whole sample of light curves from their corresponding test set. Using more than two filters does not improve the results further.}
\label{fig: different bands using in a single network}
\end{figure}

\begin{figure}
\includegraphics[width=0.49\textwidth]{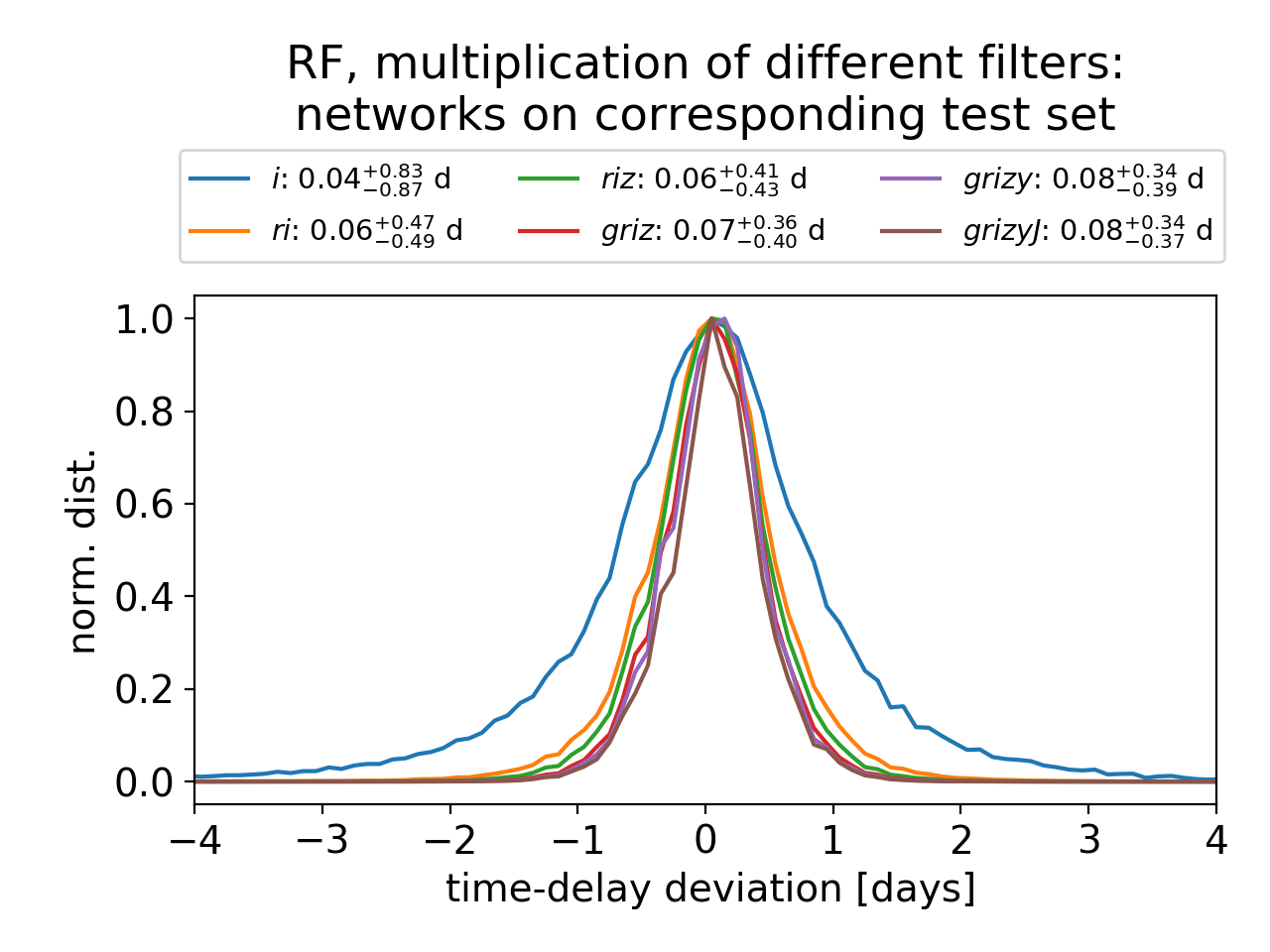}
\caption{Single RF trained per filter using a similar data structure as in Eq. (\ref{eq: data structure}) (example for the $i$ band), leading to six RF models for the six filters $g, r, i, z, y,$ and $J$. The combination
  of the filters is done by multiplying the corresponding
  distributions shown in Fig.
  \ref{fig: 8 different bands, each one seperatly}. We see that multiple filters help to
drastically  reduce the uncertainties. Therefore, observing three to four bands
  would be ideal.}
\label{fig: different bands multiplying distributions from singe networks}
\end{figure}

\section{Machine learning on further mock observations}
\label{sec: Machine learning on further mock-observation}

In this section we investigate further mock systems. We test systems with different moon phases (Sect. \ref{sec: moon phases}) and source, respectively lens redshifts (Sect. \ref{sec: source and lens redshifts}) to investigate the change of the uncertainties in comparison to our mock system from Sects. \ref{sec: Example data used for machine learning}, \ref{sec: Machine learning on example mock-observation}, and \ref{sec: Microlensing, observational noise and choice of filters}. Furthermore, we test the number of data points required before peak to achieve good time-delay measurements (Sect. \ref{sec: data points before peak}) and a quad system with various different properties in comparison to our previous studies (Sect. \ref{sec: quad lsne ia and higher microlensing uncertainties}).

\subsection{Different moon phases}
\label{sec: moon phases}

In this section we address the effect of different moon phases. We
assume the same LSN Ia as in Sects. \ref{sec: Example data used for
  machine learning}, \ref{sec: Machine learning on example
  mock-observation}, and \ref{sec: Microlensing, observational noise and choice of filters}, but place it differently in time. From Fig.
\ref{fig: appendix further bands of mock observations}, we can already
estimate that if we ignore the $u$ band, which has too low signal-to-noise anyway, mostly the $g$ band will be influenced as other bands
are significantly brighter than the 5$\sigma$ point-source depth or
there is only a minor dependence on the moon phase.

For the LSN Ia presented in Sects. \ref{sec: Example data used for
  machine learning}, \ref{sec: Machine learning on example
  mock-observation}, and \ref{sec: Microlensing, observational noise and choice of filters}, we see from Fig. \ref{fig: appendix further
  bands of mock observations} that for the $g$ band, the observations 
  before the peak are significantly affected by moon light,
  which according to Fig. \ref{fig: 8 different bands, each one
  seperatly} leads to an uncertainty around $2.1 \, \mathrm{d}$.
For a case where the peak in the $g$ band overlaps with the full moon we find a
similar uncertainty, whereas a case where the peak in the $g$ band
matches the new moon has an uncertainty around $1.7 \,
\mathrm{d}$. For cases where the peak is not significantly brighter
than the 5$\sigma$ point-source depth, the moon phase is important, but
given that our ML models work with a variable 5$\sigma$ point-source
depth, the effect of the moon phase is taken into account in our uncertainties.
In terms of follow-up observations,
one might consider to observe longer at full moon especially in the bluer bands
to reach a greater depth or resort to redder bands if the moon will likely affect the observations in the bluer bands adversely, but apart from that, we recommend in general to follow-up
all LSNe Ia independently of the moon phase.

\subsection{Source and lens redshifts}
\label{sec: source and lens redshifts}

The mock system we investigated in Sects. \ref{sec: Example data
  used for machine learning} and \ref{sec: Machine learning on example
  mock-observation} has $\sourcez = 0.76,$ which roughly corresponds to
the median source redshift of the OM10 catalog. Furthermore, we have
learned from Sect. \ref{sec: Uncertainties due to microlensing and noise} that
the observational noise is the dominant source of uncertainty and we
therefore expect a large dependence of the time-delay measurement on
$\sourcez$ (assuming a fixed exposure time during observations).

We therefore investigate in this section $\sourcez = 0.55$ and
$\sourcez = 0.99$, which correspond to the 16th and 84th percentiles, respectively, 
of the source redshift from LSNe Ia in the OM10 catalog. To probe just
the dependence on $\sourcez$, we leave all other parameters as defined
in Table \ref{tab: Example double LSNe Ia}. We do not scale the absolute time delay
with the source redshift, since this is just a hypothetical experiment to demonstrate 
how different brightnesses, related to the source redshift, influence the time-delay measurement.

 The two cases are shown in
Fig. \ref{fig: example light curve different source and lens
  redshifts}, where we see the much better quality of the light curve
for $\sourcez = 0.55$ (upper panel) in comparison to $\sourcez = 0.99$
(lower panel). Further, we also probe the lens redshift by
investigating $\lensz=0.16$ and $\lensz=0.48$, which also corresponds
to the 16th and 84th percentile of the OM10 catalog and where we also
leave other parameters unchanged.

The results are summarized in Table \ref{tab: source and lens redshift
  investigation}. We see that in comparison to $\sourcez = 0.76$, the
case with $\sourcez = 0.55$ has an improved uncertainty by $\sim$%
$0.2 \, \mathrm{d}$, where the case $\sourcez = 0.99$ has a reduced
uncertainty by $\sim$%
$0.7 \, \mathrm{d}$. This trend is expected, and
means that especially for the case of $\sourcez = 0.99$, a greater
depth would improve the results significantly. Comparing the results of
varying lens redshifts, we see a much smaller impact on the
uncertainty.  Still there is a slight trend that higher lens redshifts
correspond to larger time-delay uncertainties, which is in good agreement with
\cite{Huber:2020dxc}, who find the tendency that microlensing
uncertainties increase with higher lens redshift if everything else is
fixed. The reason for this is that the
physical size of the microlensing map decreases with higher lens redshift, which makes a SN Ia
appear larger in the microlensing map and therefore events where micro
caustics are crossed are more likely. More details are available in
\cite{Huber:2020dxc}.

The impact of the source redshift on the best filters to target is
discussed previously in Sect. \ref{sec: filters used for training}.

\begin{table}
\centering
\caption{Time-delay measurement of different LSNe Ia with varying source and lens redshifts.}
\begin{tabular}{ccc}
$\sourcez, \lensz$ & corresponding test set & \texttt{SNEMO15} data set \\
\midrule
0.76, 0.252 (Fig. \ref{fig: example light curve 187 system}) & $0.04^{+0.83}_{-0.87} \, \mathrm{d}$ & $0.02^{+1.38}_{-1.42} \, \mathrm{d}$ \\
\midrule
0.55, 0.252 (Fig. \ref{fig: example light curve different source and lens redshifts})& $0.04^{+0.59}_{-0.67} \, \mathrm{d}$ & $0.01^{+1.29}_{-1.25} \, \mathrm{d}$ \\[0.07cm]
0.99, 0.252 (Fig. \ref{fig: example light curve different source and lens redshifts})& $0.01^{+1.64}_{-1.66} \, \mathrm{d}$ & $0.02^{+2.1}_{-2.16} \, \mathrm{d}$ \\
\midrule
0.76, 0.16 & $0.04^{+0.83}_{-0.89} \, \mathrm{d}$ & $-0.09^{+1.26}_{-1.30} \, \mathrm{d}$ \\[0.07cm]

0.76, 0.48 & $0.06^{+0.97}_{-1.06} \, \mathrm{d}$ & $-0.09^{+1.45}_{-1.51} \, \mathrm{d}$ \\

\end{tabular}

\label{tab: source and lens redshift investigation}
\end{table}

\begin{figure}
\includegraphics[width=0.48\textwidth]{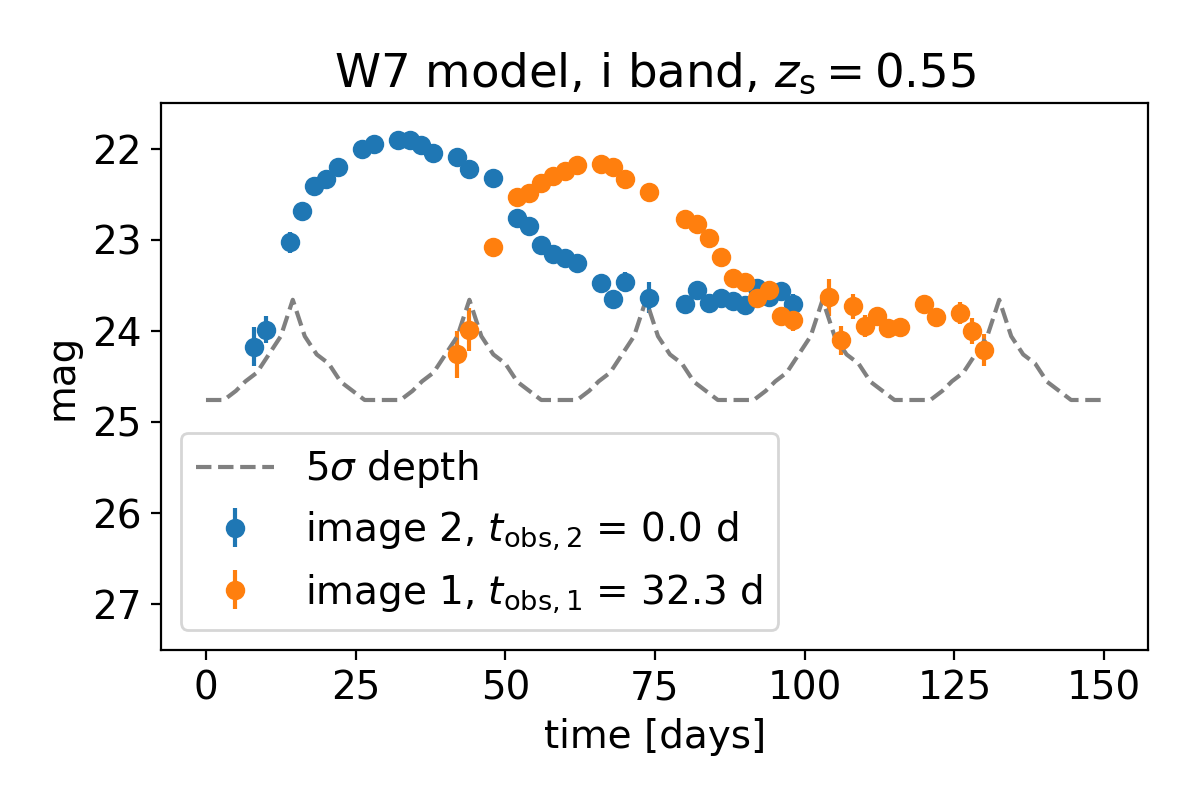}
\includegraphics[width=0.48\textwidth]{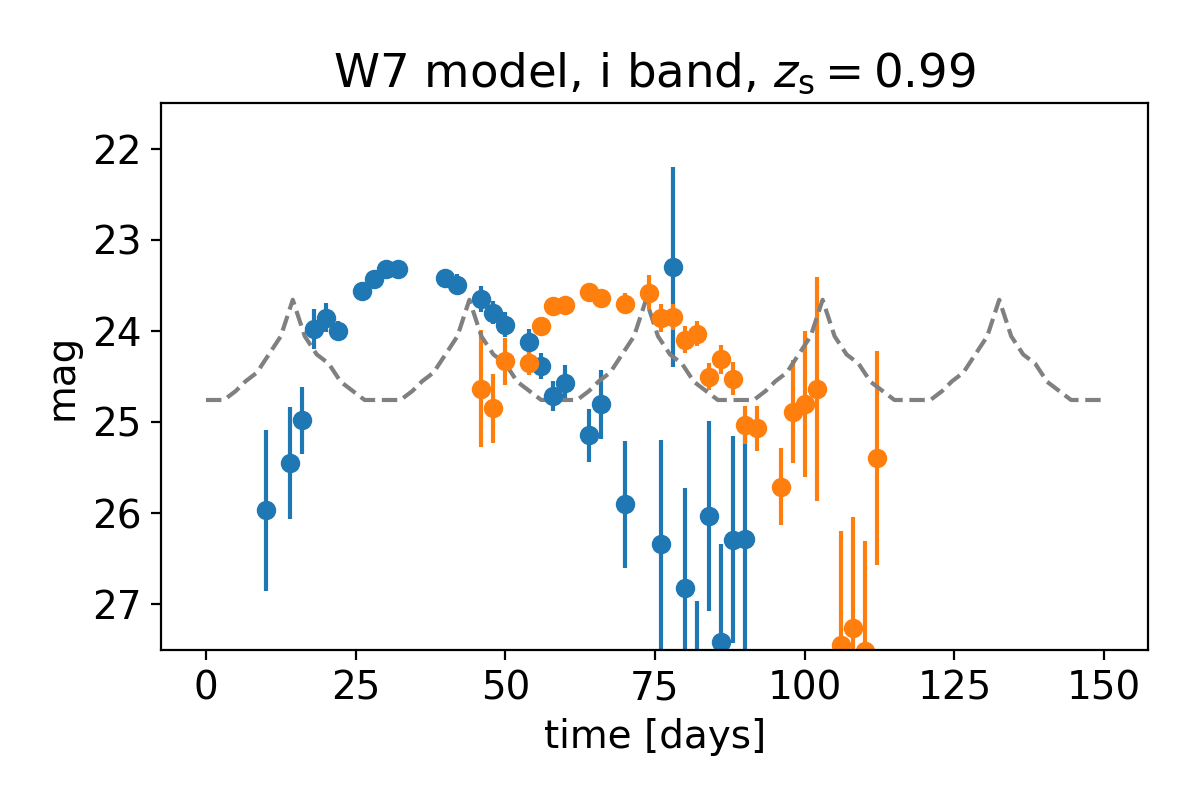}
\caption{Two LSNe Ia similar to Fig. \ref{fig: example light curve
    187 system} but with different source redshifts. The LSN Ia in
  the upper panel has $\sourcez =0.55,$ and the one in the lower panel
  has $\sourcez =0.99$.}
\label{fig: example light curve different source and lens redshifts}
\end{figure}

\subsection{Data points before peak}
\label{sec: data points before peak}

\begin{figure}
\includegraphics[width=0.49\textwidth]{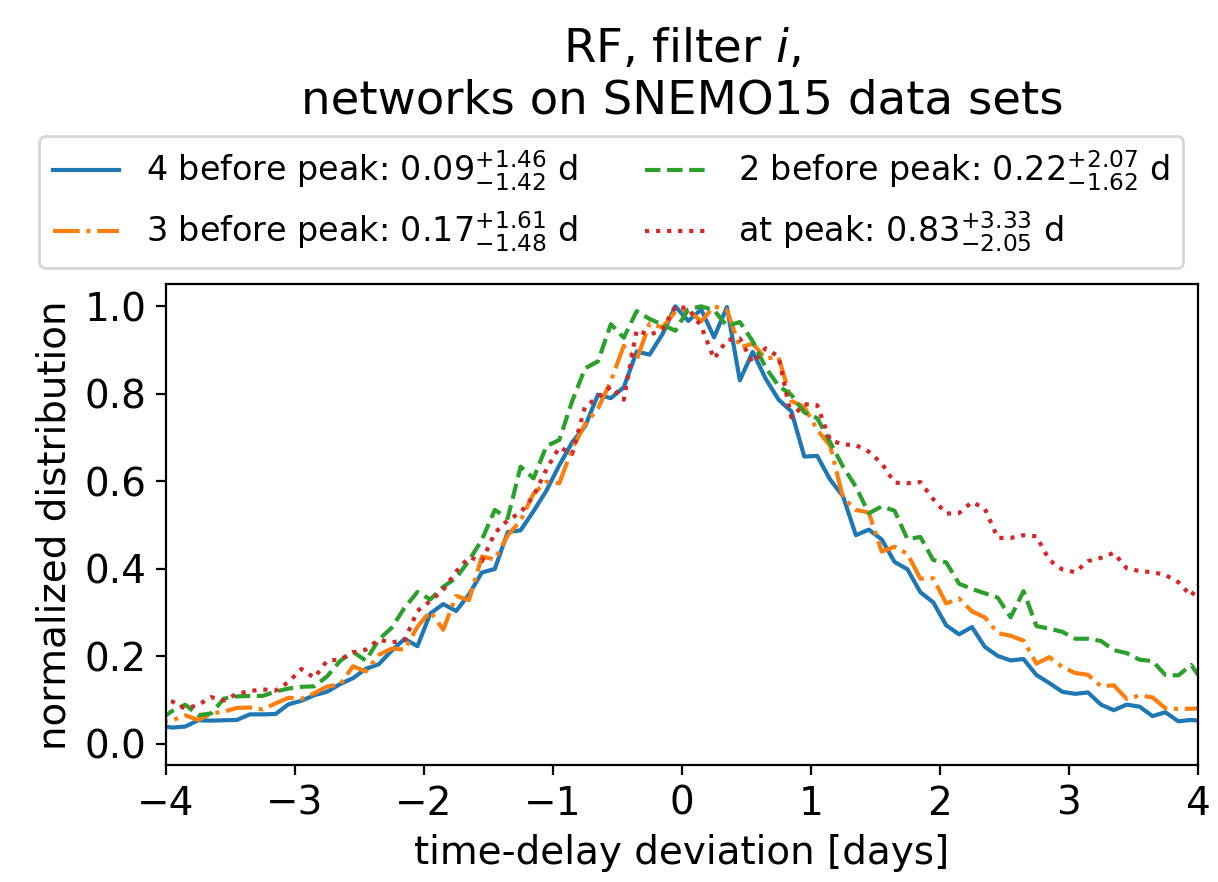}
\caption{
  Time-delay deviations of mock observations similar to Fig. \ref{fig: example light curve 187 system} but with a later detection, meaning fewer data
  points before the peak in the $i$ band of the first-appearing SN image. Each histogram is based on the whole sample of light curves from the related \texttt{SNEMO15} data set. We compare the cases where we have
  four, three, or two data points before the peak in
  comparison to an at-peak detection.}
\label{fig: data points before peak}
\end{figure}

In this section we discuss the number of data points required before
peak to achieve a good time-delay measurement. The case presented in
Sect. \ref{sec: Example data used for machine learning} has a large
number of data points before peak, which is not always achievable in
practice, especially since vetting of transient candidates and triggering of light-curve observations often require additional time.  Therefore, we investigate a similar mock system as in Fig.
\ref{fig: example light curve 187 system}, but with a later
detection in the first-appearing SN image. In Fig. \ref{fig: data points before peak}, we show a case
where we have the first data point at the peak in the $i$ band in
comparison to three other cases where we have four, three, or two data
points before the peak. The case for the at-peak detection provides as
expected the worst precision but more worrying is the large bias of
0.83 days. Already two data points before peak improve the results
significantly and allow precision cosmology for LSNe Ia with a
time delay greater than 22 days. Nevertheless, we aim for four data
points before peak as we could achieve a bias below 1 percent already
for a delay greater than 10 days; furthermore, the precision is also
improved substantially and almost at the level of the mock observation in
Fig. \ref{fig: example light curve 187 system} and corresponding results in
Fig. \ref{fig: SENOM15 test for DL and RF using
  filters iz}. This would correspond in the observer frame to a
detection about eight to ten days before the peak in the $i$ band.
 Given that a SN Ia typically peaks $\sim$18 rest-frame days after explosion and the typical lensed SN redshift is $\sim$0.7, we would need to detect and start follow-up observations of the first-appearing SN image within $\sim$%
$15$ days (observer frame) in order to measure accurate time delays.
The results presented here are in good agreement with the feature importance investigations shown in Fig. \ref{fig: appendix feature importance}, where we find that especially the rise slightly before the peak is very important for the RF.

\subsection{Quad LSNe Ia and higher microlensing uncertainties}
\label{sec: quad lsne ia and higher microlensing uncertainties}

\begin{figure}
\includegraphics[width=0.49\textwidth]{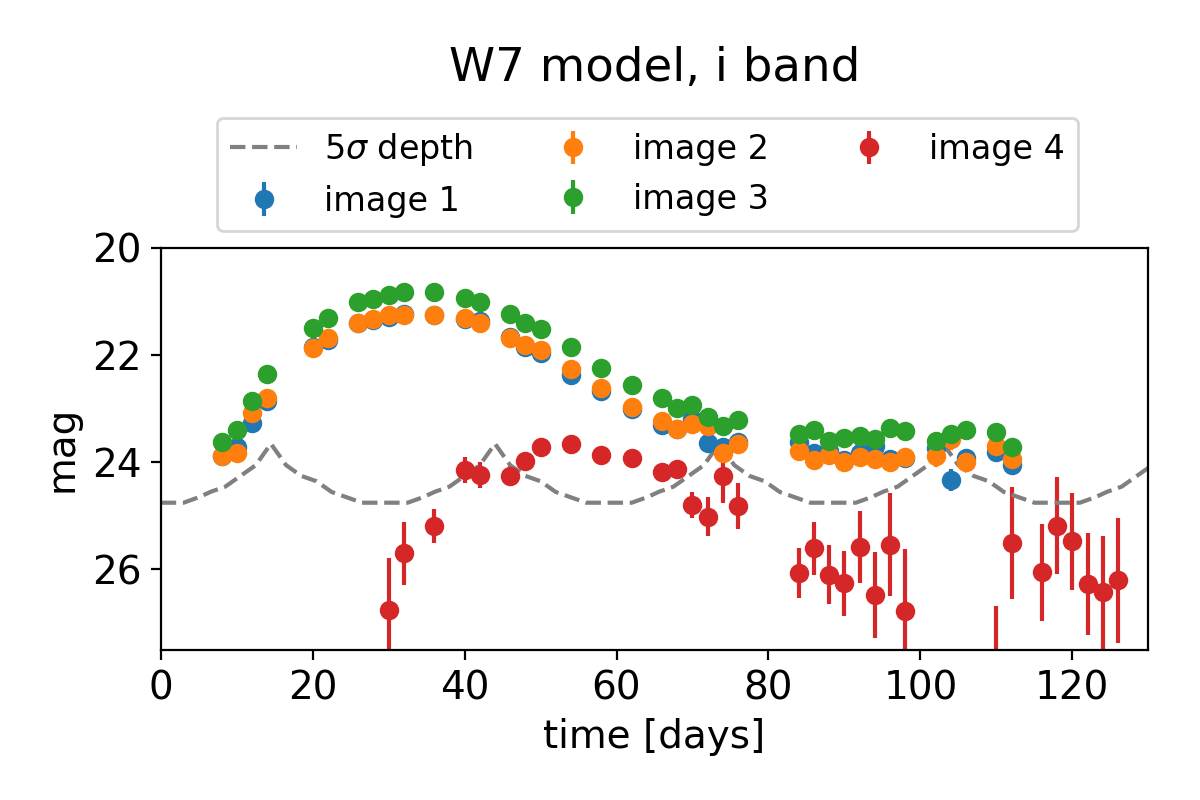}
\caption{Light curves of the mock quad LSN Ia from Table \ref{tab: quad LSN Ia mock} for the $i$ band.}
\label{fig: quad LSN Ia mock}
\end{figure}

So far we have only discussed double LSNe Ia, but in this section we
present a LSN Ia with four images. Our mock quad LSN Ia is
similar to the one presented in Sect. \ref{sec: Example data used
  for machine learning}, but we varied the source position for the
double system in the same lensing environment using the GLEE software
\citep{Suyu:2010,Suyu:2012ApJ} such that we get a quad system, where the
parameters are listed in Table \ref{tab: quad LSN Ia mock} and the light curves from the
system are shown in Fig. \ref{fig: quad LSN Ia mock}. For images one
to three, the $\kappa$ and $\gamma$ values are closer to 0.5 in
comparison to the double system from Table \ref{tab: Example double
  LSNe Ia}, which means that the macro magnification is higher but
microlensing uncertainties are increased as shown in
\cite{Huber:2020dxc}. For image four, we have $\kappa$ and $\gamma$
values far from 0.5; this leads to lower microlensing
uncertainties but therefore also to a much fainter image, which can be
seen in Fig. \ref{fig: quad LSN Ia mock}.

\begin{table}
\caption{Source redshift, $\sourcez$, lens redshift, $\lensz$, convergence, $\kappa$, shear, $\gamma$, and the time values for the four images of a mock quad LSN Ia.}
\begin{tabular}{cccccc}
& $\sourcez$ & $\lensz$  & $(\kappa,\gamma)$ & t [d] \\
\midrule
image 1 & 0.76 & 0.252 & (0.435, 0.415)  & $\equiv0.00$\\[0.07cm]
image 2 & 0.76 & 0.252 & (0.431, 0.424)  & 0.01\\[0.07cm]
image 3 & 0.76 & 0.252 & (0.567, 0.537)  & 0.34\\[0.07cm]
image 4 & 0.76 & 0.252 & (1.28, 1.253)  & 20.76
\end{tabular}
\centering

  \vspace{1ex}
     {\raggedright \textbf{Notes}. The image separation varies between 0.6 and 1.6 arcsec, and therefore it might be challenging to resolve all images with ground-based telescopes given limits due to seeing. \par}
\label{tab: quad LSN Ia mock}
\end{table}

In principle such a quad system can be investigated in two ways. The
first approach is to train a separate RF per
pair of images, leading to six RF models in total. The other way is to
train a single RF for the whole quad system that takes as input
magnitude values of four images instead of two images, similar to
Eq. (\ref{eq: data structure}). The outputs as shown in Figs.
\ref{fig: fully connected neural network} and \ref{fig: regression
  tree} are then four instead of two time values.

The results for both approaches are summarized in Table \ref{tab: quad
  LSN Ia mock uncertainties trained as double vs. quad} and the
correlation plots are shown in Appendix \ref{sec:Appendix correlation
  plots}. We find fewer correlations for the approach ``separate RF per pair of images"
than for the approach ``single RF for all images," especially for the cases where the noisy fourth
image is included in the time-delay measurement. This is because in
the first case, six RF models are trained independently from each other,
whereas the second case only uses a single RF that predicts four
time values for the four images. Still the case ``separate RF per pair
of images" is preferred because it provides lower biases and tighter
constraints. This is not surprising, as providing all the data from the
four images at once is a much more complex problem to handle in
comparison to training a RF for just two images. While the time-delay 
deviations between both approaches are almost comparable for pairs of images among the first, second and third images, for
the cases where the fourth image is included, the single RF for
the whole quad system performs much worse. This suggests that
especially handling noisy data can be treated better in the approach
of a separate RF for each pair of images and therefore it is
always preferred to train a separate RF per pair of images.

In the following we analyze the different uncertainties of
the time-delay measurements from different pairs of images as shown in Table 
\ref{tab: quad LSN Ia mock uncertainties trained as double vs. quad}. The most
precise time delay is the one between the first and second image, but
if we compare this uncertainty to the uncertainty of the lower panel
of Fig. \ref{fig: SENOM15 test for DL and RF using filters iz} for
the double LSNe Ia from Fig. \ref{fig: example light curve 187
  system}, we see that the precision is 0.2 days worse. This can be
easily explained by the higher microlensing uncertainties coming from
the $\kappa,$ and $\gamma$ values much closer to $0.5$ as shown in
Table \ref{tab: quad LSN Ia mock} in comparison to Table \ref{tab: Example double LSNe
  Ia}. Higher microlensing uncertainties are also the reason why
uncertainties of $\Delta t_{31}$ and $\Delta t_{32}$ are larger than
$\Delta t_{21}$, even though the third image is the brightest one and
therefore has the lowest amount of observational noise.  The precision
and also accuracy of the time-delay measurement where image four is
involved are the worst in Table \ref{tab: quad LSN Ia mock
  uncertainties trained as double vs. quad}, which is explained by the
very poor quality of the light curve from the fourth image. We further
see that $\Delta t_{31} $ and $\Delta t_{32}$ as well as $\Delta
t_{41}$ and $\Delta t_{42}$ have very similar uncertainties, which is
expected since light curves from image one and two are almost
identical and therefore this is a good check of consistency.

Even though the time-delay measurements between the first
three images have the lowest time-delay deviation in days, the absolute time
delay is very short, which leads to a very high relative deviation. For
this specific mock quad LSNe Ia, it would only make sense to measure
time delays with respect to the fourth image, where we would achieve a
precision around 10 percent and an accuracy of 0.7 percent.

\begin{table}
\centering
\caption{Deviations of the time-delay measurements
  ($\tau_{ij} = \Delta t_{ij} - \Delta t_{\mathrm{true},ij}$) 
  for the LSN Ia
  quad system shown in Fig. \ref{fig: quad LSN Ia mock}. }
\begin{tabular}{cccccc}
& separate RF & single RF \\
& per pair of images  & for all images \\
\midrule
Time-delay dev. of $\Delta t_{21}$ & $0.01^{+1.63}_{-1.63} \, \mathrm{d} $ & $-0.01^{+1.65}_{-1.64} \, \mathrm{d} $\\[0.07cm]
Time-delay dev. of $\Delta t_{31}$ & $-0.05^{+1.85}_{-1.85} \, \mathrm{d} $ & $0.01^{+1.89}_{-1.87} \, \mathrm{d} $\\[0.07cm]
Time-delay dev. of $\Delta t_{41}$ & $0.15^{+1.84}_{-1.96} \, \mathrm{d} $ & $0.26^{+2.24}_{-2.36} \, \mathrm{d} $\\[0.07cm]
Time-delay dev. of $\Delta t_{32}$ & $-0.03^{+1.86}_{-1.89} \, \mathrm{d} $ & $0.01^{+1.89}_{-1.88} \, \mathrm{d} $\\[0.07cm]
Time-delay dev. of $\Delta t_{42}$ & $0.14^{+1.81}_{-1.93} \, \mathrm{d} $ & $0.26^{+2.25}_{-2.33} \, \mathrm{d} $\\[0.07cm]
Time-delay dev. of $\Delta t_{43}$ & $0.15^{+2.07}_{-2.19} \, \mathrm{d} $ & $0.25^{+2.40}_{-2.55} \, \mathrm{d} $\\
\end{tabular}

   \vspace{1ex}
     {\raggedright \textbf{Notes}. The second
  column shows the case where a separate RF is trained per
  pair of images, leading to six RF models in total, in comparison
  to a single RF (third column) for the whole quad system. \par}
\label{tab: quad LSN Ia mock uncertainties trained as double vs. quad}
\end{table}

\section{Discussion}
\label{sec: Discussion}

We train a FCNN with two hidden layers
and a RF using four theoretical SN Ia models, to measure time delays in LSNe
Ia. We find that both ML models work very well on a test set based on the same four
theoretical models used in the training process, providing uncertainties around 0.7 to 0.9 days for
the $i$ band almost without any bias. Applying the trained ML models to
the \texttt{SNEMO15} data set, which is composed of empirical SN Ia light 
curves not used in the training process, we find that the uncertainties increase
by about 0.5 days, but this is not surprising as such light curves have
never been used in the training process and a measurement with a 1.5-day uncertainty on a single band is still a very good measurement. 

However, when applied to the \texttt{SNEMO15} data set, the FCNN yields 
biased results. The biases are mostly within 0.4 days, but larger 
ones are also possible, making our FCNN approach not suitable for precision cosmology.
Furthermore, this shows that the generalizability to light-curve shapes 
not used in the training process is not working for our FCNN approach,
since biases on the corresponding test set composed of four theoretical models as used in 
the training process are negligible.
This was already suggested by results presented in 
Fig. \ref{fig: three models for training evaluating on single model test sets},
where the training on three theoretical models was not general enough to 
perform well on the fourth model not used in the training process.
However, we introduced random shifts in time of the light curves,
which reduced the bias significantly and motivated us to apply our FCNN trained on four theoretical models (with random shifts in time to reduce the bias), to the \texttt{SNEMO15} data set as a final test, where we find unfortunately significant biases.
Deeper and larger fully connected networks will not solve this problem as they will just fit
the training data better and do not guarantee the generalizability.
To overcome this, regularization, dropout or uncertainty estimates as additional output to the time values might help. However, this would be some kind of fine tuning to our 
\texttt{SNEMO15} data set, because our investigations up to that 
stage
(which shows that our FCNN, trained using three theoretical models, generalizes well to a test set composed of four theoretical models, with negligible resulting biases),
were very encouraging to apply our FCNN
to the final test (\texttt{SNEMO15} data set), which it failed. However, we defer further 
investigations of FCNN to future work, 
especially since more complex ML approaches such as recurrent neural networks or
long short-term memory networks \citep{Sherstinsky_2020} might fit the problem even better.

The RF provides significantly lower biases on the \texttt{SNEMO15} data set
 -- with 4 or more data points before peak, which means a detection of the first LSNe Ia image
about eight to ten days before peak, the bias can be kept within 0.10
days. If one of the images is very faint as shown in Fig. \ref{fig:
  quad LSN Ia mock}, we still can reach an accuracy of 0.15 days and
therefore a delay longer than 15 days provides already a time-delay
measurement better than 1 percent. Given the low bias in the RF
especially in comparison to the FCNN, the RF is the one to
use for a real application.

\begin{figure}
\includegraphics[width=0.49\textwidth]{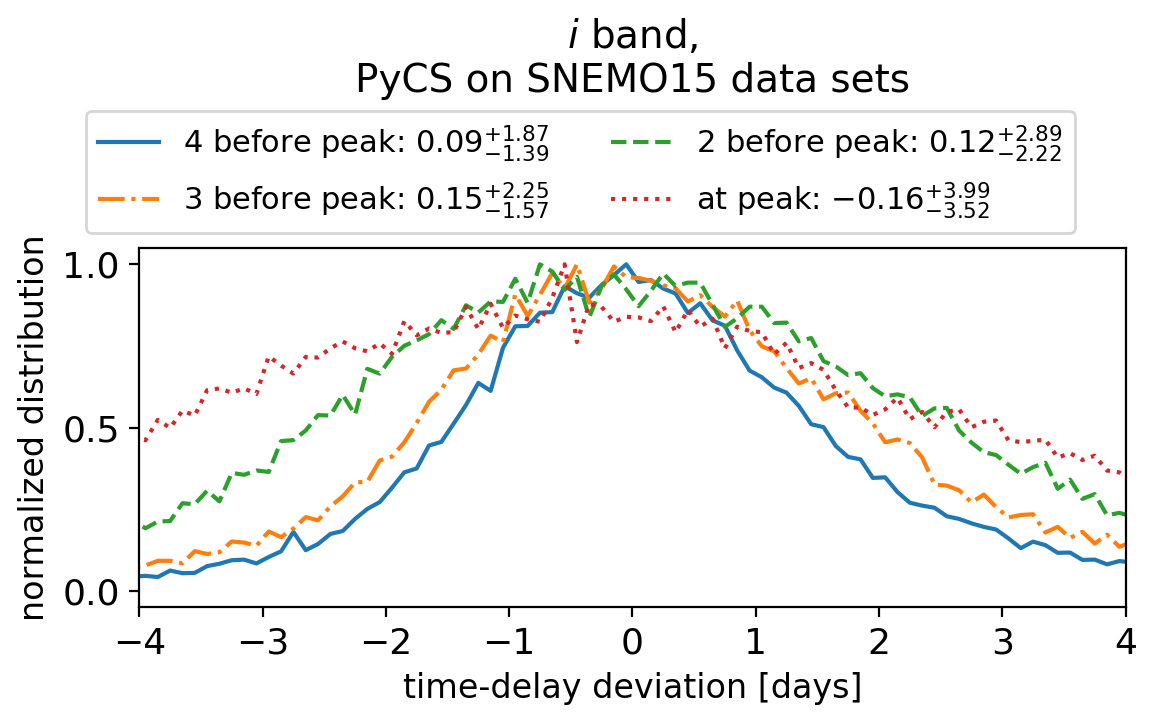}
\caption{Same as Fig. \ref{fig: data points before peak} but this time using \texttt{PyCS} on all samples from the \texttt{SNEMO15} data set.}
\label{fig: data points before peak using PyCS}
\end{figure}

\cite{Huber:2019ljb} used the free-knot spline estimator from
\texttt{PyCS} \citep{2013:Tewesb,Bonvin:2015jia} to measure time
delays for LSNe Ia. To compare this approach to our results, we apply
\texttt{PyCS} as used in \cite{Huber:2019ljb} to the \texttt{SNEMO15}
data set. For the system shown in Fig. \ref{fig: example light curve
  187 system} with a very well sampled light curve, we achieve similar
uncertainties as the RF shown in Fig. \ref{fig: SENOM15 test for DL
  and RF using filters iz}. However, as soon as we look at cases,
where we have a reduced number of data points before peak as shown in
Fig. \ref{fig: data points before peak using PyCS} (in comparison to
the RF results in Fig. \ref{fig: data points before peak}), we see
that the RF approach achieves a much higher precision. In terms of the
bias, as long as we provide two data points or more before peak the RF
and \texttt{PyCS} provide sufficient results. For the case where the
first data point is at the peak of the $i$ band, even though
\texttt{PyCS} provides a much better bias than the RF, the
measurement has substantially poorer
precision. Overall the RF works better to measure time delays
in LSNe Ia in most cases in comparison to \texttt{PyCS}. However in a
real application, both approaches could be used to cross-check the time-delay measurements.

\section{Summary}
\label{sec: Summary}

In this work we have introduced two ML techniques, namely
a deep learning network using a FCNN with two hidden layers
and a RF to measure time delays of LSNe Ia. We
simulated LSN Ia light curves for the training process, including
observational noise and microlensing uncertainties using four different
theoretical models. Our training set is composed of 400000 LSNe Ia
coming from 4 theoretical models, 10000 microlensing map positions, and
10 noise realizations. Our test set has a size of 40000 LSNe Ia, and
we drew 1000 microlensing map positions instead of 10000 as for the
training set. We constructed a further data set based on the empirical
\texttt{SNEMO15} model to create realistic LSN Ia light curves not
used in the training process to check if our approach is general
enough to handle real observations of LSNe Ia. To add microlensing to
the \texttt{SNEMO15} model, we used the microlensed light curves from
the theoretical models, subtracting the macrolensed light curve
to get the microlensing contribution.

To summarize our results, we looked at the more realistic results from
the empirical \texttt{SNEMO15} data set. From the
investigation of the RF and the FCNN, we find that only the RF provides
sufficiently low bias and is therefore the approach to use in a real
application. From all investigated systems where we assumed a two-day cadence with
a few random gaps, we found that we can
achieve an accuracy better than 1\% for the RF if we restrict ourselves
to LSN Ia systems with a delay longer than 15 days, where we obtain
the first data point around eight to ten days before the peak in the light curve of the first-appearing 
SN image. In terms of precision, we can achieve an uncertainty of 1.5 days from the 
$i$ band alone for the median source redshift $\sim$%
$0.76$ of LSNe Ia in
OM10. Using three bands where the time delay is measured separately
for each RF and combined afterward, we can reach an approximately $1.0$-day uncertainty. The three most promising filters to target are $g,
r,$ and $i$ for $\sourcez \lesssim 0.6$ and $r, i,$ and $z$ for higher
source redshifts. As a fourth and fifth band, the $z$ and $y$  for $\sourcez \lesssim 0.6$
and the $g$ and $y$  for $\sourcez \gtrsim 0.6$ might be considered. 
We find that the gain from multiple filters
is the best if a ML model is trained individually per band. The other
bands investigated in this work ($u, J,$ and $H$) provide very poor-quality
light curves and are therefore not useful.

From our investigations, we find mainly that the observational noise is
the dominant source of uncertainty in measuring time delays, and to improve the results
presented here, a greater depth would be required. The depth we assumed
for follow-up observations is one magnitude deeper than the
single-epoch LSST-like 5$\sigma$ depth, meaning 25.7, 25.3, 24.7, 23.8,
and 23.0 for $g, r, i, z,$ and $y$, respectively.  From the investigation of the
source redshifts, we find that in comparison to the median source redshift of $\sim$%
$0.76$ of LSNe Ia in OM10, $\sourcez = 0.55$ can
improve the precision in the $i$ band by 0.2 days, but $\sourcez =
0.99$ might lower the uncertainty by 0.7 days, which suggests that,
especially for higher source redshifts, a greater depth might be
required. Although a greater depth could also compensate for the moon
phase, the impact on the uncertainty is weaker (an at most 0.4 days worse
uncertainty in our investigation) and becomes even less relevant the
redder the bands are. We further find that typical uncertainties in
the microlensing parameters ($\kappa, \gamma,$ and $s$) are not
relevant for our training process. Only a significantly overestimated $s$ value
would lead to an underestimation of the uncertainties. Furthermore, we
find that our approach works best if an individual RF is trained per
pair of images.

In comparison to the free-knot spline estimator from \texttt{PyCS}
\citep{2013:Tewesb,Bonvin:2015jia} as used in \cite{Huber:2019ljb}, our
approach works better overall, providing an improved precision of up to
$\sim$0.8 days. We therefore can expect slightly more LSNe Ia with
well-measured time delays than the number predicted by \cite{Huber:2019ljb}.

In this work we have developed a new method to measure time delays of
LSNe Ia. The RF provides accurate and precise time-delay measurements
that are comparable with or better than current methods and is therefore an
important tool to pave the way for LSNe Ia as cosmological probes.
The downsides of our approach are: that a RF needs to be trained separately for
each individual system's observing pattern; the dependence on the SN Ia models used
in the training process; and that our approach cannot for the moment be applied to
other types of LSNe. A highly promising approach to overcoming this and building a ML network that
is more general is the use of recurrent neural networks or
long short-term memory networks \citep{Sherstinsky_2020}, which will be investigated in a future study.

\FloatBarrier

\begin{acknowledgements}
We thank F.~Courbin, S.~Schuldt and R. Cañameras for useful discussions. We also would like to thank the anonymous
referee for helpful feedback, which strengthened this work.
SH and SHS thank the Max Planck Society for support through the Max
Planck Research Group for SHS. This project has received funding from
the European Research Council (ERC) under the European Union’s Horizon
2020 research and innovation programme (grant agreement No
771776). 
This research is supported in part by the Excellence Cluster ORIGINS which is funded by the Deutsche Forschungsgemeinschaft (DFG, German Research Foundation) under Germany's Excellence Strategy -- EXC-2094 -- 390783311.
DG acknowledges support from the Baden-Württemberg Foundation through the Baden-Württemberg Eliteprogramm for Postdocs.
UMN has been supported by the Transregional Collaborative Research
Center TRR33 ‘The Dark Universe’ of the Deutsche
Forschungsgemeinschaft.
JHHC acknowledges support from the Swiss National Science
Foundation and through European Research Council (ERC) under the European
Union's Horizon 2020 research and innovation programme (COSMICLENS:
grant agreement No 787866).
MK acknowledges support from
the Klaus Tschira Foundation. 
%

\end{acknowledgements}

\bibliographystyle{aa}
\bibliography{Time_delay_measurement}

\FloatBarrier

\clearpage
\begin{appendix}
\onecolumn
\section{Photometric uncertainties}
\label{sec:Appendix LSST uncertainty}

The photometric uncertainty $\sigma_1$ from Eq. \ref{eq:noise realization random mag including error LSST science book} is defined as:
\begin{equation}
\sigma_1^2 = \sigma_\mathrm{sys}^2+\sigma_\mathrm{rand}^2$,
where $\sigma_\mathrm{sys}=0.005 \, \mathrm{mag}
\label{eq:noise calculation appendix formula 1}
\end{equation}
and 
\begin{equation}
\sigma_\mathrm{rand}^2=(0.04-\gamma^c) x + \gamma^c x^2 (\mathrm{mag}^2).
\label{eq:noise calculation appendix formula 2}
\end{equation}
The parameter $\gamma^c$ varies from 0.037 to 0.040 for different filters and
$x=10^{0.4(m_\mathrm{AB,X}-m_5)}$, where $m_\mathrm{AB,X}$ is the AB magnitude in filter $X$ from Eq. (\ref{eq: microlensed light curves for ab magnitudes}) of the SN data
point and $m_5$ is the 5$\sigma$ point-source depth (for more details, see \cite{2009:LSSTscience}, Sect. 3.5, p. 67). 

A magnitude value $m_\mathrm{AB,X}$, which is much higher (fainter) than the 5$\sigma$ point-source depth $m_5$, can lead to unrealistic $m_\mathrm{data,X}$ in Eq. (\ref{eq:noise realization random mag including error LSST science book}). Normally one would just delete such a data point, but for our ML model this is not possible, given that we always need the same number of data points as input. Therefore, to avoid such outliers in our data set used for training, validation and testing, we ensure that all magnitude values lower than $m_5$ cannot exceed $m_5$ (bright data point just due to large error) or be much fainter than this value (data point not observable). Specifically, if the SN image brightness $m_\mathrm{AB,X}$ (see Eq. (\ref{eq:noise realization random mag including error LSST science book})) is fainter than $m_5$ and the calculated uncertainties would lead to a $m_\mathrm{data,X}$ that is smaller (brighter) than $m_5$ or larger (fainter) than $\mathrm{max}_t(m_\mathrm{AB,X}(t)) + (\mathrm{max_t}(m_\mathrm{AB,X}(t)) - m_5)$, we replace that data point with a uniform random number between $m_5$ and $m_\mathrm{AB,X} + (m_\mathrm{AB,X} - m_5)$, where the term $(m_\mathrm{AB,X} - m_5)$ ensures that the uniform random number can be fainter than $m_\mathrm{AB,X}$,
but not by more than the magnitude difference between $m_\mathrm{AB,X}$ and $m_5$.

\FloatBarrier

\section{Light curves of mock observation in multiple bands}
\label{sec:Appendix further bands of mock observation}

\begin{figure}[h!]
\subfigure{\includegraphics[width=0.24\textwidth]{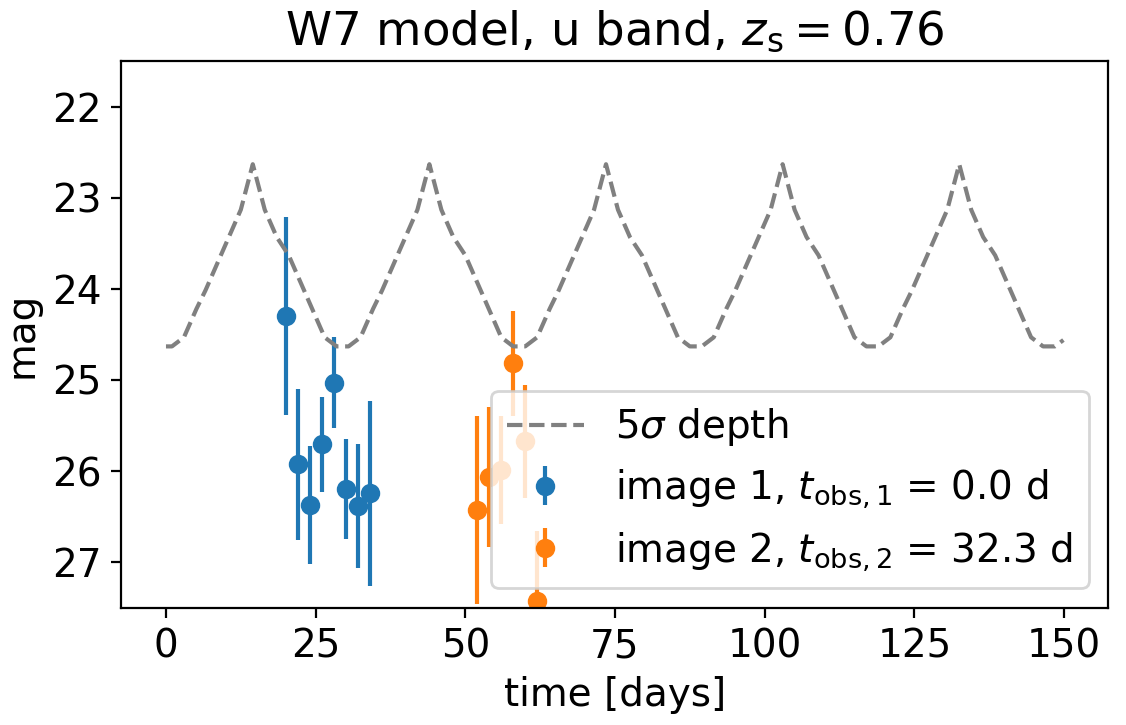}}
\subfigure{\includegraphics[width=0.24\textwidth]{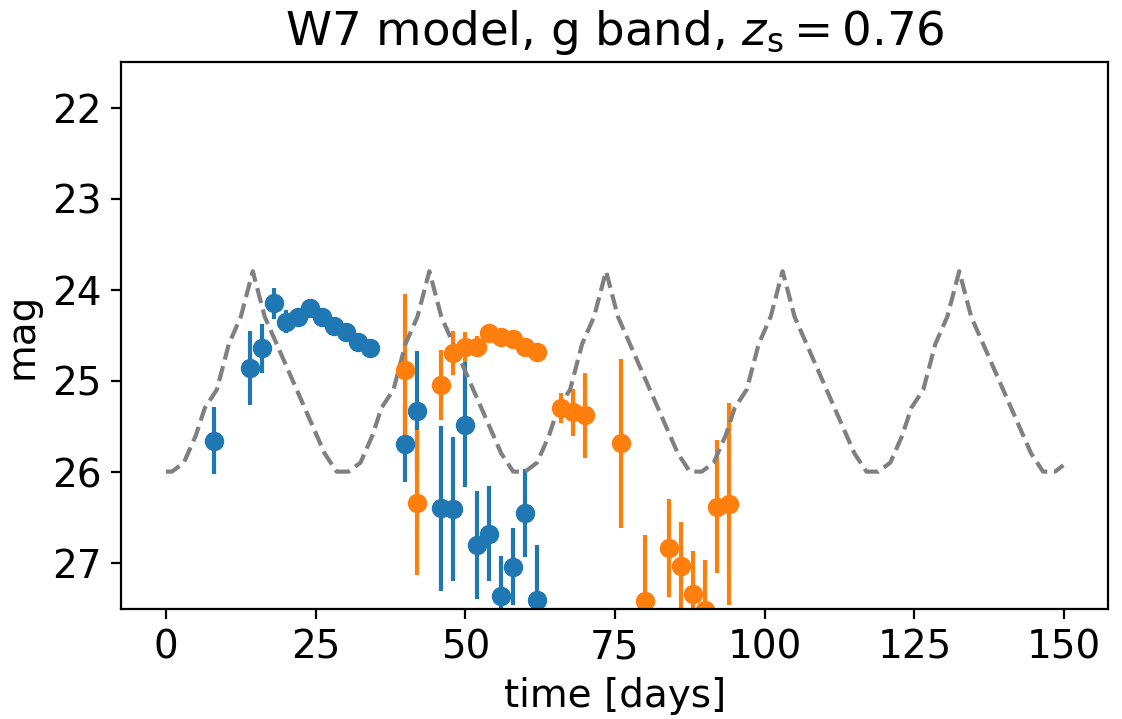}}
\subfigure{\includegraphics[width=0.24\textwidth]{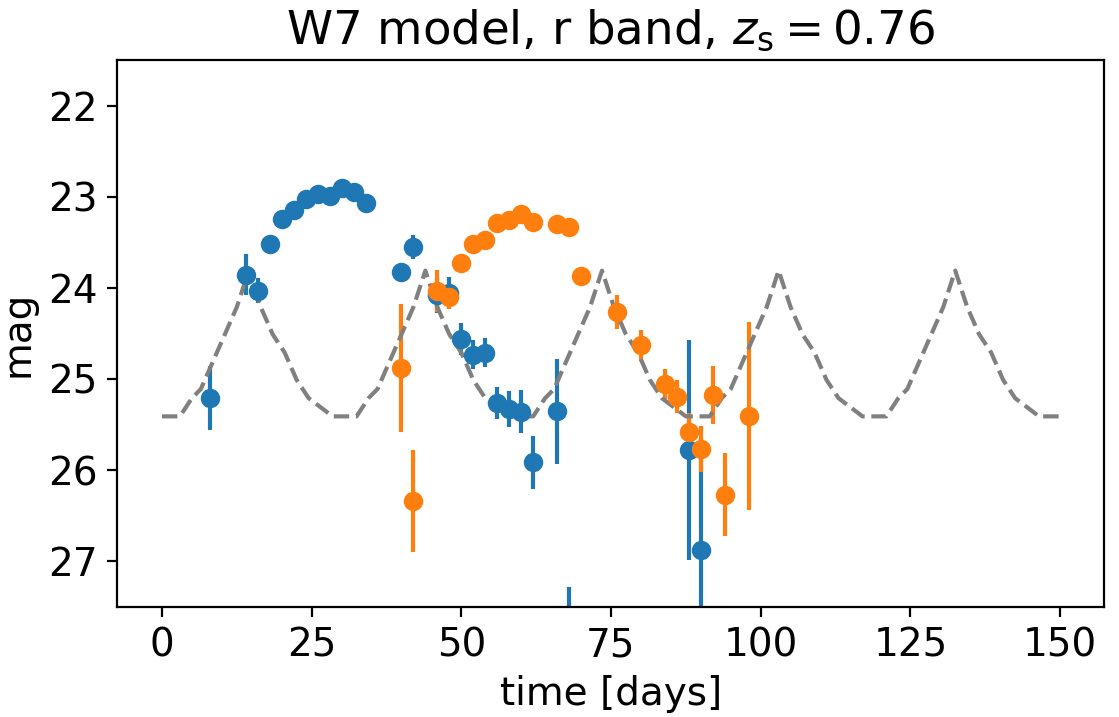}}
\subfigure{\includegraphics[width=0.24\textwidth]{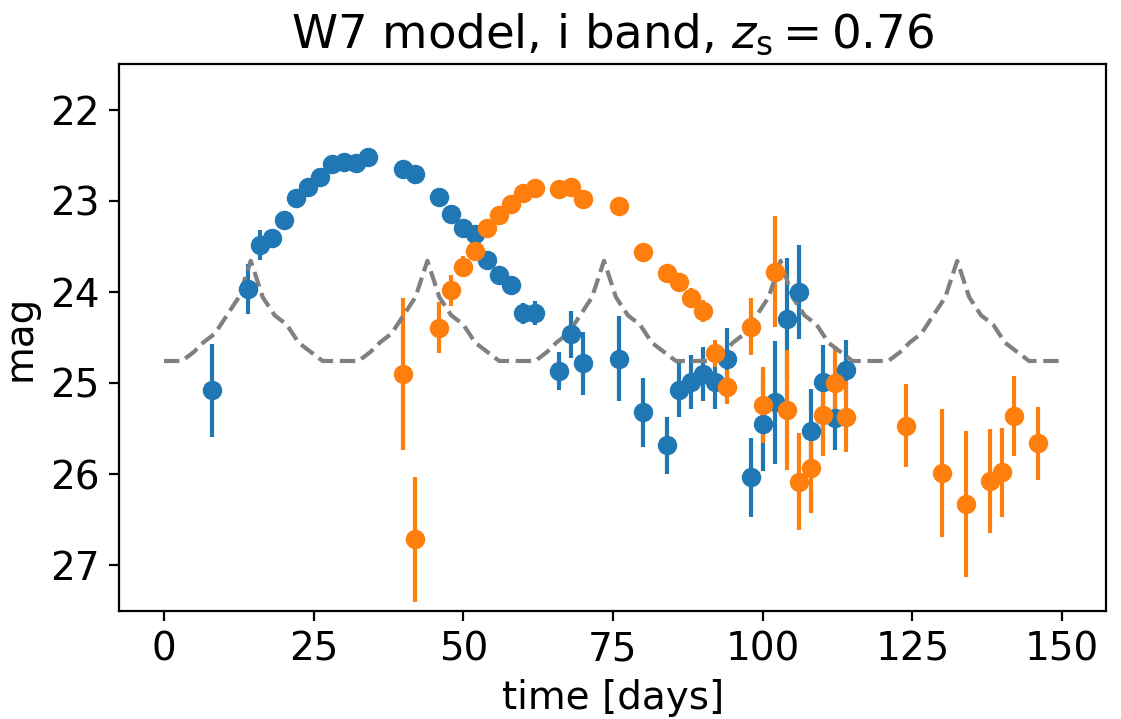}}
\subfigure{\includegraphics[width=0.24\textwidth]{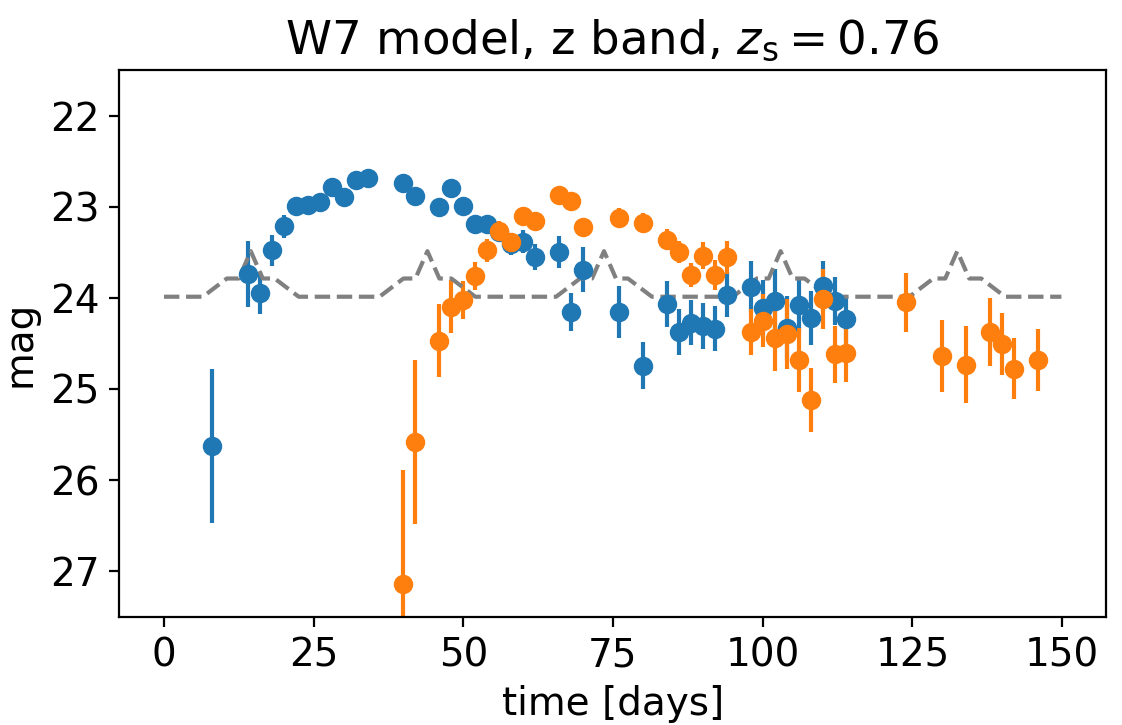}}
\subfigure{\includegraphics[width=0.25\textwidth]{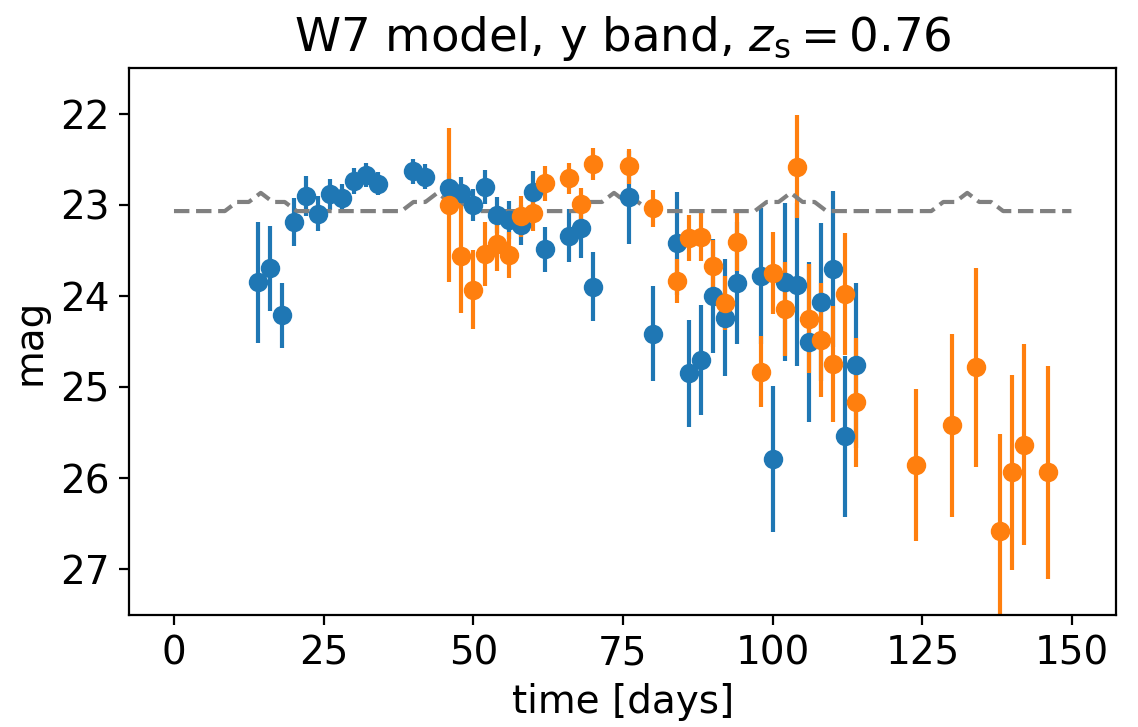}}
\subfigure{\includegraphics[width=0.25\textwidth]{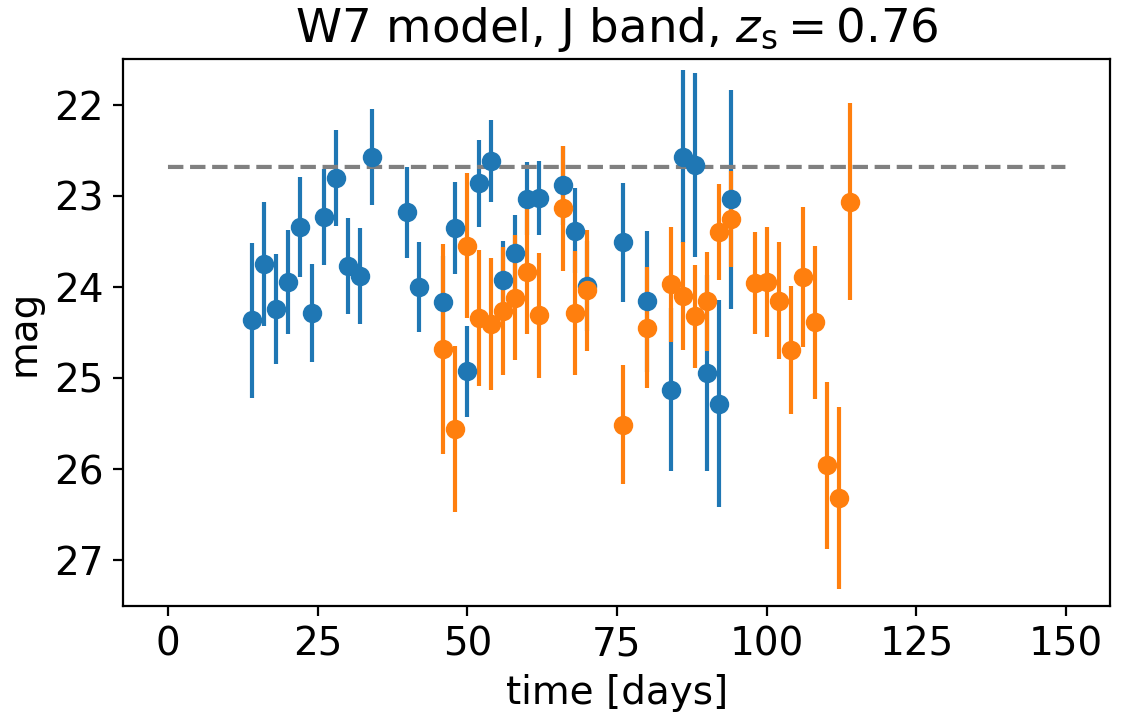}}
\subfigure{\includegraphics[width=0.25\textwidth]{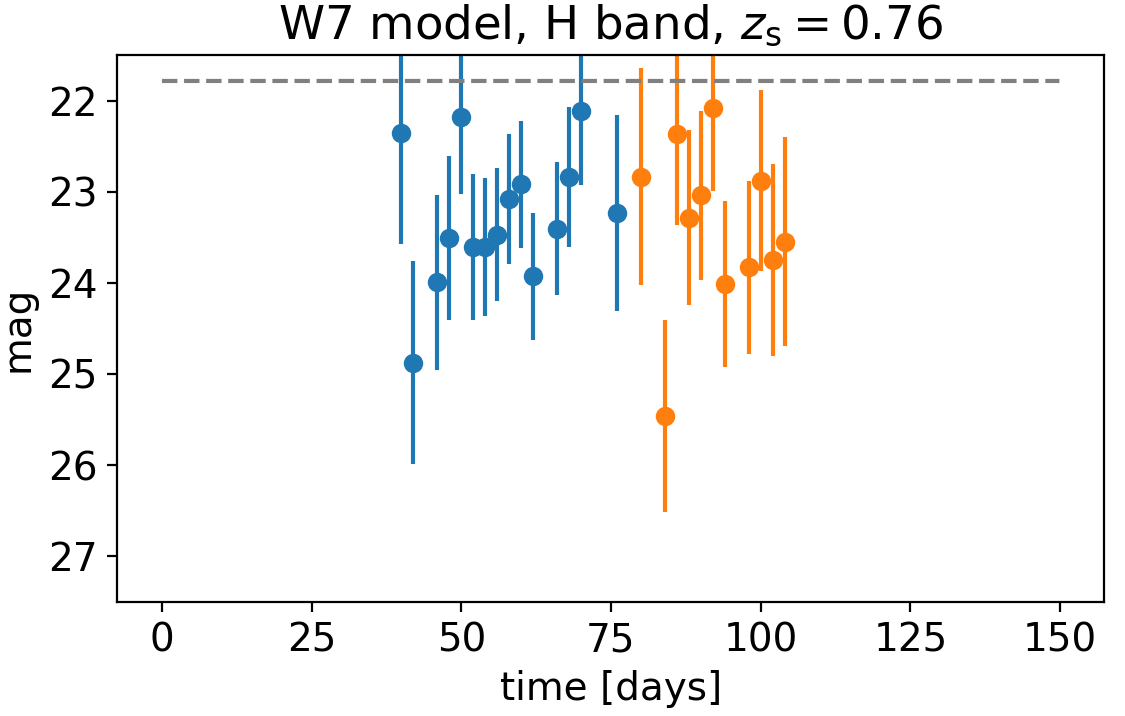}}
\caption{Further bands for the mock observation of Sect. \ref{sec: Example data used for machine learning}. The dashed gray line marks the 5$\sigma$ point-source depth that accounts for the moon phase.}
\label{fig: appendix further bands of mock observations}
\end{figure}

Figure \ref{fig: appendix further bands of mock observations} shows all the bands from the mock observation discussed in Sects. \ref{sec: Example data used for machine learning}, \ref{sec: Machine learning on example mock-observation}, and \ref{sec: Microlensing, observational noise and choice of filters}.

\section{Bias reduction:  Training on three models}
\label{sec:Appendix Bias training just on 3 models}

In this section we discuss some hypothetical cases where we assume
that only three of the four theoretical models are available for the
training of the ML method.

We start with the FCNN as described in Sect. \ref{sec: Deep
  Learning Network} and investigate four different cases
namely: (merger, N100, sub-Ch), (merger, N100, W7), (merger, sub-Ch,
W7), and (N100, sub-Ch, W7) for the training process. The black solid
line in each panel of Fig. \ref{fig: three models for training
  evaluating on single model test sets} shows the case where the
trained model is evaluated on the corresponding test set, meaning that
the test set contains light curves from all three models in the same
fraction as in the training data. The other four distributions, shown
in each panel, correspond to the evaluation of the FCNN trained
on 3 models, on a test set that contains just light curves from a
single SN Ia model. The shown results are the median (50th percentile), with the 84th-50th percentile (superscript) and 16th-50th percentile (subscript). The left column contains the normalized light
curves as displayed in Fig. \ref{fig: SN Ia models vs SNEMO} and the
right column contains the same models but during the training process
we allow for a random shift in time from [$-5$, 5] days and a random
shift in magnitude from [$-0.4$, 0.4]. We only apply the shifts in the
training set and not in the test set although the results look
similar. While the shift in magnitude is only applied to make the
noise level more flexible, the shifts in time help to increase the
variety of the light curves, especially so that the locations of the peak
from different models do overlap.

From the left column of Fig. \ref{fig: three models for training
  evaluating on single model test sets}, we learn that as soon as a
model was included in the training process the network performs very
well on the test set of the model. For the case of (merger, N100,
sub-Ch) for the training process, the network works also very well on
the test set from W7, even though it has never seen light curves from
that model. The reason for that seems to be that light curves from
the W7 model are close to the sub-Ch model for early times and close
to the N100 model in later times (especially for the rest-frame $g$
band) as shown in Fig. \ref{fig: SN Ia models vs SNEMO}. The other
three cases still work on the model it was not trained on, but we
detect biases on the order of almost a day, especially for the sub-Ch
and merger model (left column, rows two and four). From the right
column of Fig. \ref{fig: three models for training evaluating on
  single model test sets} we find that our applied random shifts in
time and magnitude significantly help to overcome these biases. Even
though this broadens the distributions of the models it was trained on
slightly, it tightens the distribution of the models it was not trained
on and therefore helps to generalize to real observations.

\begin{figure*}[h!]
\centering
\subfigure{\includegraphics[trim=4 17 4 16,width=0.44\textwidth]{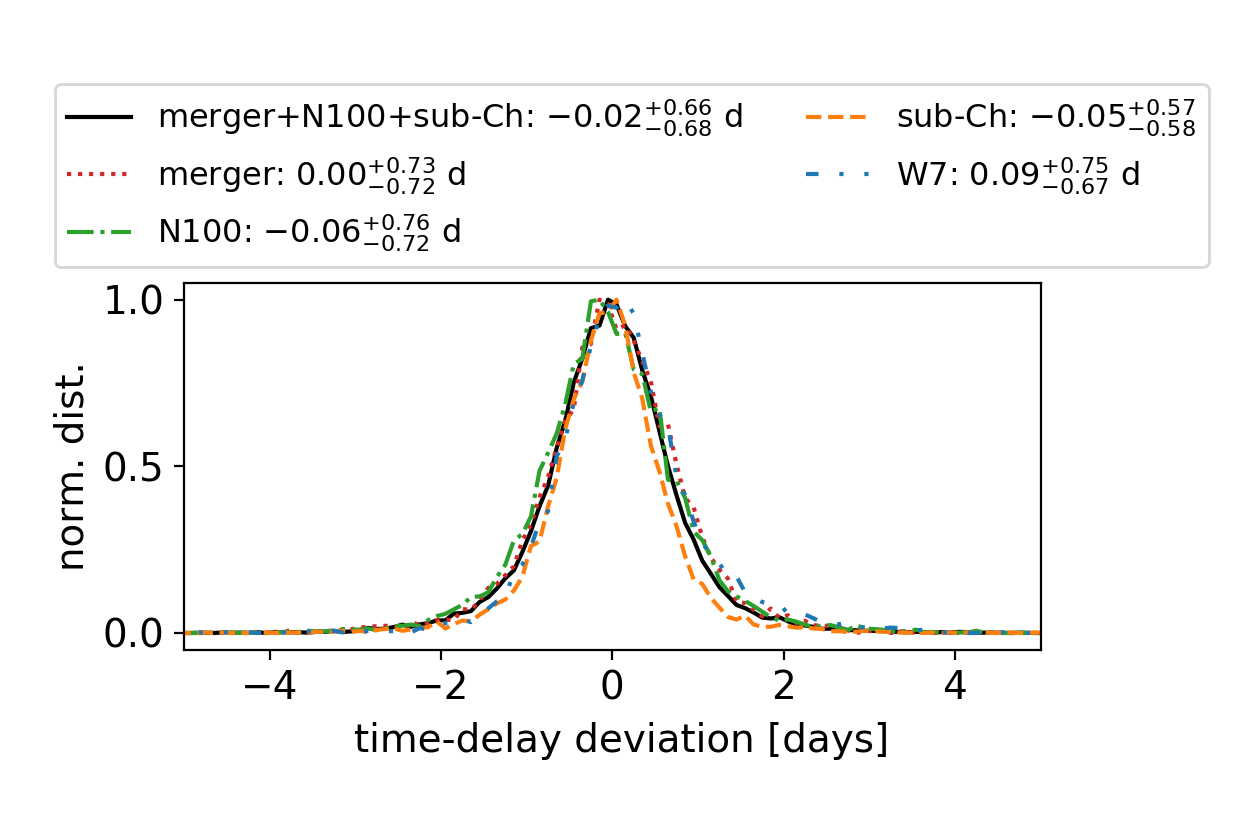}}
\subfigure{\includegraphics[trim=4 17 4 16,width=0.44\textwidth]{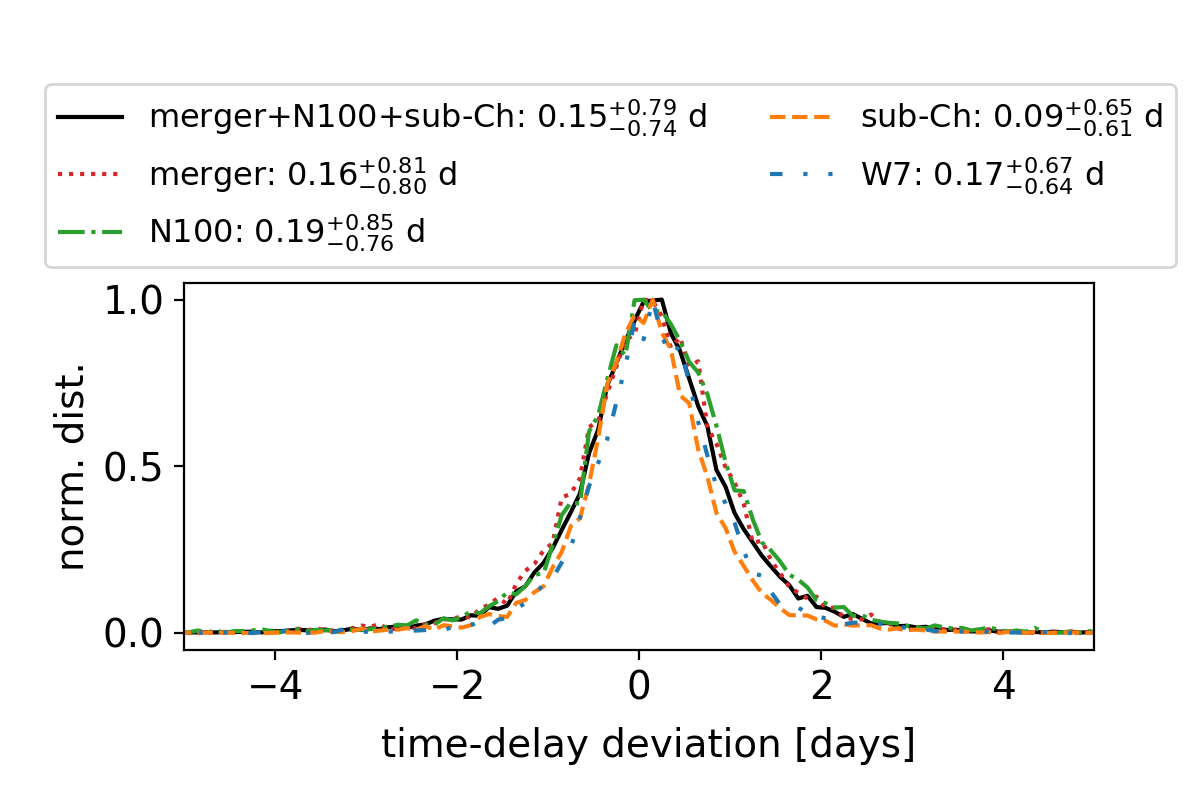}}
\subfigure{\includegraphics[trim=4 17 4 16,width=0.44\textwidth]{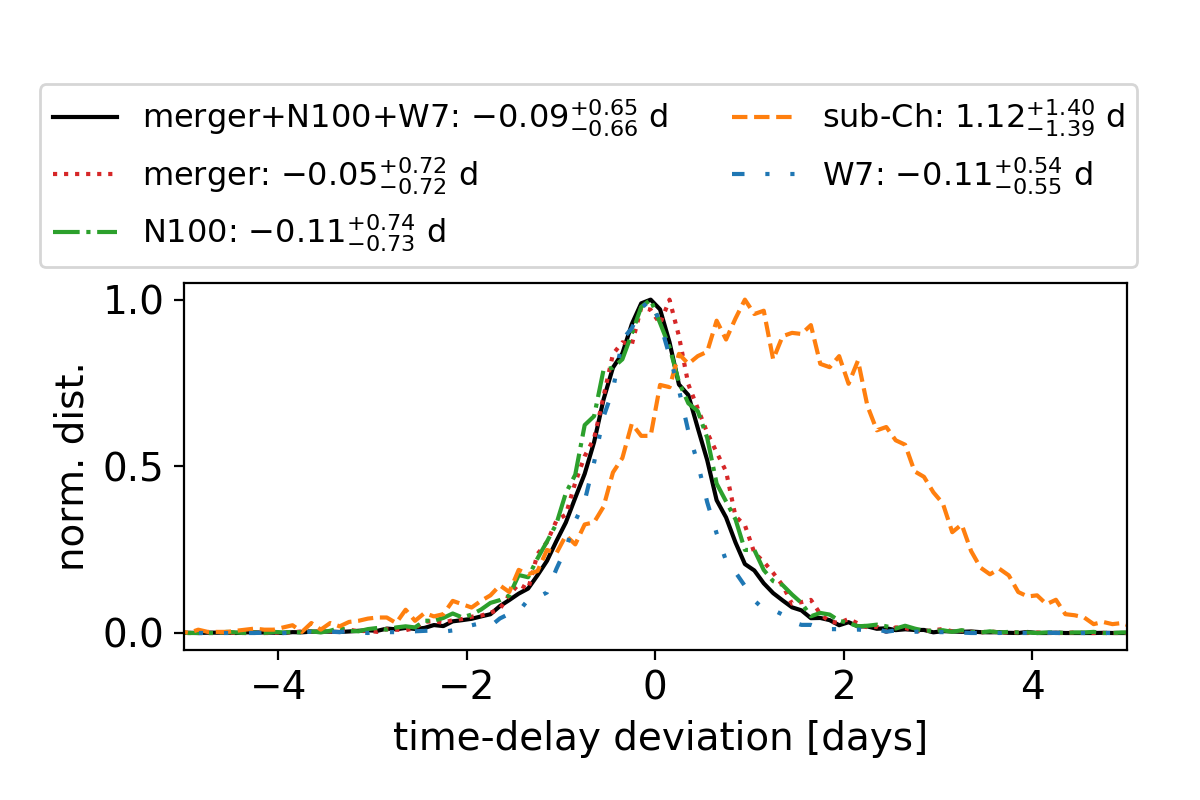}}
\subfigure{\includegraphics[trim=4 17 4 16,width=0.44\textwidth]{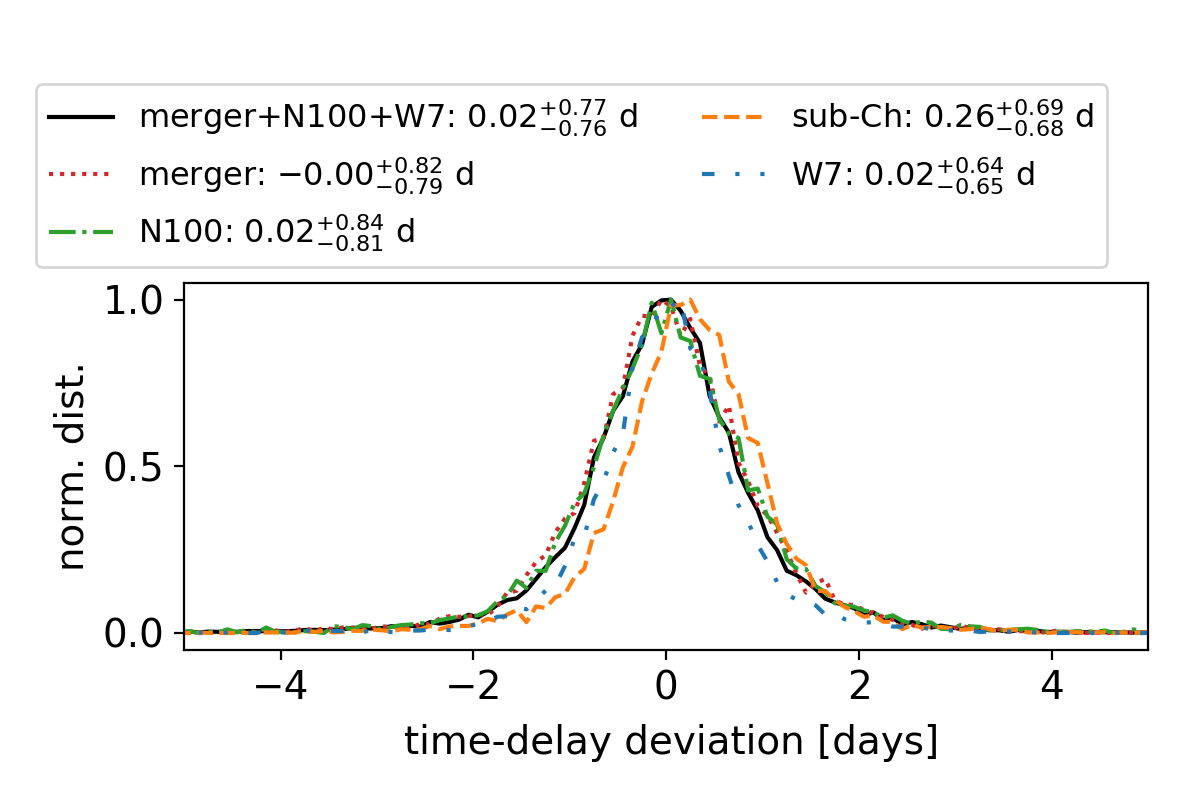}}
\subfigure{\includegraphics[trim=4 17 4 16,width=0.44\textwidth]{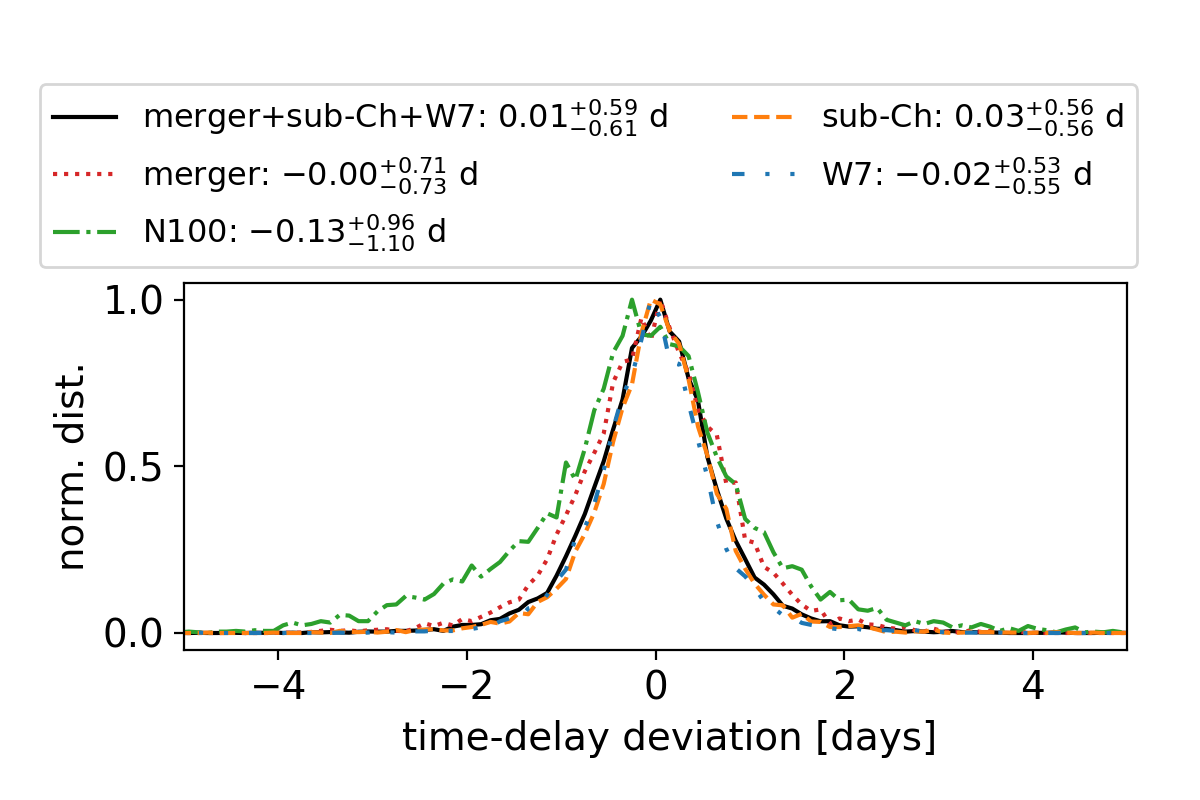}}
\subfigure{\includegraphics[trim=4 17 4 16,width=0.44\textwidth]{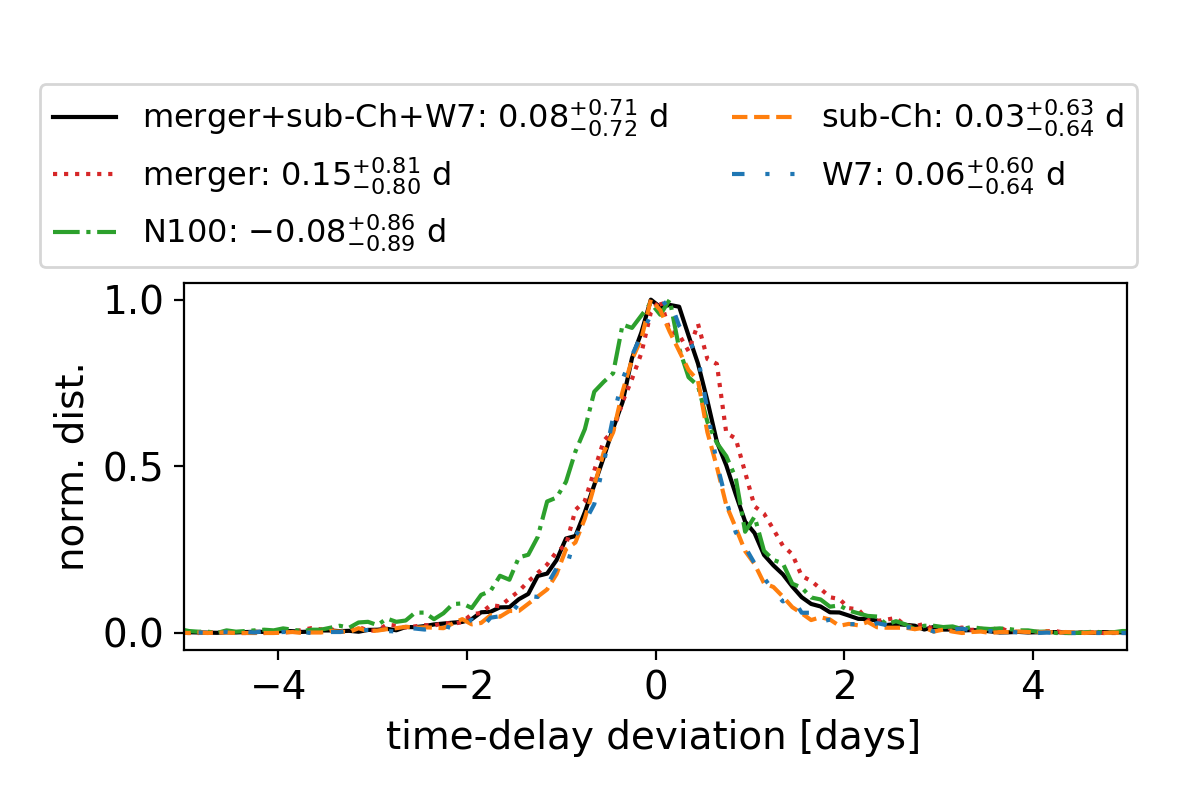}}
\subfigure{\includegraphics[trim=4 17 4 16,width=0.44\textwidth]{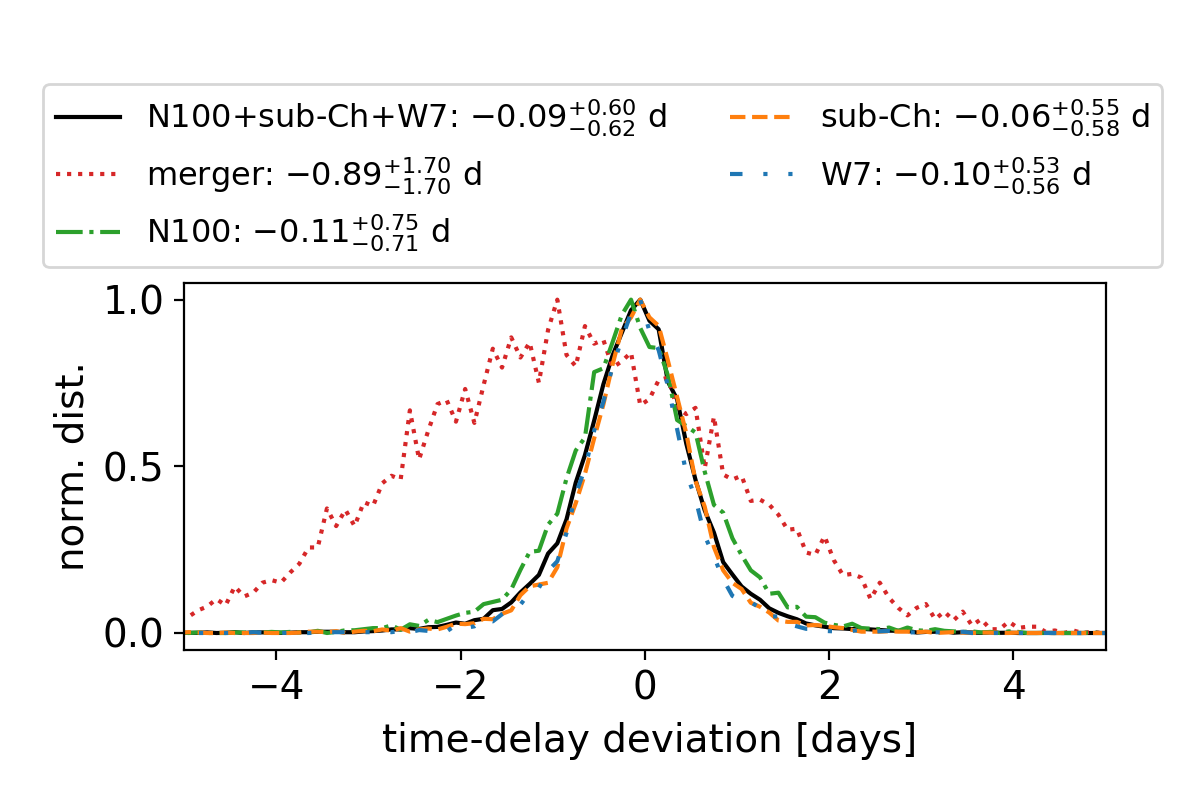}}
\subfigure{\includegraphics[trim=4 17 4 16,width=0.44\textwidth]{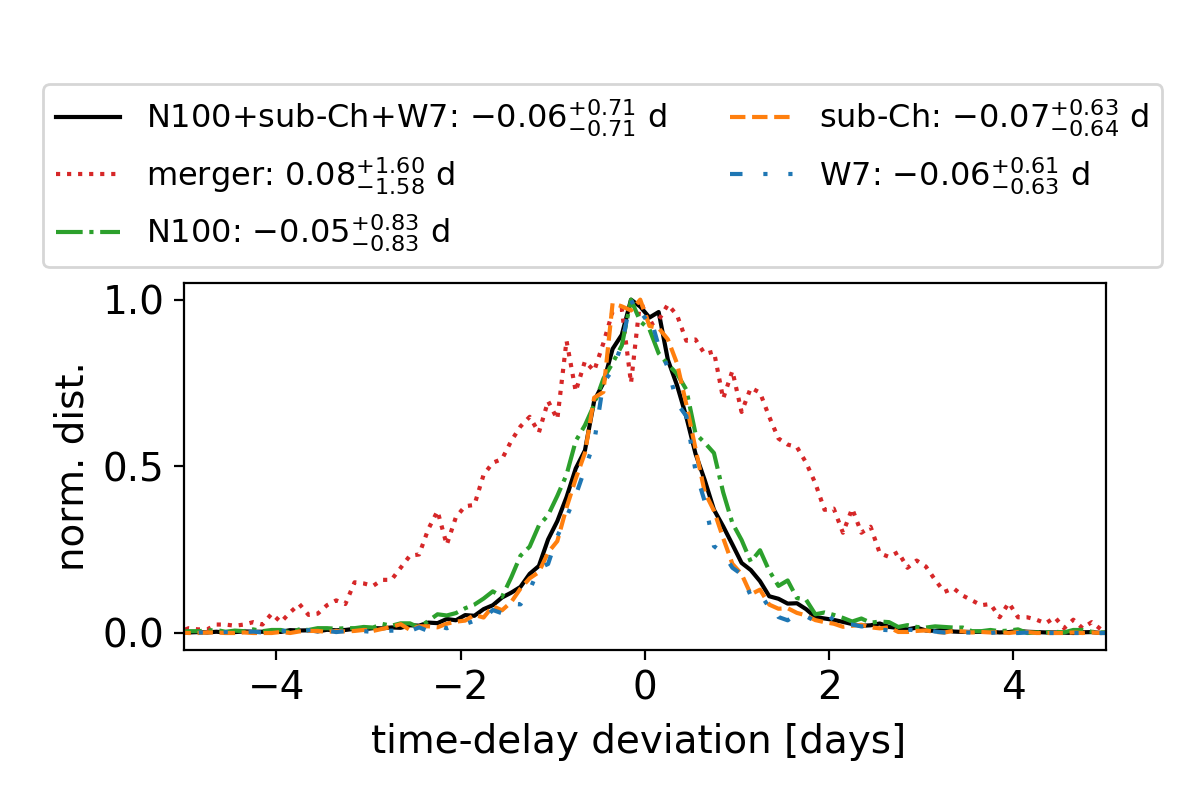}}
\caption{FCNNs trained on three SN Ia models and evaluated on all samples of its corresponding test set, in comparison to the evaluation on all samples from four test sets from the four individual SN Ia models. The left column shows the case where the SN Ia models have been used as shown in Fig. \ref{fig: SN Ia models vs SNEMO}. The right column contains plots in which the SN Ia models are randomly shifted within $\pm 5 \mathrm{d}$ in time and $\pm 0.4$ in magnitude.  The random shifts make the training set generalize better to test sets that are not represented by the training set.}
\label{fig: three models for training evaluating on single model test sets}
\end{figure*}

\begin{figure*}[h!]
\centering
\subfigure{\includegraphics[trim=0 5 0 12,width=0.41\textwidth]{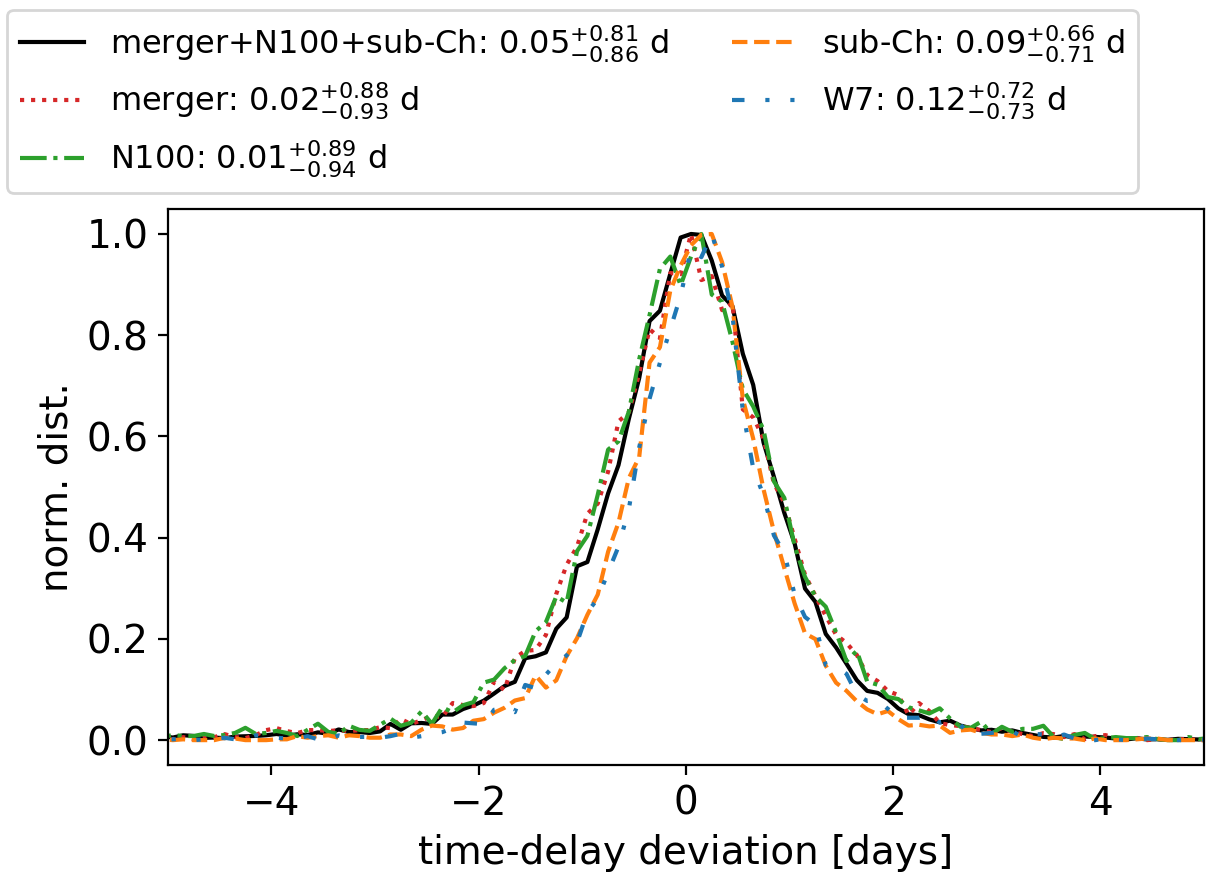}}
\subfigure{\includegraphics[trim=0 5 0 12,width=0.41\textwidth]{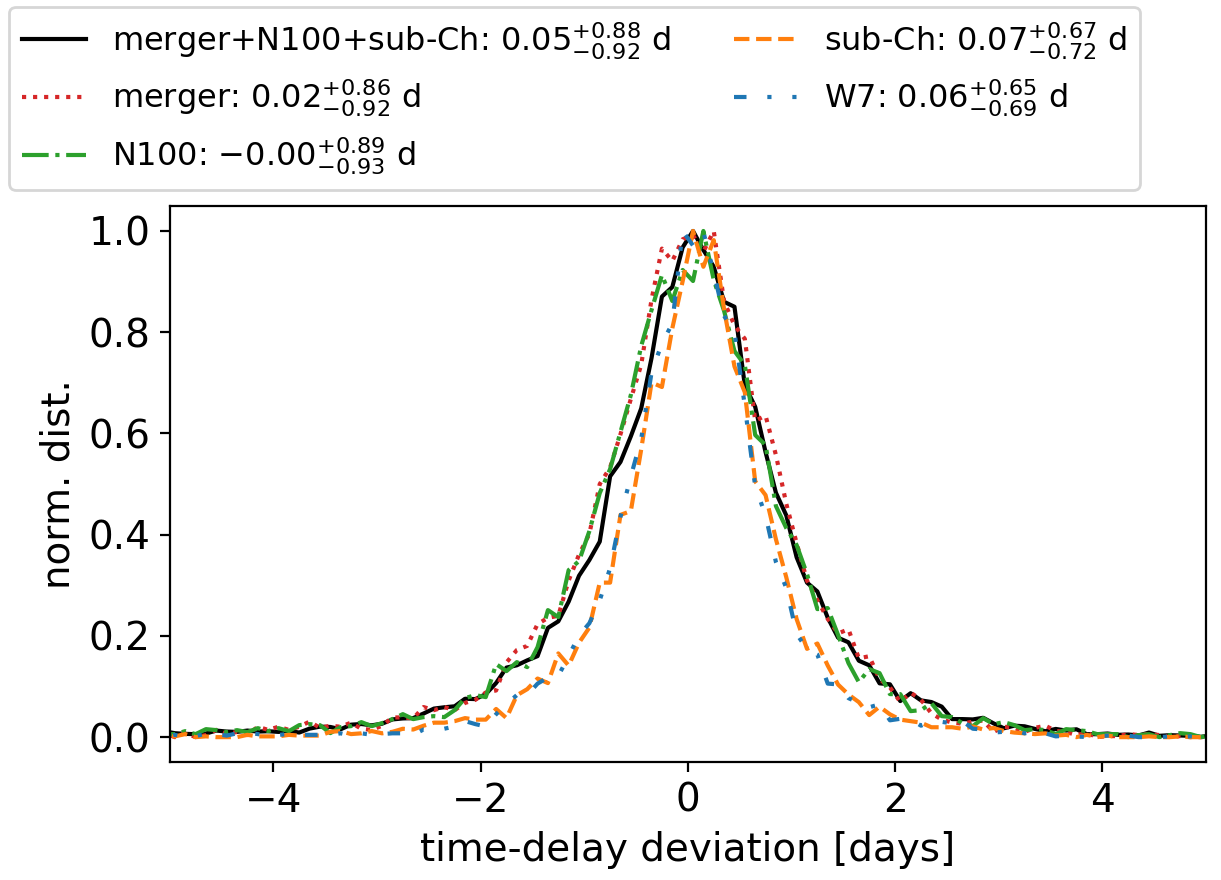}}
\subfigure{\includegraphics[trim=0 5 0 12,width=0.41\textwidth]{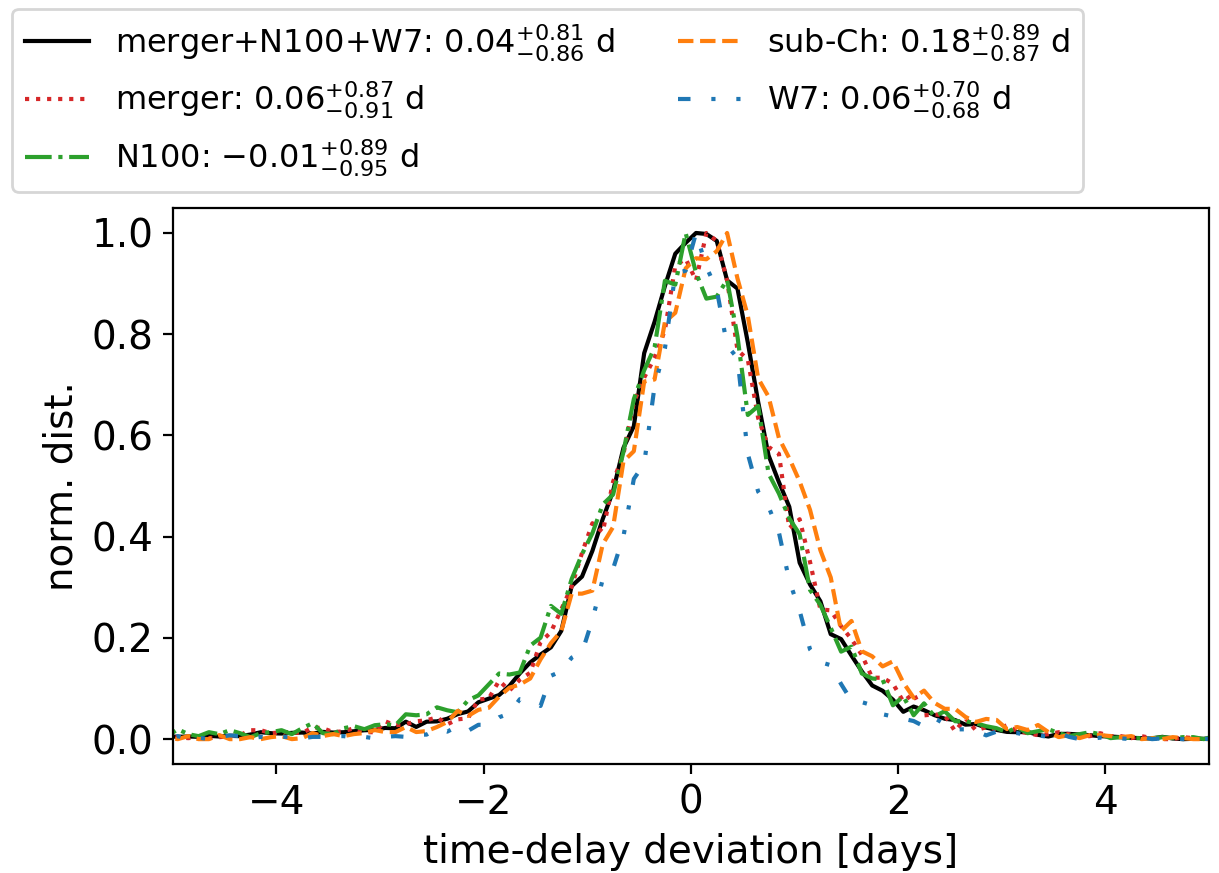}}
\subfigure{\includegraphics[trim=0 5 0 12,width=0.41\textwidth]{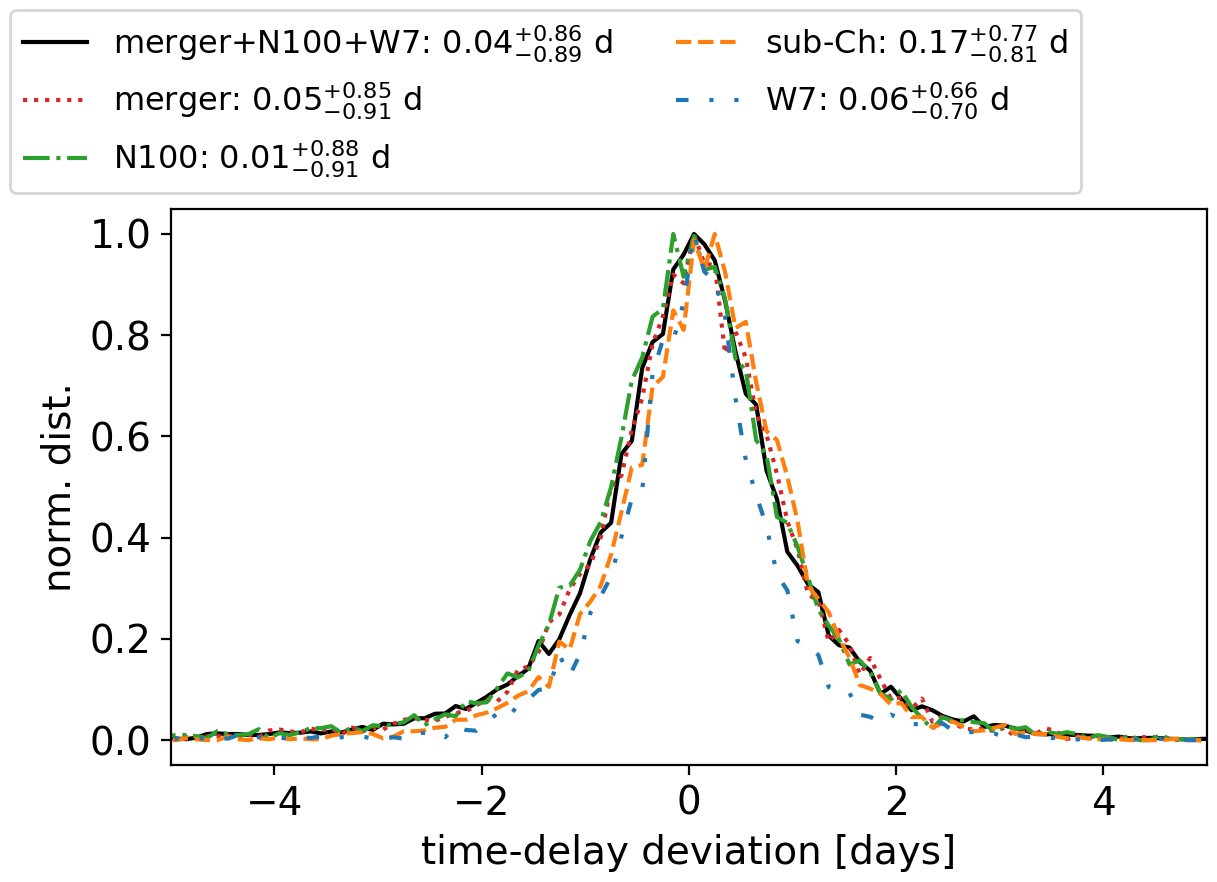}}
\subfigure{\includegraphics[trim=0 5 0 12,width=0.41\textwidth]{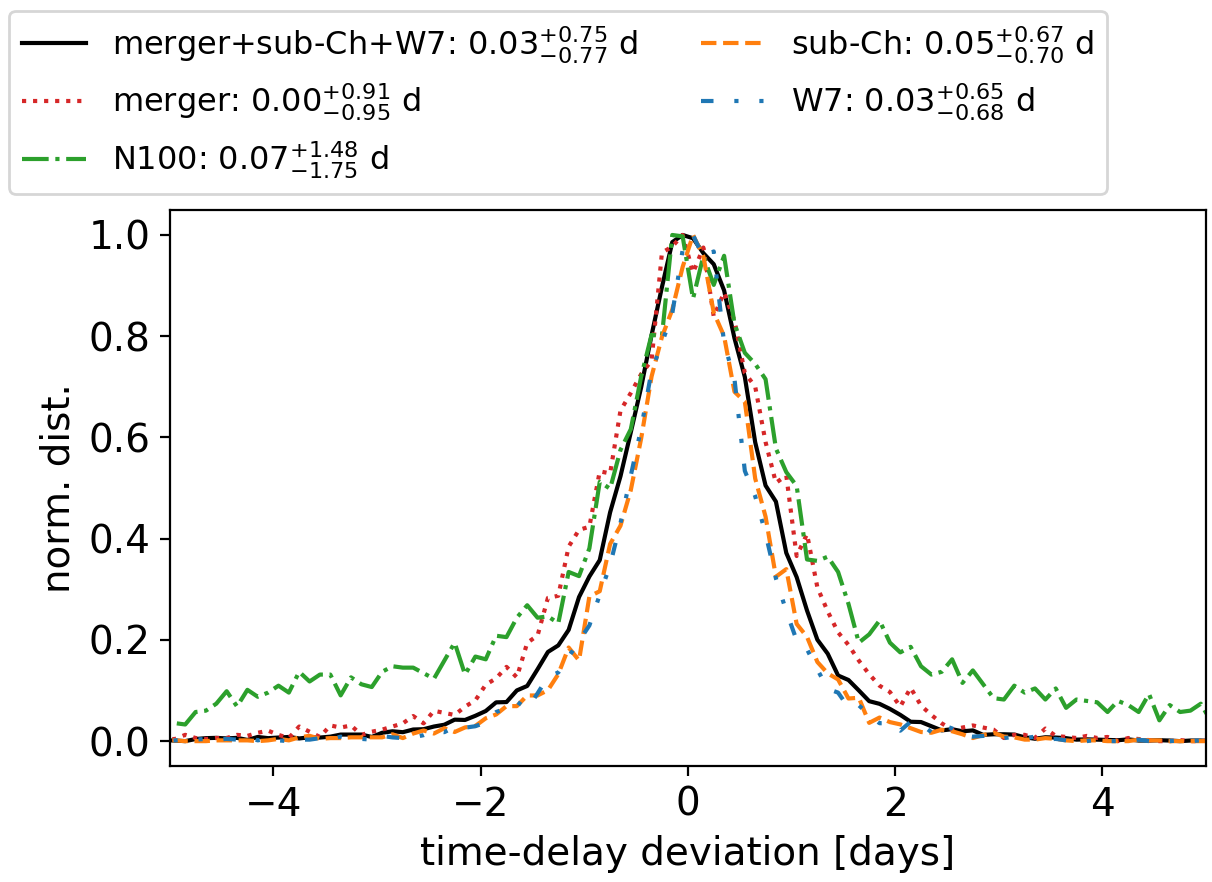}}
\subfigure{\includegraphics[trim=0 5 0 12,width=0.41\textwidth]{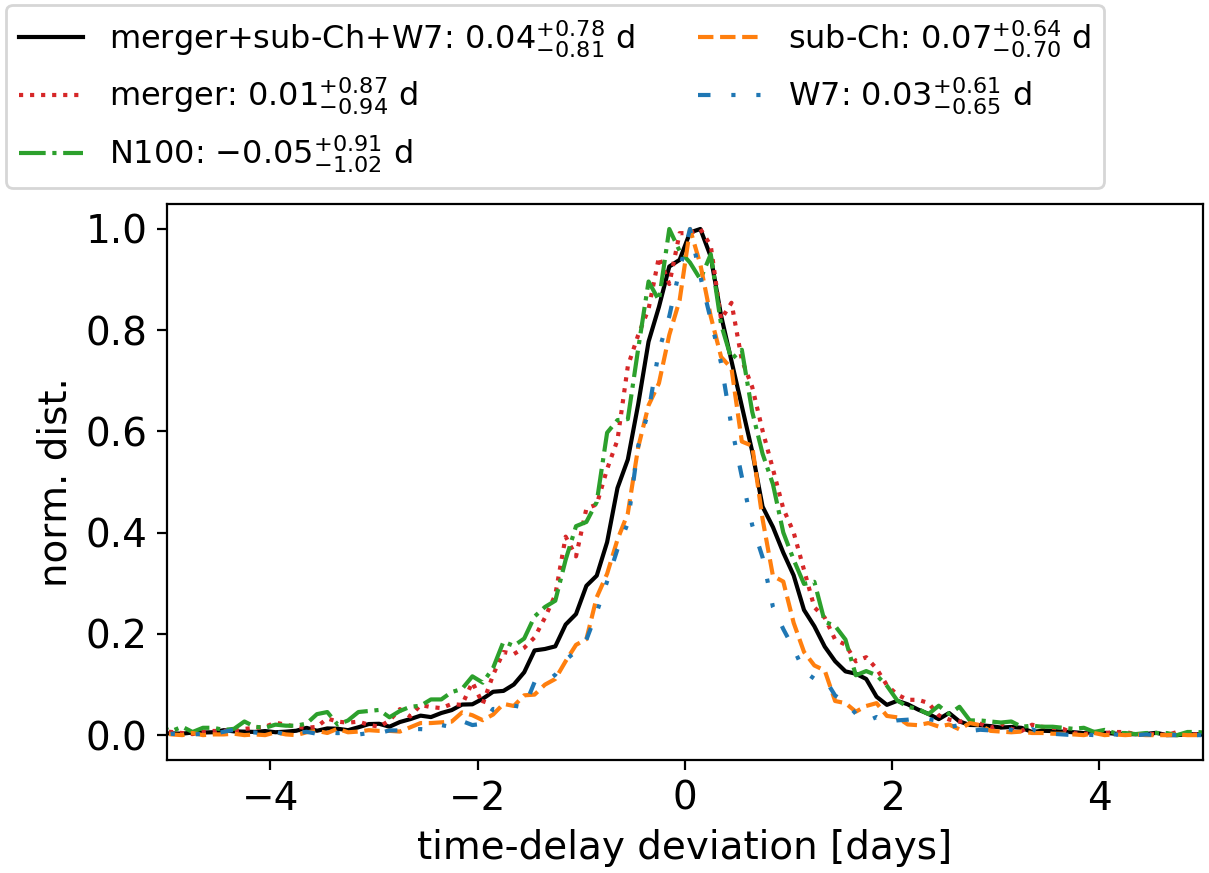}}
\subfigure{\includegraphics[trim=0 5 0 12,width=0.41\textwidth]{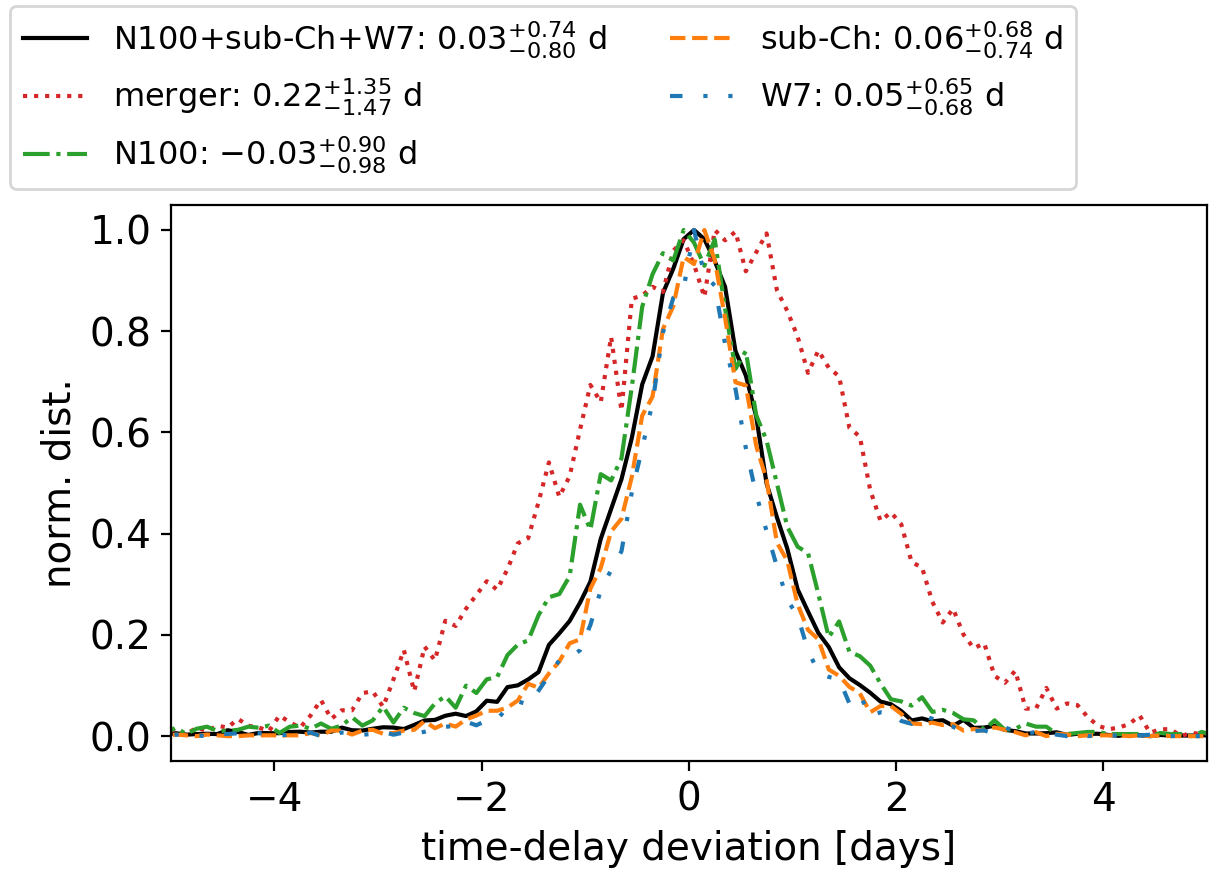}}
\subfigure{\includegraphics[trim=0 5 0 12,width=0.41\textwidth]{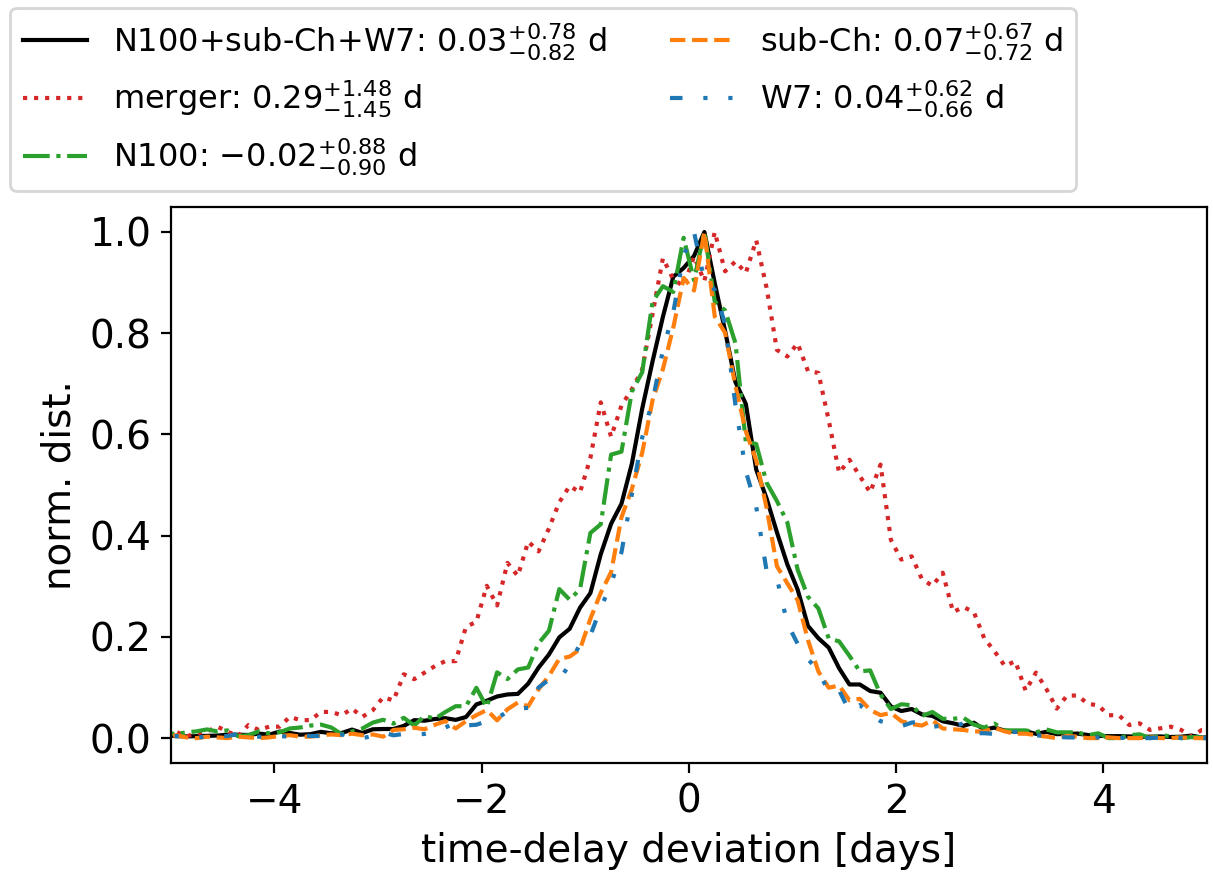}}
\caption{RF networks trained on three SN Ia models and evaluated on all samples of their corresponding test set, in comparison to the evaluation on all samples from four test sets from the four individual SN Ia models. The left column shows the case where the SN Ia models have been used as shown in Fig. \ref{fig: SN Ia models vs SNEMO}. The right column contains plots where the SN Ia models are randomly shifted within $\pm 5 \mathrm{d}$ in time and $\pm 0.4$ in magnitude.  The random shifts are far less important than for the FCNN approach shown in Fig. \ref{fig: three models for training evaluating on single model test sets}.}
\label{fig: three models for training evaluating on single model test sets RF results}
\end{figure*}

In Fig. \ref{fig: three models for training evaluating on single model test sets RF results}
we do the same experiment but this time using the RF. For the cases (merger, N100, sub-Ch),
(merger, N100, W7), and (merger, sub-Ch, W7) we find that applying the random shifts
 in the training process improves the bias and precision on the test set based on
the model not used in the training process. Only for the case of using (N100, sub-Ch, W7)
in the training process the merger results get worse. This suggests that especially the merger model, which deviates most from the other three models as shown in 
Fig. \ref{fig: SN Ia models vs SNEMO} and therefore helps to increase the variaty, 
is required in the training process. However, the comparison between Figs. \ref{fig: three models for training
  evaluating on single model test sets} and \ref{fig: three models for training evaluating on single model test sets RF results}
  shows that the RF does already a pretty good job without the random shifts,
  and therefore the RF generalizes better to slightly different light-curve shapes
   in comparison to our 
  FCNN approach. Nevertheless, the results from the right column of Fig. 
  \ref{fig: three models for training evaluating on single model test sets} are an encouraging sign that with 
  the random shifts in time our FCNN approach can also be generalized better to SN Ia light curves, which
were not used in the training process.

Instead of random shifts in time, one might argue that different random multiplication factors that stretch or squeeze the light curves in time might work even better. Therefore, we tested different ranges for the random factors, namely, (0.95 to 1.05), (0.9 to 1.1), (0.85 to 1.15), (0.8 to 1.2), (0.7 to 1.3), (0.6 to 1.4), (0.5 to 1.5), and (0.2 to 1.8), where the factor 1 provides the light curves as shown in Fig. \ref{fig: SN Ia models vs SNEMO}. We find that (0.7 to 1.3) works best to reduce biases in a similar way as shown in Fig. \ref{fig: three models for training evaluating on single model test sets}. Nevertheless applying a FCNN and RF model trained on four theoretical models with the random multiplication factors to an empirical data set based on \texttt{SNEMO15} similar as in Sect. \ref{sec: Evaluation on SNEMO test set}, we find that even though the precision for the FCNN is roughly 0.05 days better as in Fig. \ref{fig: SENOM15 test for DL and RF using filters iz} the FCNN still predicts substantial biases for the \texttt{SNEMO15} data set up to 0.5 days and is therefore not useful. Furthermore, the precision of the RF on the \texttt{SNEMO15} data set for the random shifts in time between [$-5$, 5] days (as shown in Fig. \ref{fig: SENOM15 test for DL and RF using filters iz}) is roughly 0.2 days better than results from using random multiplication factors and therefore using the random shifts in time was a valuable choice.

\FloatBarrier

\section{Train and validation loss}
\label{sec:Appendix train and validation loss}

\begin{figure}[h!]
\centering
\includegraphics[width=0.4\textwidth]{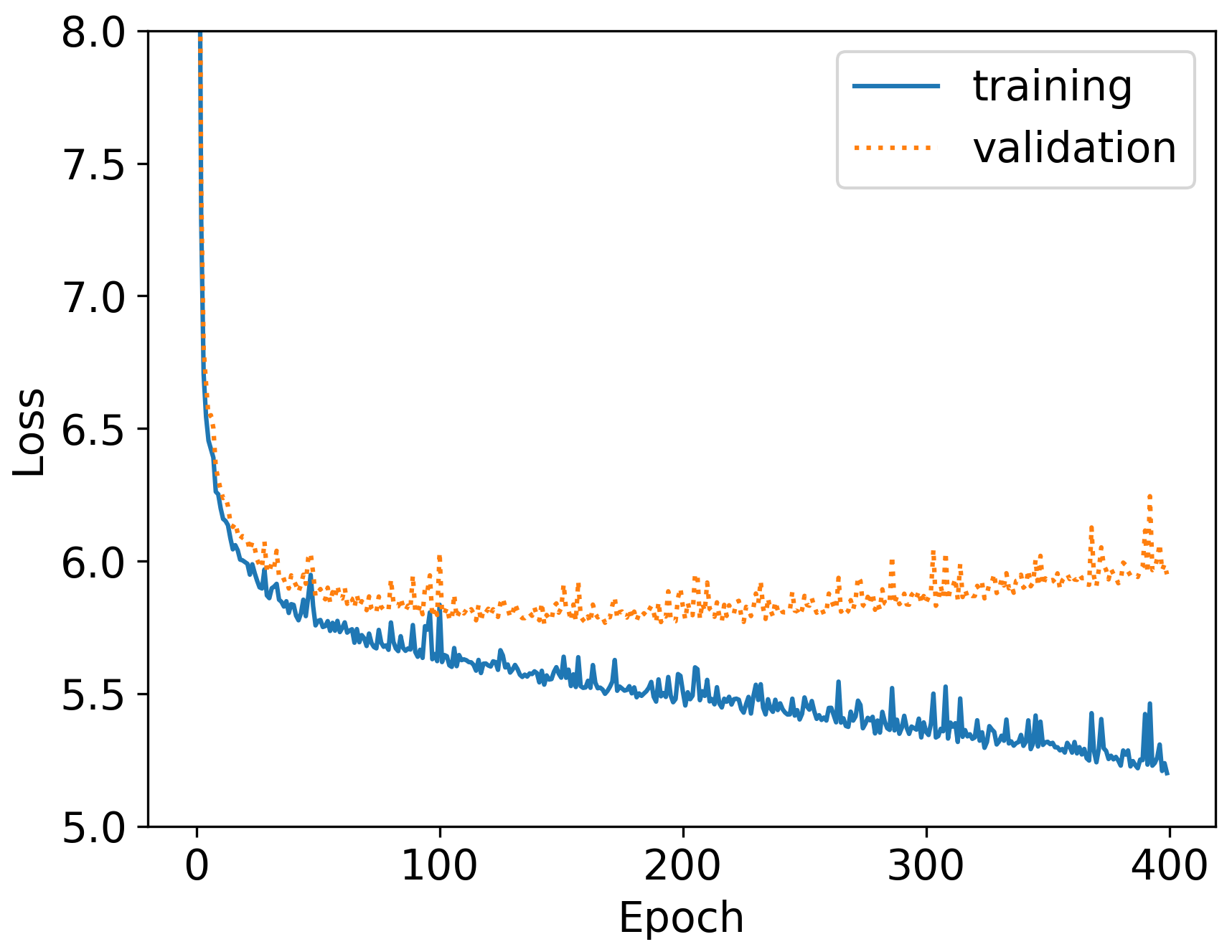}
\caption{Training and validation loss of the FCNN for 400 training epochs.}
\label{fig: train and validation loss}
\end{figure}

In Fig. \ref{fig: train and validation loss} we see for the FCNN the training loss in comparison to the validation loss for 400 training epochs. The network that provides the lowest validation loss is the one that will be stored to reduce the chance of overfitting the training data.

\section{Time-delay deviation as a function of time delay}
\label{sec:Time-delay deviation as function of time delay}

\begin{figure}[h!]
\includegraphics[width=0.48\textwidth]{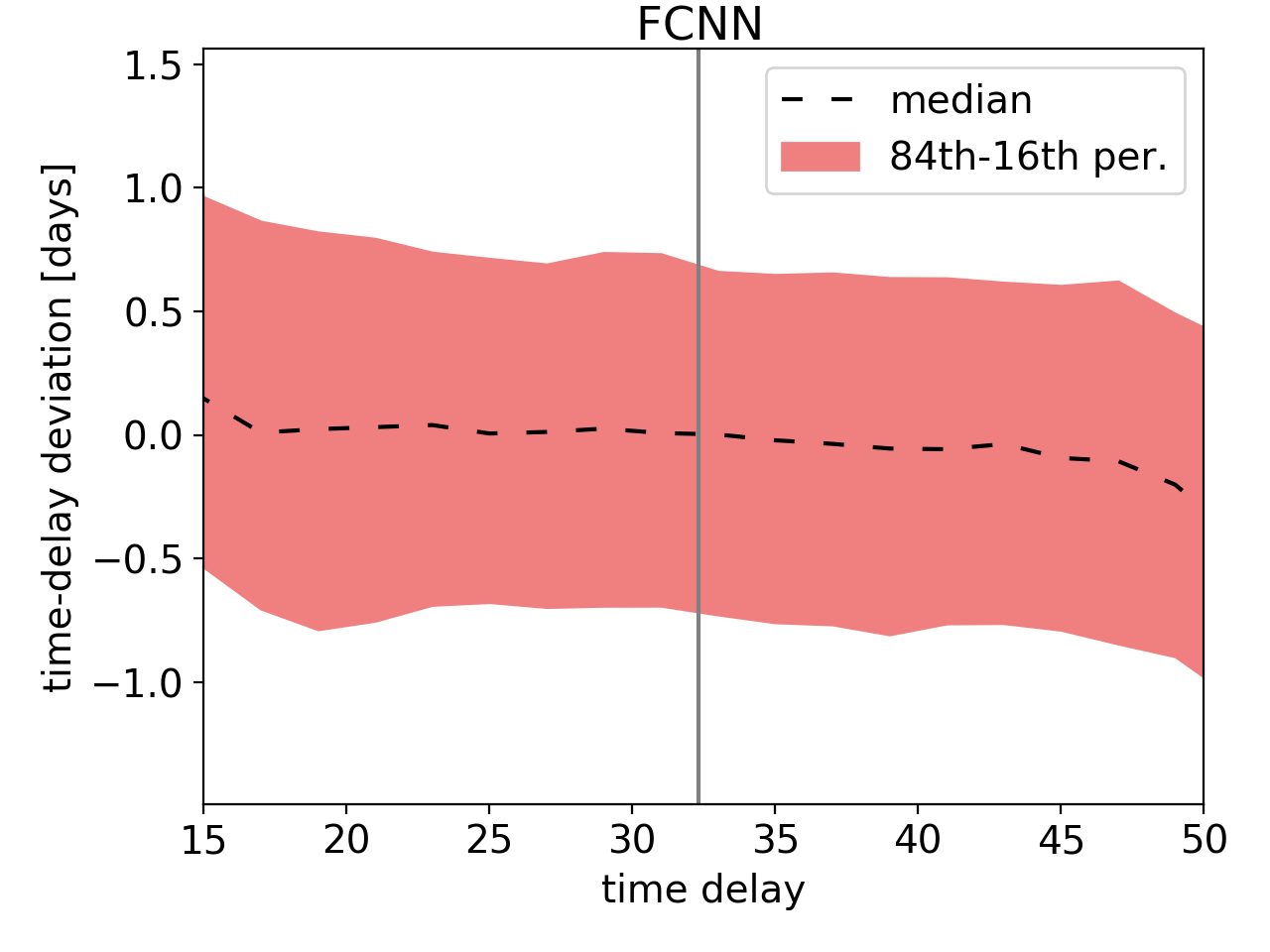}
\includegraphics[width=0.48\textwidth]{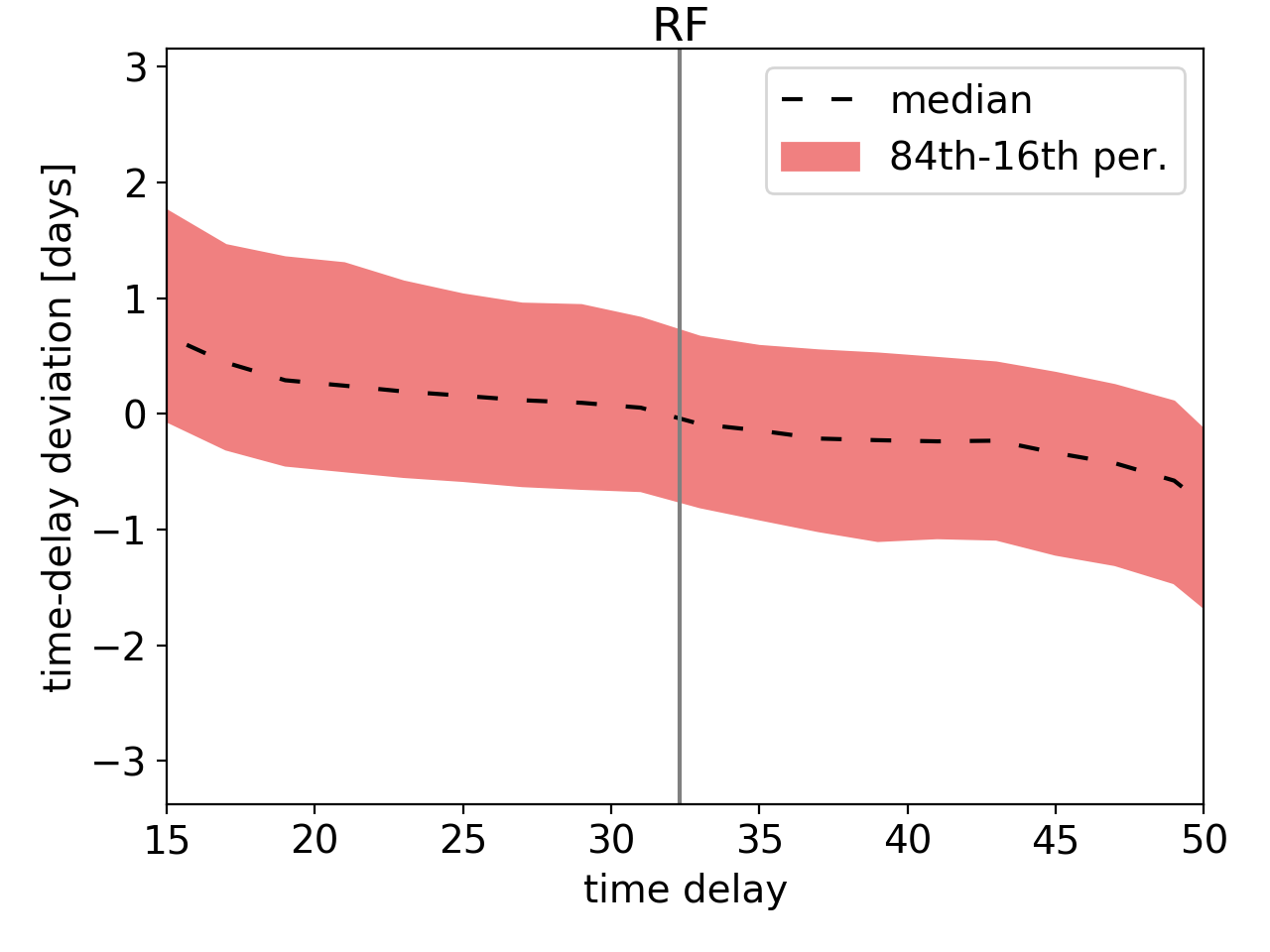}
\caption{Time-delay deviation, $\tau_i$, from Eq. (\ref{eq:td_deviation}) as a function of the time delay, where we have binned up the samples from the corresponding test set with similar time delays (bin size: two days) for the LSN Ia from Fig. \ref{fig: data to train NN}. The vertical gray line marks the true absolute time delay from the mock LSN Ia system. The corresponding test set has LSN Ia systems that span a range of input time-delay values, which include the vertical gray line.  Results from the FCNN network are shown in the left panel, and the RF is presented in the right panel.}
\label{fig: appendix time-delay deviation as function of time}
\end{figure}

Figure \ref{fig: appendix time-delay deviation as function of time} shows the time-delay deviation as a function of time for the LSN Ia from Fig. \ref{fig: data to train NN}.

\section{Feature importance of FCNN and RF}
\label{sec:Appendix Feature importance}

\begin{figure}[h!]
\subfigure{\includegraphics[width=0.47\textwidth]{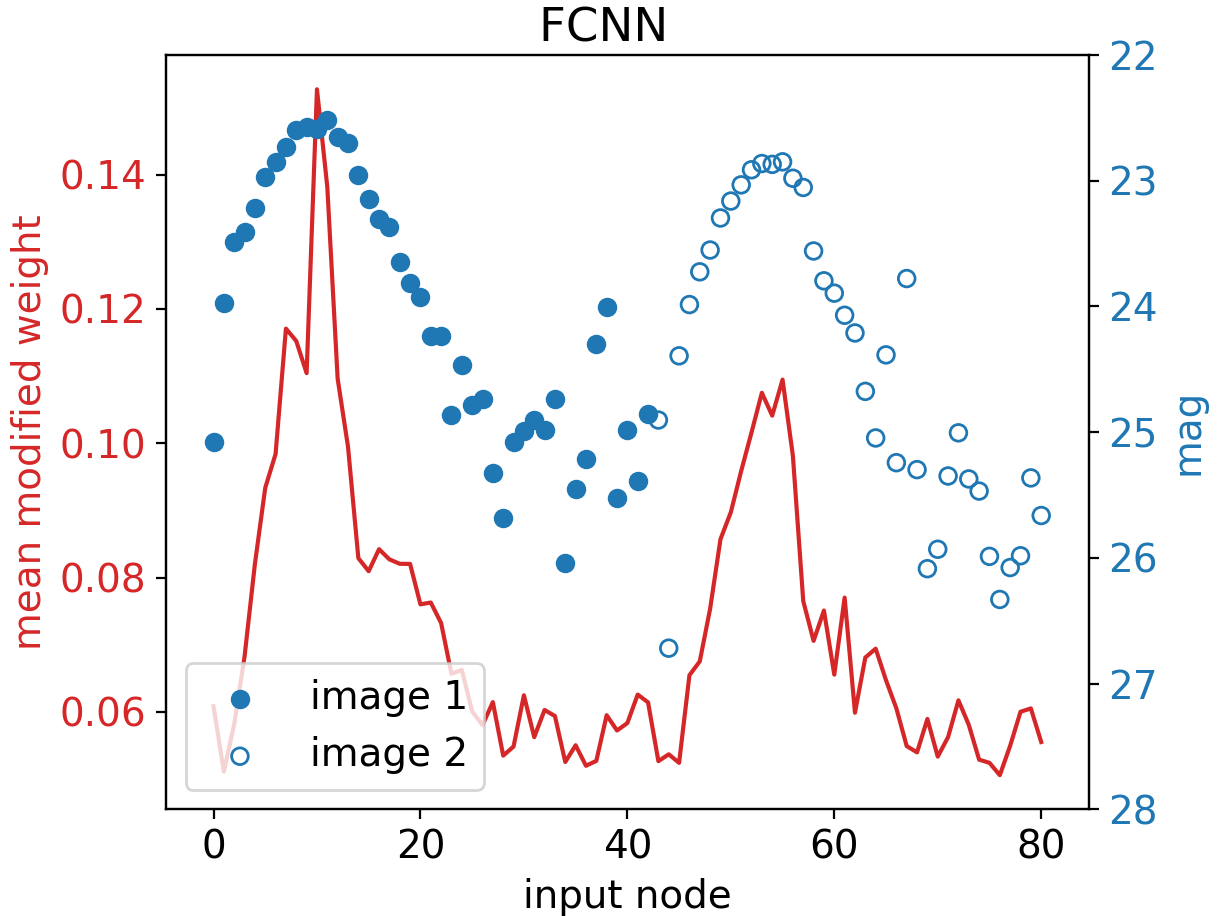}
}
\subfigure{\includegraphics[width=0.49\textwidth]{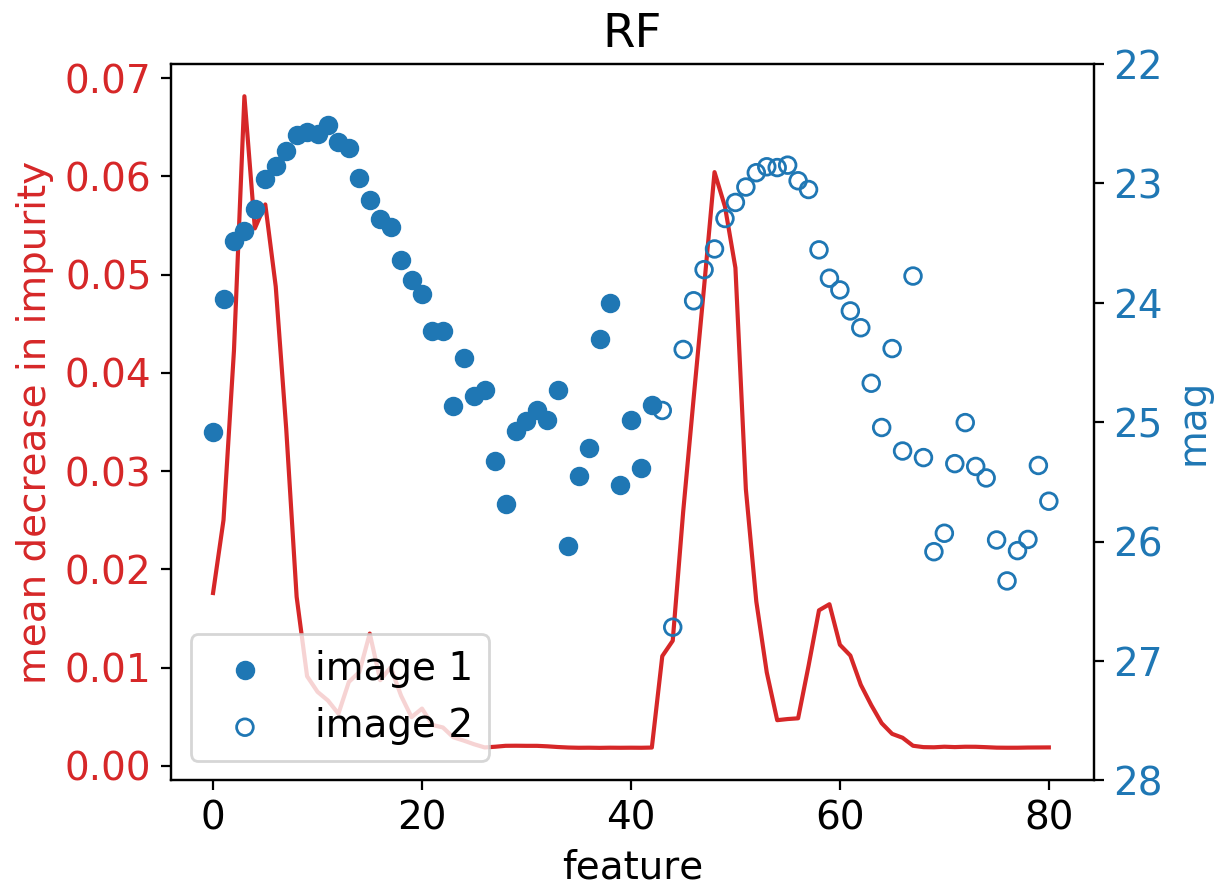}
}
\caption{Evaluation of the features (input nodes) for the FCNN (left panel) and the RF (right panel). The features (input nodes) are listed in Eq. (\ref{eq: data structure}), where feature 1 stands for $m_{i1,1}$ and the last feature (81) is $m_{i2,N_{i2}}$.}
\label{fig: appendix feature importance}
\end{figure}

In this section we investigate which of the features in Eq. (\ref{eq: data structure}) are the most important ones for the FCNN and the RF. For the FCNN the estimate is difficult and therefore we consider as an approximation just the input layer, where we calculate for each feature (input node) the mean of all the weights connected to that feature (negative weights are removed because a ReLU activation function is used). The results are summarized in the upper panel of Fig. \ref{fig: appendix feature importance}, where we see that the FCNN focuses mainly on the region of the peak.

To estimate the importance of the features for the RF we use from the software \texttt{scikit-learn} the \texttt{feature\_importances\_} tool\footnote{\url{https://scikit-learn.org/stable/auto_examples/ensemble/plot_forest_importances.html}} \citep{scikit-learn,sklearn_api}. This tool basically measures the decrease in performance of a RF if a specific feature would be removed. The results are shown in the lower panel of Fig. \ref{fig: appendix feature importance}. Comparing these results to the FCNN, we find that the RF is mainly focusing on the rise before the peak but also the decline afterward is important. The peak itself does not matter much. We further see that the noisy part of the light curves is almost not considered in comparison to the FCNN.
Focusing on higher signal-to-noise data points, less on the peak and more on the rise and decline seems to help the RF to perform better on the empirical \texttt{SNEMO15} data set in comparison to the FCNN.

\section{Filters for different redshifts}

\begin{figure}[htbp]
\centering
\subfigure{\includegraphics[trim=0 0 4 0,width=0.49\textwidth]{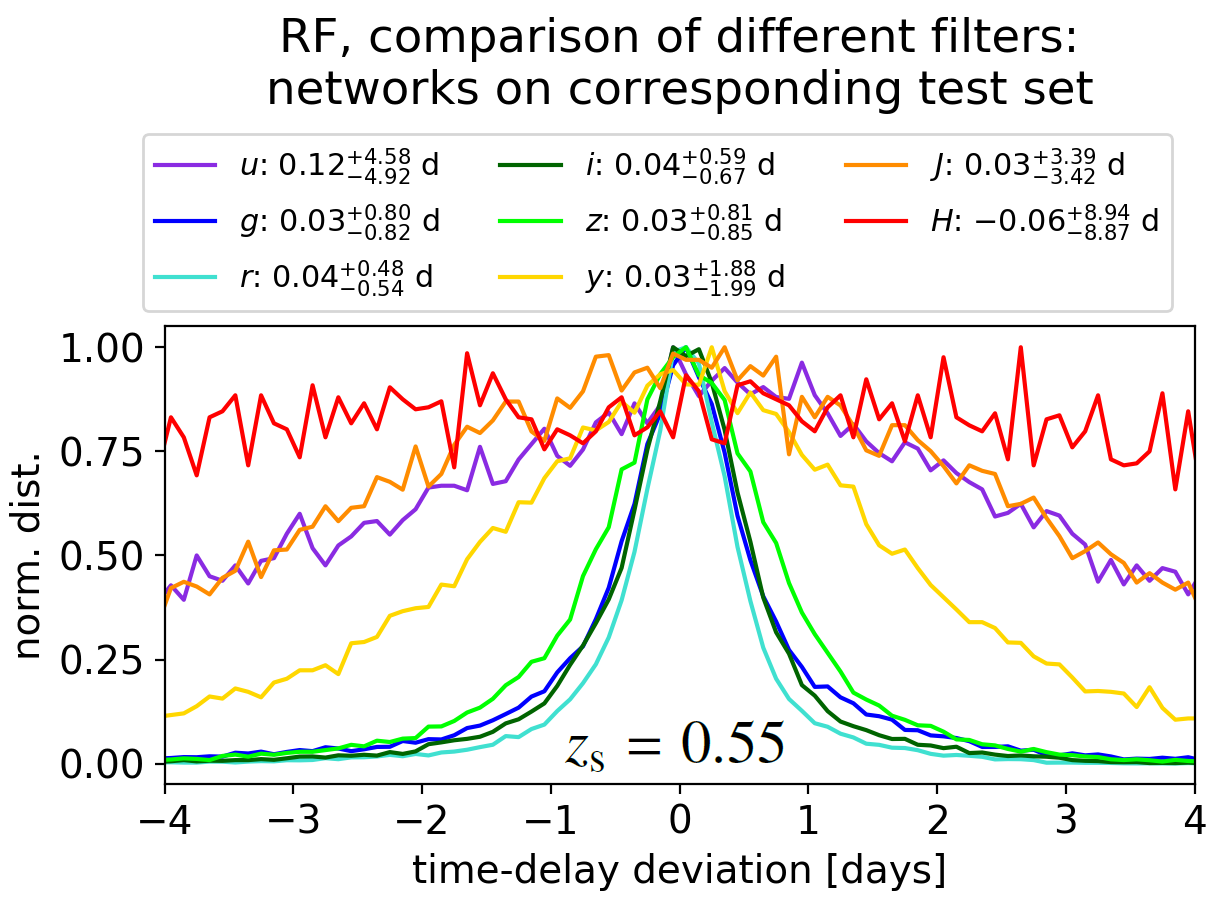}}
\subfigure{\includegraphics[trim=0 0 4 0,width=0.48\textwidth]{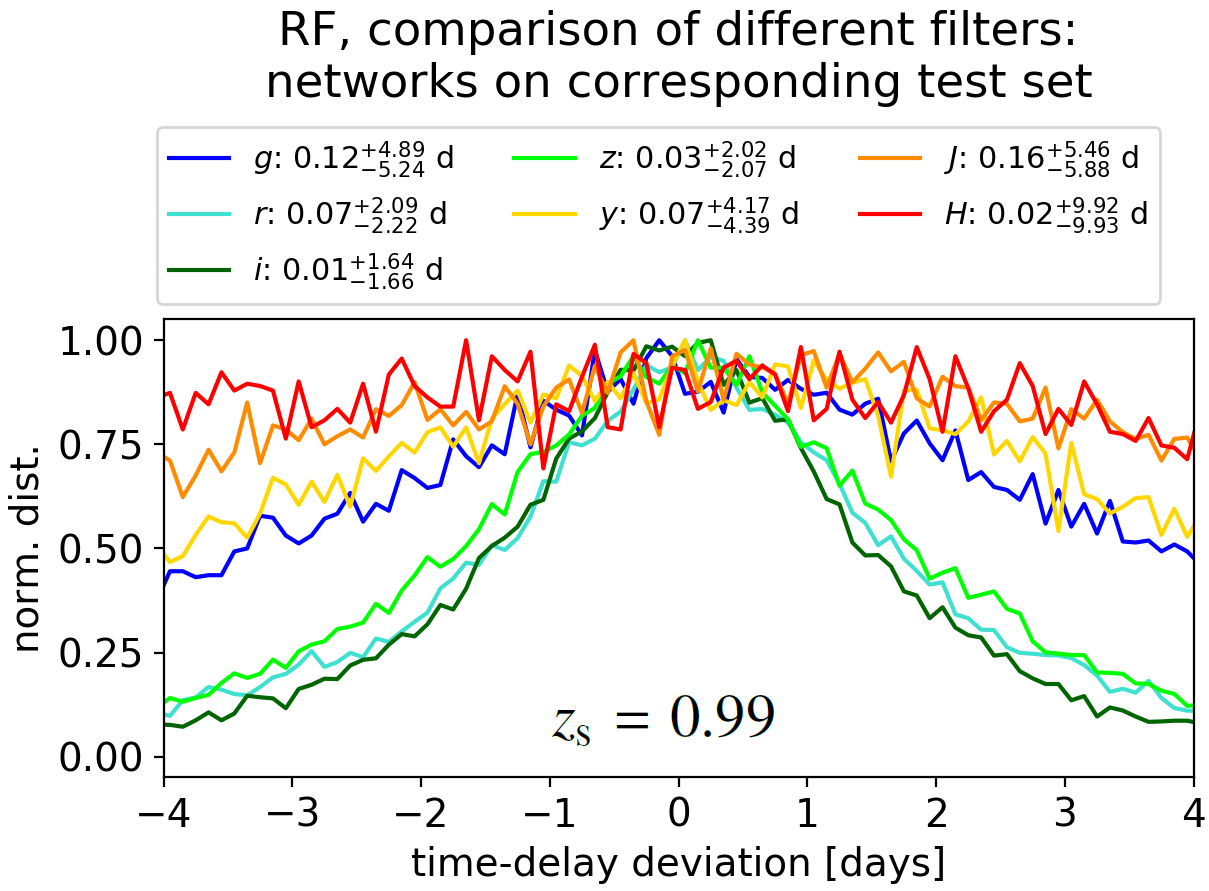}}
\caption{Both panels show different RF models, each trained on a data set from a single band (as indicated in the legend) and evaluated on all samples from the corresponding test set. The left panel shows the case for $\sourcez = 0.55,$ and the right panel shows $\sourcez = 0.99$, without the $u$ band, which is already too faint for such a source redshift.}
\label{fig: appendix filters for different redshifts}
\end{figure}

Figure \ref{fig: appendix filters for different redshifts} shows the performance of the RF on different filters for $\sourcez = 0.55$ and $\sourcez = 0.99$.

\section{Correlation plots}
\label{sec:Appendix correlation plots}

\begin{figure*}[htbp]
\centering
\subfigure{\includegraphics[trim=4 4 4 16,width=0.95\textwidth]{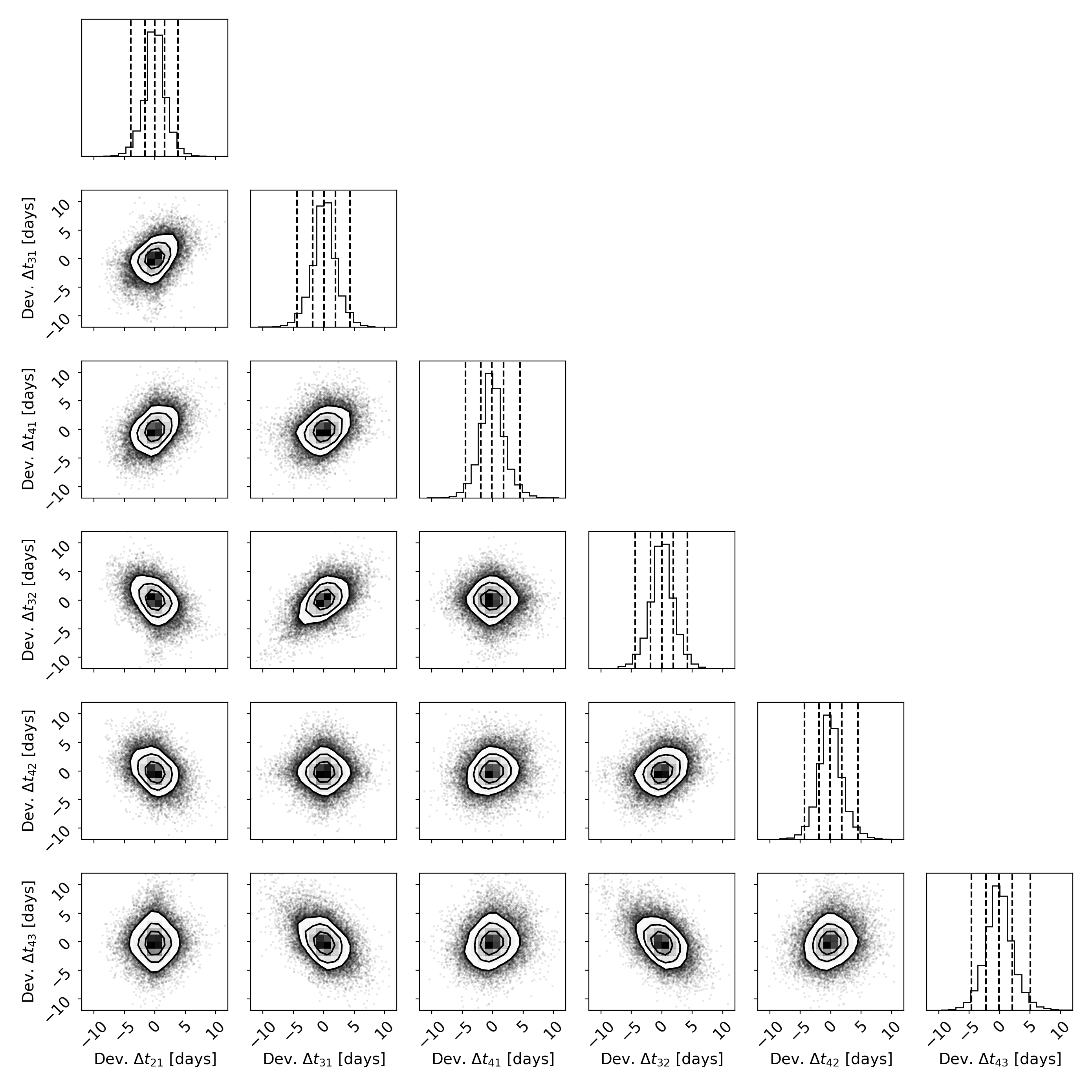}}
\caption{Correlation plots using separate RF models per pair of images for the LSN Ia quad system shown in Fig. \ref{fig: quad LSN Ia mock}. The contour plot shows the $1\sigma$, $2\sigma$, and $3\sigma$ contours. The dashed lines in the histograms correspond to the median and the $1\sigma$ and $2\sigma$ range.}
\label{fig: correlation plot quad as double}
\end{figure*}

\begin{figure*}[htbp]
\centering
\subfigure{\includegraphics[trim=4 17 4 16,width=0.95\textwidth]{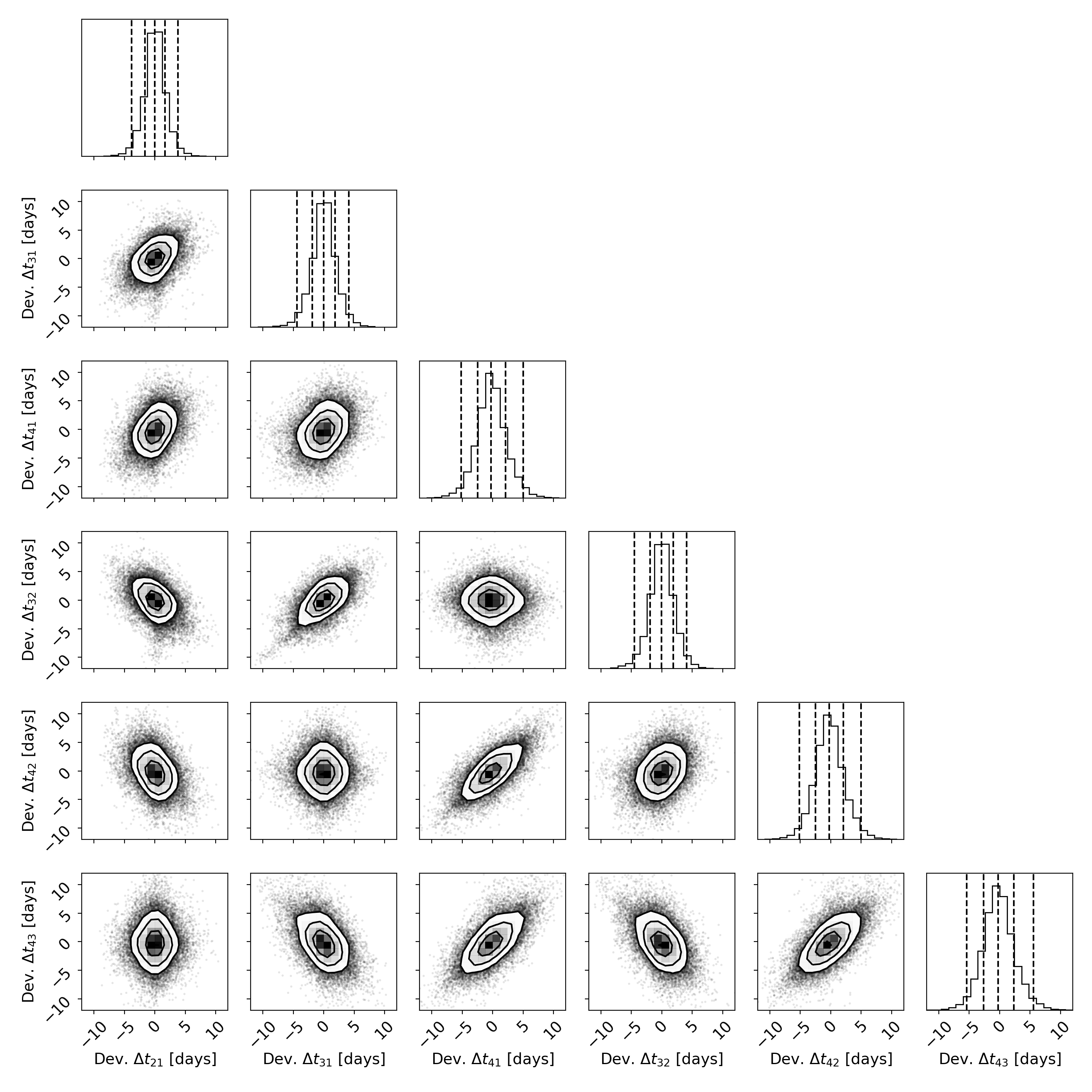}}
\caption{Correlation plots using a single RF model for all images for the LSN Ia quad system shown in Fig. \ref{fig: quad LSN Ia mock}. The contour plot shows the $1\sigma$, $2\sigma$, and $3\sigma$ contours. The dashed lines in the histograms correspond to the median and the $1\sigma$ and $2\sigma$ range.}
\label{fig: correlation plot quad}
\end{figure*}

Figures \ref{fig: correlation plot quad as double} and \ref{fig: correlation plot quad} show the correlation plots \citep{corner} for a quad system of a LSN Ia using a separate RF model per pair of images in comparison to a single RF for all images.

\end{appendix}

\end{document}

%% file: Time_delay_measurement_macro.tex
\newcommand{\bd}{\begin{displaymath}}
\newcommand{\ed}{\end{displaymath}}
\newcommand{\be}{\begin{equation}}
\newcommand{\ee}{\end{equation}}
\newcommand{\beaa}{\begin{eqnarray*}}
\newcommand{\eeaa}{\end{eqnarray*}}
\newcommand{\bea}{\begin{eqnarray}}
\newcommand{\eea}{\end{eqnarray}}

\newcommand{\erg}{\mathrm{erg}}

\newcommand{\cm}{\mathrm{cm}}

\newcommand{\sourcez}{z_{\rm s}}
\newcommand{\lensz}{z_{\rm d}}
\newcommand{\lum}{{D_\mathrm{lum}}}

\newcommand{\Rein}{R_\mathrm{Ein}}

\newcommand{\dd}{\mathrm{d}}

\DeclareSIUnit\parsec{pc}
\DeclareSIUnit\lightyear{ly}
\DeclareSIUnit\year{yr}
\DeclareSIUnit\erg{erg}
\DeclareSIUnit\ster{ster}
\DeclareSIUnit\arcsec{arcsec}
\DeclareSIUnit\rad{rad}
\DeclareSIUnit\mag{mag}

\usepackage{color} 
\definecolor{gruen}{cmyk}{0.35,0.01,0.80,0.1}

\definecolor{blue}{rgb}{0.0,0.0,1.0}
\definecolor{magenta}{rgb}{1.0,0.0,1.0}
\definecolor{orange}{rgb}{1.0,0.5,0.0}